\documentclass[aps,nofootinbib,preprintnumbers,twocolumn,superscriptaddress,prd]{revtex4-1}

\usepackage{amsmath}
\usepackage{amssymb}
\usepackage{slashed}
\usepackage{graphics}
\usepackage{epsfig}
\usepackage{bbm,bm}
\usepackage{psfrag}
\usepackage{hyperref}
\usepackage{color}
\usepackage{comment}
\usepackage{amsfonts}
\usepackage{dsfont}
\usepackage{booktabs}
\usepackage[markup=nocolor]{changes}
\usepackage{slashed}

\graphicspath{{Pictures/}}
\usepackage[utf8]{inputenc}

\usepackage[english]{babel}

\newcommand{\gs}{g_\mathrm{s}}

\newcommand{\gsL}{g_{\mathrm{s}\Lambda}}
\newcommand{\I}{\mathrm{i}}

\newcommand{\Nc}{N_{\mathrm{c}}}
\newcommand{\Nf}{N_{\mathrm{f}}}
\newcommand{\SU}{\mathrm{SU}}

\newcommand{\rmF}{\mathrm{F}}
\newcommand{\rmB}{\mathrm{B}}

\newcommand{\be}{\begin{eqnarray}}
\newcommand{\de}{\partial}
\newcommand{\ee}{\end{eqnarray}}
\newcommand{\Eqref}[1]{Eq.~\eqref{#1}}
\newcommand{\Figref}[1]{Fig.~\ref{#1}}
\newcommand{\Secref}[1]{Sec.~\ref{#1}}
\newcommand{\Appref}[1]{App.~\ref{#1}}

  {\left\lbrace\begin{array}{@{}l@{}}}%
  {\end{array}\right.}

\newcommand{\widesim}[2][1.5]{
	\mathrel{\underset{#2}{\scalebox{#1}[1]{$\sim$}}}
}

\newcommand{\kap}{\hat{\kappa}}

\newcommand{\hresc}{\hat{h}}
\newcommand{\lresc}{\hat{\lambda}}

\newcommand{\HTQCD}{$\mathbbm{Z}_2$-Yukawa-QCD}

\begin{document}

\title{Asymptotic freedom in \HTQCD\ models}

\author{Holger Gies}
\email{holger.gies@uni-jena.de}
\affiliation{\mbox{\it Theoretisch-Physikalisches Institut, Friedrich-Schiller-Universit{\"a}t Jena,}
	\mbox{\it D-07743 Jena, Germany}}
\affiliation{\mbox{\it Abbe Center of Photonics, Friedrich-Schiller-Universit{\"a}t Jena,}
	\mbox{\it D-07743 Jena, Germany}}
\affiliation{Helmholtz-Institut Jena, Fr\"obelstieg 3, D-07743 Jena, Germany}

\author{Ren\'e Sondenheimer}
\email{rene.sondenheimer@uni-jena.de}
\affiliation{\mbox{\it Theoretisch-Physikalisches Institut, Friedrich-Schiller-Universit{\"a}t Jena,}
	\mbox{\it D-07743 Jena, Germany}}

\author{Alessandro Ugolotti}
\email{alessandro.ugolotti@uni-jena.de}
\affiliation{\mbox{\it Theoretisch-Physikalisches Institut, Friedrich-Schiller-Universit{\"a}t Jena,}
	\mbox{\it D-07743 Jena, Germany}}

\author{Luca Zambelli}
\email{luca.zambelli@uni-jena.de}
\affiliation{\mbox{\it Theoretisch-Physikalisches Institut, Friedrich-Schiller-Universit{\"a}t Jena,}
	\mbox{\it D-07743 Jena, Germany}}

%\pacs{}

%%%%%%%%%%%%%%%%%%%%%%%%%%%%%%%%%%%%%%%%

\begin{abstract}

  \HTQCD\ models are a minimalistic model class with a Yukawa and a
  QCD-like gauge sector that exhibits a regime with asymptotic freedom
  in all its marginal couplings in standard perturbation theory.  We 
  discover the existence of further asymptotically free trajectories for
these models by exploiting generalized boundary conditions.
We construct such trajectories as quasi-fixed points for the Higgs potential
within different approximation schemes.  We substantiate
  our findings first in an effective-field-theory approach, and obtain
  a comprehensive picture using the functional renormalization group.
  We infer the existence of scaling solutions also by means of a
  weak-Yukawa-coupling expansion in the ultraviolet. In the same
  regime, we discuss the stability of the quasi-fixed point solutions
  for large field amplitudes.  We provide further evidence for such
  asymptotically free theories by numerical studies using
  pseudo-spectral and shooting methods.

\end{abstract}

\maketitle

%%%%%%%%%%%%%%%%%%%%%%%%%%%%%%%%%%%%%%%%
\section{Introduction}
\label{sec:intro}
%%%%%%%%%%%%%%%%%%%%%%%%%%%%%%%%%%%%%%%%

Gauged Yukawa models form the backbone of our description of
elementary particle physics: they provide mechanisms for mass
generation of gauge bosons as well as for chiral fermions via the Brout-Englert-Higgs mechanism. Many
suggestions of even more fundamental theories beyond the standard
model, such as grand unification, models of dark matter,
supersymmetric models, etc., also involve the structures of gauged Yukawa
systems. A comprehensive understanding of such systems is thus clearly
indispensable.

Despite their fundamental relevance, gauged Yukawa systems can also
exhibit a genuine conceptual deficiency. Many generic models develop
Landau-pole singularities in their perturbative renormalization group
(RG) flow towards high energies, indicating that these models may not
be ultraviolet (UV) complete. If so, such models do not constitute
quantum field theories which are fully consistent at any energy
scale. Insisting on UV completeness by enforcing a UV cutoff to be
sent to infinity typically requires to send the renormalized coupling
to zero. This problem is also called \textit{triviality}.

An important class of UV-complete nontrivial theories are those
featuring asymptotic freedom \cite{Gross:1973id,Politzer:1973fx}
which allow to send the cutoff to infinity at the expense of a
vanishing bare coupling while keeping the renormalized coupling at a
finite value. In fact, a conventional perturbative analysis
\cite{Gross:1973ju,Cheng:1973nv,Gross:1974cs,Politzer:1974fr,Chang:1974bv,Chang:1978nu,Fradkin:1975yt,Salam:1978dk,Bais:1978fv,Salam:1980ss,Callaway:1988ya}
is capable of revealing the existence of asymptotically free gauged Yukawa models,
and allows a classification in terms of their matter content and
corresponding representations. Recent studies of aspects of such
models \cite{Giudice:2014tma,Holdom:2014hla,Hansen:2017pwe} 
and constructions of phenomenologically acceptable models
\cite{Hetzel:2015bla,Pelaggi:2015kna,Pica:2016krb,Molgaard:2016bqf,Heikinheimo:2017nth}
have been performed; however, a unique route to an unequivocal model
appears not obvious. Phenomenological constraints on the gauge and
matter side typically require an appropriately designed scalar sector,
as UV Landau poles often show up in the Higgs self-coupling.

The standard model is, in fact, not asymptotically free because of the
perturbative Landau pole singularity in the U(1) gauge sector. Still,
all other gauge couplings as well as the dominant top-Yukawa coupling
and the Higgs self-coupling decrease towards higher
energies. In fact, the value of the Higgs boson mass and the top quark
mass are \textit{near-critical} \cite{Buttazzo:2013uya} in the sense
that the perturbative potential approaches flatness towards the
UV. Whereas a substantial amount of effort has been devoted to clarify
whether the potential is exactly critical or overcritical (metastable
and long-lived) in recent years
\cite{Buttazzo:2013uya,Bednyakov:2015sca,DiLuzio:2015iua,Andreassen:2017rzq},
a conclusive answer depends on the precise value of the strong
coupling and the top Yukawa coupling
\cite{Alekhin:2012py,Bezrukov:2014ina} as well as on the details of
the microscopic higher-order interactions
\cite{Gies:2013fua,Branchina:2013jra,Hegde:2013mks,Gies:2014xha,Eichhorn:2015kea,Chu:2015nha,Chu:2015ula,Akerlund:2015fya,Sondenheimer:2017jin}. 
In summary, we interpret the present data as being compatible with the
critical case of the Higgs interaction potential approaching flatness
towards the UV. This viewpoint is also a common ground for the
search for conformal extensions of the standard model
\cite{Holthausen:2013ota,Helmboldt:2016mpi,Ahriche:2016ixu,Shaposhnikov:2018xkv}.

For the present work, this viewpoint serves as a strong motivation to
study asymptotically free gauged Yukawa systems. Whereas perturbation
theory seems ideally suited for this, conventionally made implicit
assumptions may reduce the set of asymptotically free RG trajectories
visible to perturbation theory. In fact, new asymptotically free
trajectories in gauged-Higgs models have been discovered
with the aid of generalized boundary
conditions imposed on the renormalized action \cite{Gies:2015lia,Gies:2016kkk}. 
This result has also
been astonishing as it was obtained in a class of models which does
not exhibit asymptotic freedom in naive perturbation theory. Still,
the existence of these new trajectories has been confirmed by
weak-coupling approximations, effective-field-theory approaches,
large-$N$ methods, as well as more comprehensively with the functional
RG \cite{Gies:2016kkk}.

As such dramatic conclusions about the existence of new UV-complete
theories requires substantiation and confirmation, the purpose of this
work is to study the emergence of these new RG trajectories in a model
that also exhibits asymptotic freedom already in standard perturbation
theory. This allows to understand the novel features of the RG
trajectories in greater detail. For this, we use the simplest gauged
Yukawa system that exhibits asymptotic freedom perturbatively, it
consists of a QCD-like matter sector with nonabelian SU($\Nc$) gauge
symmetry Yukawa-coupled to a single real scalar field. This \HTQCD\
model can be viewed as a subset of the standard model
\cite{Eichhorn:2015kea,Reichert:2017puo}, with the Yukawa sector
representing the Higgs boson and the top quark. 
In this model, the existence of asymptotically free trajectories has
already been known since the seminal work of Cheng, Eichten, and Li 
\cite{Cheng:1973nv} based on standard perturbation theory. 

In the present work, we discover the existence of new asymptotically
free trajectories in addition to the standard perturbative
solution. For this, we follow the strategy of
\cite{Gies:2015lia,Gies:2016kkk} using effective-field-theory methods
and the functional RG in order to get a handle on the global
properties of the Higgs potential. We generalize the approach to an
inclusion of a fermionic sector and also identify a new approximation
technique ($\phi^4$-dominance) that allows to get deeper analytical
insight into the functional flow equations.

While the existence of new asymptotically free trajectories as well as
some of their properties are reminiscent to the conclusions already
found for the gauged-Higgs models \cite{Gies:2015lia,Gies:2016kkk}, we
also find some interesting differences. Again, the class of new
solutions has free parameters, such as a field- or coupling-rescaling
exponent and the location of the (rescaled) minimum of the potential
during the approach to the UV. For the present \HTQCD\ model, we find
that the exponent is more tightly constraint by the requirement of a
globally stable potential. Also the rescaled potential minimum has to
remain nonzero towards the UV, exemplifying the fact that the model
develops a non-trivial UV structure which is not visible in the deep
Euclidean region (DER). The present work thus pays special attention
to the difference between working in the DER, as is often implicitly
done in standard perturbation theory, and a more general analysis.

As our methods can address the global behavior of the potential, our
work also adds new knowledge to the results known from standard
perturbation theory: for the asymptotically free Cheng-Eichten-Li
solution, we demonstrate that the potential is and remains globally
stable when running the RG towards the UV; an analytic approximation
of the potential can be given in terms of hypergeometric functions.

In \Secref{sec:perturbativeoneloop}, we review the standard analysis 
of asymptotic freedom for perturbatively renormalizable \HTQCD\ models,
for a generic number of colors and fermion flavors.
We then specify our analysis to three colors and six flavors, to get closer
to the standard model and only in \Secref{sec:weakhexpansion}, while summarizing most of our findings,
we will generalize them to an arbitrary number of colors.
In \Secref{sec:FRG}, we present the functional renormalization group (FRG) approach by which we derive the RG flow equations for our model.
In \Secref{sec:effectiveFTdeepEuclidean} and \Secref{sec:effectiveFTthresholds}, we generalize the treatment of \Secref{sec:perturbativeoneloop} and include perturbatively 
nonrenormalizable Higgs self-interactions by polynomially truncating the FRG
equations, as in effective field theory (EFT) approaches,
within and beyond the deep Euclidean region.
In the subsequent sections we then address the task of solving the FRG
equation for a generic scalar potential.
In \Secref{sec:EFT_for_f}, we construct functional approximations of
asymptotically free solutions by inspecting a regime where the scalar fluctuations are dominated by a quartic interaction.
Another description is then obtained from the expansion in powers of the weak Yukawa coupling in \Secref{sec:weakhexpansion}.
Finally in \Secref{sec:numeric_sol}, we substantiate our analytical results by using numerical tools, in particular pseudo-spectral and shooting methods.
Conclusions are presented in \Secref{sec:conclusions}.

%%%%%%%%%%%%%%%%%%%%%%%%%%%%%%%%%%%%%%%%
\section{Asymptotic freedom within perturbative renormalizability}
\label{sec:perturbativeoneloop}
%%%%%%%%%%%%%%%%%%%%%%%%%%%%%%%%%%%%%%%%

In the present work, we focus on a Yukawa model containing a real scalar field $\phi$ and a Dirac fermion $\psi$ which is in the fundamental representation of an $\SU(\Nc)$ gauge group. 
This can be viewed as a toy model for the standard-model subsector retaining only the Higgs, the top quark, and the 
gluon degrees of freedom for $\Nc=3$.
Its gauge-fixed classical Euclidean action reads
\begin{align}
	S &=\int_x \left [ \frac{1}{2}\de_\mu\phi\de^\mu\phi + \frac{\bar m}{2} \phi^2 + \frac{\bar\lambda}{8}\phi^4
	 +\bar\psi \I\slashed{D} \psi + \frac{\I \bar h}{\sqrt 2}\phi\bar\psi\psi \right. \notag\\
	 &\quad \left.
	 +\frac{1}{4}F^i_{\mu\nu}F^{i\mu\nu}+\frac{1}{2{\alpha }}(\de_\mu A_i^\mu)^2+\bar\eta^i\de^\mu\nabla^{ij}_\mu\eta^j\right].
	\label{eq:Sclassical}
\end{align}
Note that this model exhibits a discrete chiral symmetry mimicking the electroweak symmetry of the standard-model Higgs sector such that a mass term for the fermion is forbidden. 
The top quark is coupled to the gluons through the covariant derivative $D_\mu = \de_\mu + \I \bar{g}_{\mathrm{s}} A^i_\mu\tau^i$, with $\tau^i$ the generators of the $\mathrm{su}(\Nc)$ Lie algebra, and to the Higgs field via the Yukawa coupling $\bar{h}$.
The field strength tensor for the $\mathrm{SU}(\Nc)$ gauge bosons $A^i_\mu$ is given by $F^i_{\mu\nu}=\de_\mu A^i_\nu - \de_\nu A^i_\mu - \bar{g}_\mathrm{s} f^{ijk}A^j_\mu A^k_\nu$ and $\nabla^{ij}_\mu=\delta^{ij}\de_\mu + \bar{g}_\mathrm{s} f^{ijk}A^k_\mu$ is the covariant derivative in the adjoint representation.
We adopt a Lorenz gauge with an arbitrary parameter $\alpha$ in the computation
of the RG equations. We will take the Landau gauge limit ${\alpha }\to 0$  
as far as the analysis of asymptotically free (AF) solutions is concerned, also because the Landau gauge is a fixed point of the RG flow of the gauge-fixing parameter \cite{Ellwanger:1995qf,Litim:1998qi}.
The gauge fixing is complemented by the use of Faddeev-Popov ghost fields $\eta^i$ and $\bar{\eta}^i$.

Let us first review the standard analysis of this model at 
one loop, considering only the perturbatively renormalizable couplings \cite{Cheng:1973nv}.
The latter are the scalar mass $\bar m$, the Higgs self-interaction $\bar\lambda$, 
the Yukawa coupling $\bar h$ and the strong gauge coupling $\bar{g}_{\mathrm{s}}$.
In particular, we address the UV behavior of this model, and look for 
totally AF trajectories. To this end, one focuses on the RG equations
for the renormalized dimensionless couplings $\gs$, $h$, $m$, and $\lambda$.
Their definition in terms of the bare couplings and wave function
renormalizations is the usual one, which we postpone to \Secref{sec:FRG} for the moment.

As the scalar field is not charged under the gauge group, the beta function of $\gs$ reads~\cite{Gross:1973id}
\begin{align}
	\de_t \gs^2=\eta_A \gs^{2}, \quad\quad \eta_A= - \frac{\gs^2}{8\pi^2} \left(\frac{11}{3}\Nc - \frac{2}{3} \Nf \right),
	\label{eq:gs-oneloop}
\end{align}
where we have allowed for in total $\Nf$ Dirac fermions in the fundamental representation. This slightly generalizes \Eqref{eq:Sclassical}, where
we have displayed only one Dirac field. In fact, we focus in this work
on the case, where only one flavor is coupled to the scalar field via
a Yukawa interaction.  This is motivated by the fact that the
top-Yukawa coupling plays a dominant role in the RG running of the
Higgs potential and all other Yukawa couplings are negligibly
small. Allowing for the presence of further Dirac fermions charged
under SU($\Nc$) as in \Eqref{eq:gs-oneloop} does not modify the
Yukawa structure.  In the present section, we retain generic $\Nc$ and
$\Nf$, while the following sections will specifically address $\Nc =
3$ and $\Nf=6$, to mimic the standard model.  In the latter case, the
one-loop $\beta$ function for $\gs$ is negative and therefore the
strong coupling is AF, i.e., $\gs^2 \to 0$ in the UV limit.

\begin{figure}[t!]
	%	\begin{center}
	\includegraphics[width=0.49\columnwidth]{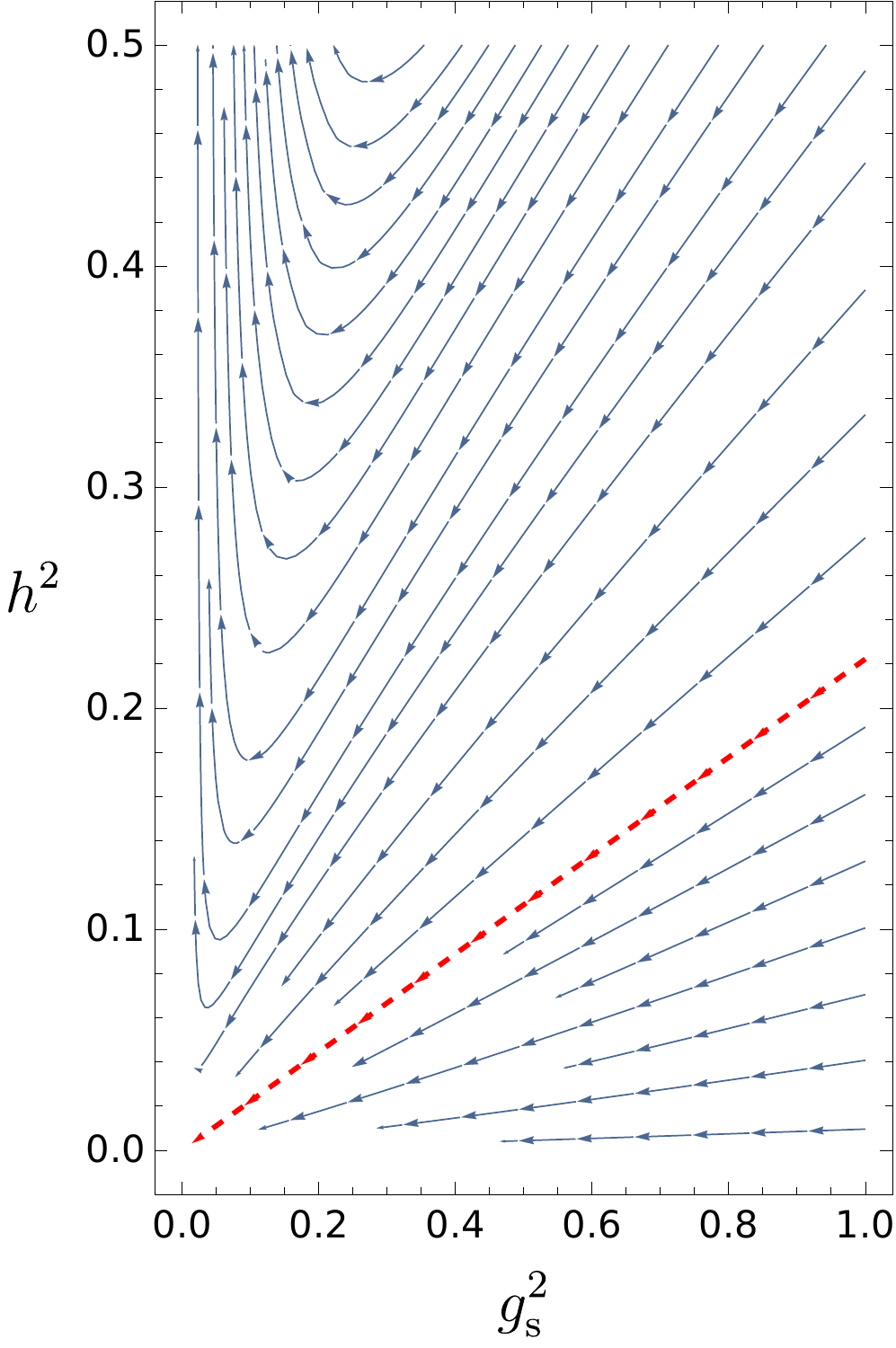}\hfill
	\includegraphics[width=0.49\columnwidth]{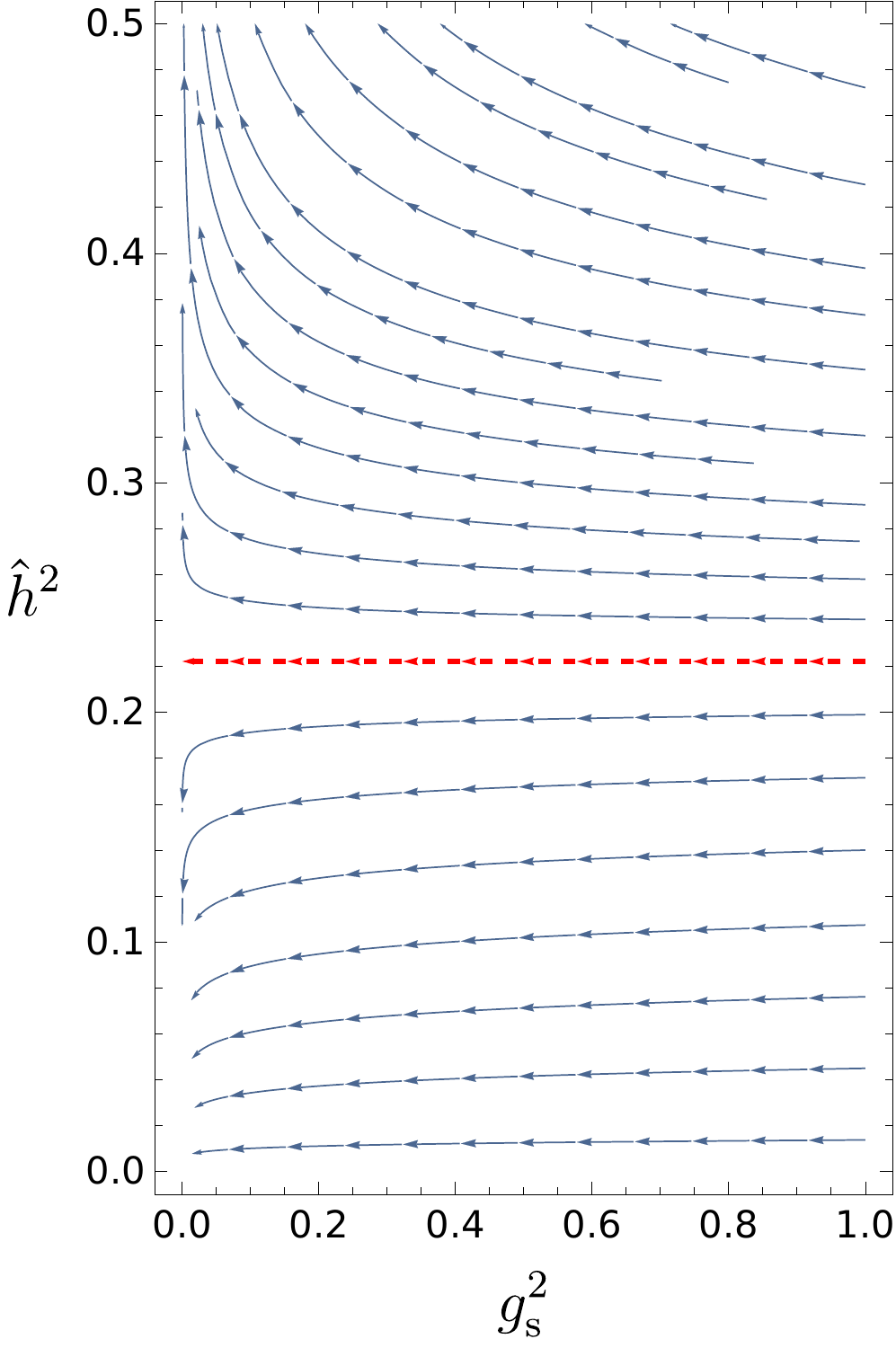}
	\caption{\emph{Left Panel}: the one-loop RG flow of the perturbatively renormalizable model
		projected on the plane of the Yukawa coupling $h^2$ and the strong gauge coupling $\gs^2$. The color and flavor numbers are fixed to $\Nc=3$ and $\Nf=6$.
		\emph{Right panel}: the same flow plotted in terms of the rescaled Yukawa coupling $\hresc^2$ defined in \Eqref{eq:xihdef}. The UV repulsive QFP is highlighted by a red dashed line that corresponds to the solution of \Eqref{eq:xihFPSM}.}
	\label{fig:StreamPloth2g2}
	%	\end{center}
\end{figure}

The RG flow equation for the Yukawa coupling $h^2$ in this model is
\begin{align}
	\de_t h^2=\frac{h^2}{16\pi^2}\left[ (3+2\Nc) h^2 - 6 \frac{\Nc^{2}-1}{\Nc} \gs^2 \right].		
	\label{eq:hsquaredot}
\end{align}
The latter two equations entail that AF trajectories exist in the
$(\gs^2,h^2)$ plane, as it is visible in the left panel of
\Figref{fig:StreamPloth2g2}, where the RG flow is represented
with 
arrows pointing towards the UV.  The dashed red line
highlights a special AF trajectory, along which $h^2$ exhibits an
asymptotic scaling proportional to $\gs^2$.  This behavior is best
characterized in terms of the rescaled coupling
\be
 \hresc^2=\frac{h^2}{\gs^2} \label{eq:xihdef}.
\ee
When this ratio at some initialization scale 
 takes the particular value 
\be
\hresc_{*}^{2}=\frac{1}{3+2\Nc}\left\{\frac{4}{3}\left(\Nf-\Nc\right)-\frac{6}{\Nc}\right\},
\label{eq:xihFP}
\ee
it is frozen at any RG time.
Indeed the $\beta$ function of $\hresc^{2}$ reads
\begin{align}
	\de_t \hresc^{2} = \frac{3+2\Nc}{16\pi^2}\gs^2\hresc^{2}\left(\hresc^{2} -  \hresc_{*}^{2}\right),
	\label{eq:betaXih}
\end{align}
and it has only one nontrivial zero at $\hresc^2=\hresc_{*}^2$ for
$\gs^2\neq0$.  We observe that this AF trajectory exists within a
finite window for $\Nf$ at fixed $\Nc$.  The
upper bound of the window is given by the requirement that the strong coupling
constant stays AF which is essential for the considered mechanism.
Beyond that upper bound, gauged-Yukawa models can still be UV complete 
through the mechanism of asymptotic safety, provided they feature a suitable 
matter content \cite{Litim:2014uca,Bond:2016dvk,Codello:2016muj,Molgaard:2016bqf,Bajc:2016efj}.
The lower bound can be obtained from Eq.~\eqref{eq:xihFP} by demanding
$\hresc_{*}^{2}>0$ such that $h^{2}>0$ to preserve unitarity, or
reflection positivity in Euclidean signature.  Thus,
we obtain 
\be \Nc + \frac{9}{2\Nc} < \Nf < \frac{11}{2}\Nc.
\label{eq:AFyukawawindow}
\ee
The standard-model case with $\Nc=3$ and $\Nf=6$
is inside this window, resulting in a fixed point at 
\begin{align}
 %\hresc^{2}
  \hresc_{*}^{2}=\frac{2}{9}.
 \label{eq:xihFPSM}
\end{align}
A partial fixed point for a ratio of AF couplings has been called \emph{quasi-fixed point} (QFP)
in Ref.~\cite{Gies:2016kkk}. It is a defining condition for AF scaling solutions and a
useful tool to search for such trajectories~\cite{Gross:1973ju,Chang:1974bv,Callaway:1988ya,Giudice:2014tma}.

The fixed-point nature of \Eqref{eq:xihFPSM} and its stability
properties are best appreciated in the right panel of
\Figref{fig:StreamPloth2g2}, where the QFP
corresponds again to the dashed red trajectory.
Using the flow in theory space in terms of $\hresc^{2}$,
this trajectory classifies as UV unstable. UV-complete trajectories hence have to emanate from the QFP. In turn, these trajectories are IR attractive, hence the low-energy behavior is governed by the QFP, enhancing the predictive power of the model.

From another perspective, the AF trajectory defined by
\Eqref{eq:xihFPSM} can be viewed as an upper bound on the ratio of the
Yukawa coupling and the gauge coupling at some initializing scale. For
$\hresc^{2}>\hresc_{*}^{2}$, asymptotic freedom is lost and the Yukawa coupling hits a
Landau pole at a finite RG time towards the UV. The Yukawa coupling
becomes AF only for $\hresc^{2} \leq \hresc_{*}^{2}$. Throughout the
main text of this work, we will concentrate on the implications of the
RG flow for the particular ratio defined by this upper bound where the
flow of the Yukawa coupling is locked to the running of $\gs$. For
$\hresc^{2} < \hresc_{*}^{2}$, the Yukawa coupling is driven faster
than the gauge coupling towards the Gau\ss ian fixed point for high
energies. These scaling solutions are sketched in
App.~\ref{app:moreAFYukawa}.

\begin{figure}[t!]
	\includegraphics[width=0.49\columnwidth]{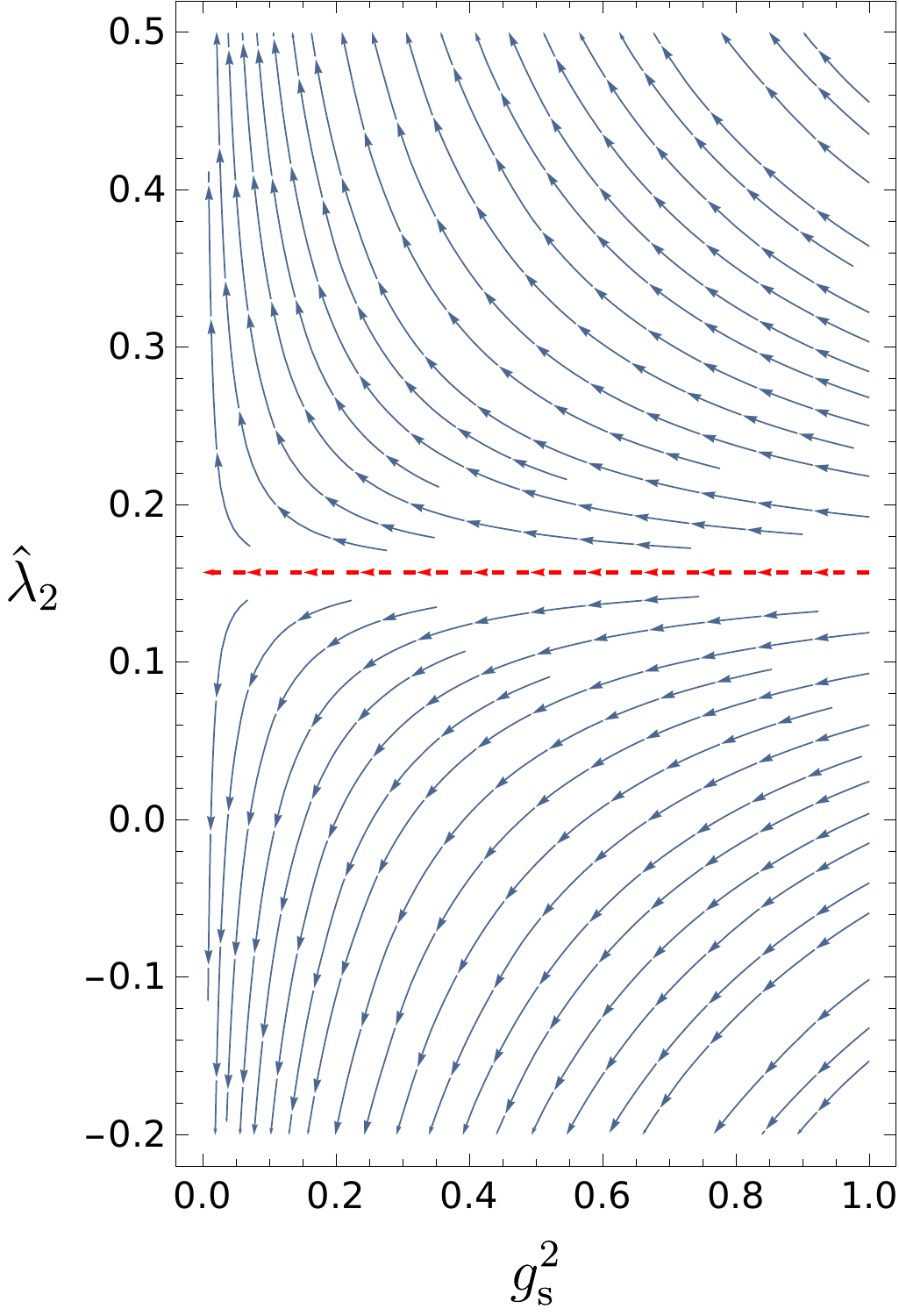}\hfill
	\includegraphics[width=0.49\columnwidth]{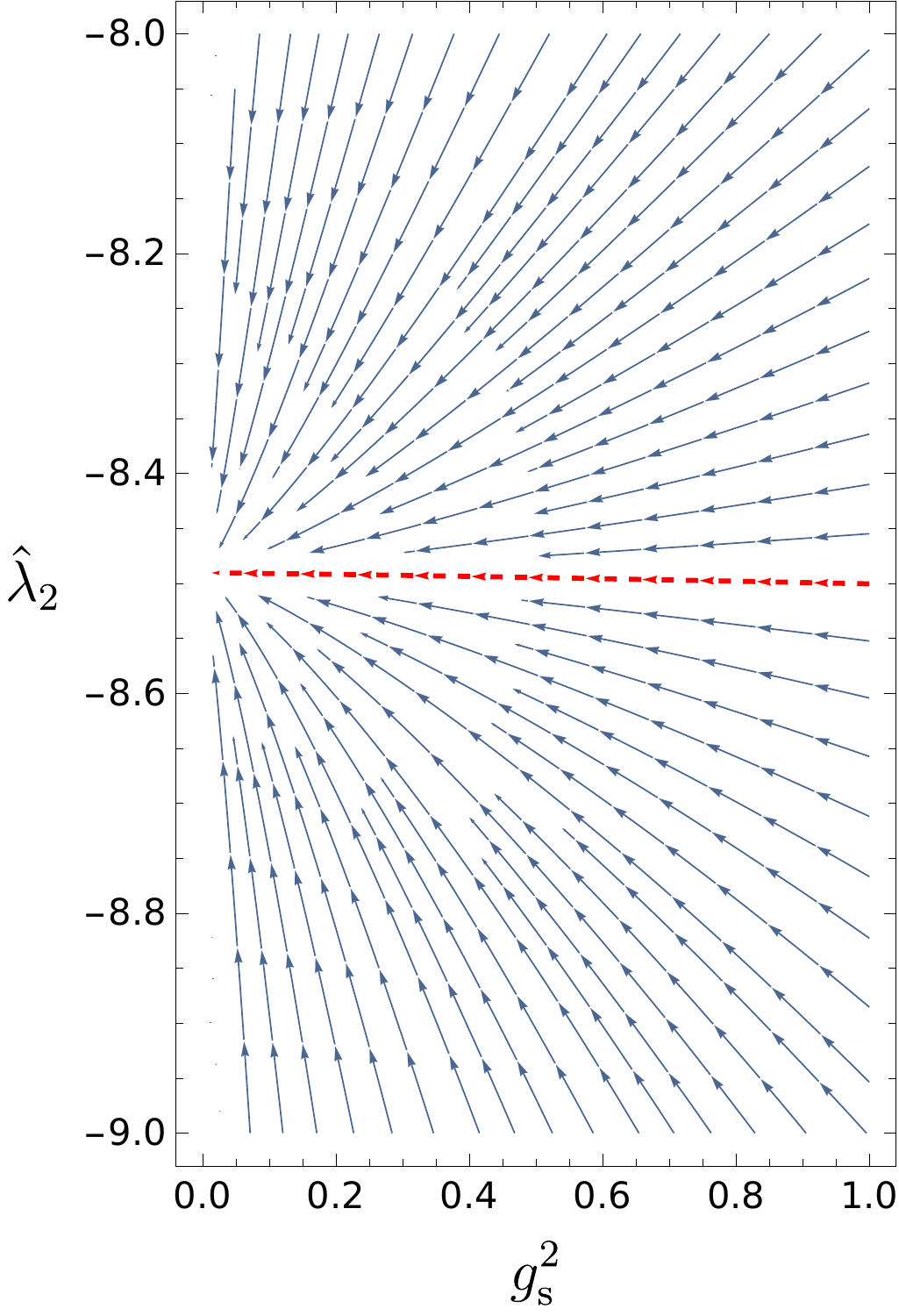}
	\caption{The one-loop RG flow of the perturbatively renormalizable model
		projected on the plane of the rescaled Higgs quartic coupling $\lresc_2$ (for $P=1/2$ as defined in \Eqref{eq:L2def}) and the strong gauge coupling $\gs^2$. The color and flavor numbers are fixed to $\Nc=3$ and $\Nf=6$.
	The UV repulsive QFP $\lresc_2^+$ in the left panel and the UV attractive QFP $\lresc_2^-$ in the right panel are highlighted by red dashed lines that correspond to the solutions in \Eqref{eq:L2rootCEL}.}
	\label{fig:StreamPlotPerturbative1Loop}
	%\end{center}
\end{figure}

In order to investigate the implications for the Higgs sector, we first study the $\beta$ function for the renormalized quartic coupling at the one-loop level
\be
\de_t\lambda&=&\frac{9}{16\pi^2}\lambda^2-\frac{\Nc}{4\pi^2}h^4+2\eta_\phi \lambda, \label{lambdadot}\\
\eta_\phi&=&\frac{\Nc}{8\pi^2}h^2,\label{eq:etaphi_1loop}
\ee
where $\eta_{\phi}$ is the anomalous dimension of the scalar field. 
We would like to emphasize at this point that we restrict the discussion to the \emph{deep Euclidean region} (DER) here, where all the masses are negligible compared to the RG scale. This implies in particular that any threshold effect given by the mass parameter $m$ of the scalar field is neglected. 
In case the system is in the symmetry-broken regime, 
effects from a nonvanishing vacuum expectation value  on the properties of the top quark are also ignored for the moment,
as they would alter the beta functions for the Yukawa coupling and the gauge coupling as well.

The $\beta$ function for the quartic coupling is a parabola with two
roots that are proportional to $h^2$.  As before, we classify AF
trajectories by a QFP condition for a suitable ratio 
\be
\lresc_{2}=\frac{\lambda}{h^{4P}}, \quad P>0, \label{eq:L2def} 
\ee
where the power $P$ is determined by the requirement that $\lresc_{2}$
achieves a finite positive value in the UV.  The flow equation for
this rescaled Higgs coupling then receives contributions from the
$\beta$ function of $h^2$.  As already stated,
we focus on the AF trajectories with
$\hresc^2=\hresc_{*}^2$.  
In this case it is convenient to define an
anomalous dimension for the Yukawa coupling by \be \eta_{h^2} \equiv
\left. -\frac{\partial_t h^2}{h^2}\right|_{h^2=\gs^2
  \hresc_{*}^{2}}=-\eta_A,
 \label{eq:defetah2}
 \ee
 which is related to the anomalous dimension of the gauge field as $h^{2} \sim \gs^{2}$ for this specific trajectory. 
 Moreover, it is useful to introduce two rescaled anomalous dimensions, by factoring out the Yukawa coupling
 \be
 \hat{\eta}_{h^2}&=&\frac{\eta_{h^2}}{h^2}
 =\frac{1}{8\pi^2\, \hresc_{*}^{2}}\left(\frac{11}{3}\Nc-\frac{2}{3}\Nf\right)\, ,
 \label{eq:etahath2}\\
 \hat{\eta}_{\phi}&=&\frac{\eta_{\phi}}{h^2}
 =\frac{\Nc}{8\pi^2}.
  \label{eq:etahatphi}
 \ee
 It turns out that the only possible QFP
 occurs at $P=1/2$, as suggested by the scaling of the two roots of
 \Eqref{lambdadot}.  In this case the $\beta$
 function of the rescaled Higgs coupling reads
 \be
 \de_t \lresc_{2}=h^2\!\left(\frac{9}{16\pi^2} \lresc_{2}^2
 +\left(2\hat{\eta}_\phi+\hat{\eta}_{h^2}\right) \lresc_{2}
 -\frac{\Nc}{4\pi^2}\right).\ \ 
 \label{eq:simplestbetaL_2}
 \ee
In fact for $P=1/2$ the QFP equation $\partial_t \lresc_{2}=0$
admits two real roots, one positive and one negative.
For instance, choosing $\Nc=3$ and $\Nf=6$ results in
\be
	\lresc_{2}^{\pm}=\frac{1}{6}\left(-25\pm\sqrt{673}\right).
	\label{eq:L2rootCEL}
\ee
The $\beta$ function for $\lresc_{2}$ is a convex parabola, therefore
the positive (negative) root corresponds to a UV repulsive
(attractive) QFP.  The phase diagram is depicted in
\Figref{fig:StreamPlotPerturbative1Loop}.  Exactly on top of
$\lresc_{2}^+$ the Yukawa coupling drives the Higgs coupling to zero
towards the UV.  For an initial condition such that the rescaled
scalar coupling is smaller than $\lresc_{2}^+$, $\lresc_{2}$ is
attracted in the UV towards the negative root and the perturbative
potential appears to become unstable.  For an initial value bigger
than $\lresc_{2}^+$, the scalar coupling hits a Landau pole in the 
UV. Hence, the requirement of a stable and UV-complete theory enforces
$\lresc_2=\lresc_{2}^+$. As for the Yukawa coupling, this trajectory is
IR attractive, hence the low-energy behavior is governed by the QFP
$\lresc_{2}^+$. Thus, the theory exhibits a higher degree of predictivity.

In the remaining part of this paper, we restrict ourselves to an
asymptotic UV running of the Yukawa coupling described by
\Eqref{eq:xihdef} and \Eqref{eq:xihFP}.  We will refer to the AF
solution described by \Eqref{eq:xihFP} and by the positive root in
\Eqref{eq:L2rootCEL} as the \emph{Cheng--Eichten--Li} (CEL) solution,
since it was first described in Ref.~\cite{Cheng:1973nv}. The further
AF solutions with $\hresc^2<\hresc_{*}^2$ have also already been
discussed in Ref.~\cite{Cheng:1973nv} as well as in later analyses
\cite{Chang:1974bv,Giudice:2014tma}; for completeness, we review them
in \Appref{app:moreAFYukawa}. For the remainder of the paper, we
consider the asymptotic UV running of the Yukawa coupling of
Eqs.~\eqref{eq:xihdef} and \eqref{eq:xihFP}, because it is most
predictive: whereas classically the gauge coupling $\gs$, the Yukawa
coupling $h$ and the scalar self-interaction $\lambda$ are
independent, our AF trajectory locks the running of $h$ and $\lambda$
to that of $\gs$. Physically, this implies that the mass of the
fermion (top quark) as well as that of the Higgs boson will be
determined in terms of the initial conditions for the gauge sector and
the scalar mass-like parameter, i.e., the Fermi scale. This maximally
predictive point in theory space is also called the Pendleton-Ross
point \cite{Pendleton:1980as}. Let us finally emphasize that we focus
exclusively on the UV behavior of our model class in the present
work. The low-energy behavior will be characterized by possible
top-mass generation from $\mathbbm{Z}_2$ symmetry breaking and a
QCD-like low-energy sector for the remaining fermion flavors and gauge
degrees of freedom. In models with a gauged Higgs field, a distinction
of Higgs- and QCD-like phases as well as details of the particle
spectrum might be much more intricate
\cite{Maas:2012tj,Maas:2014pba,Maas:2017xzh,Maas:2017wzi}.

The question that is left open by the preceding standard perturbative UV analysis
is as to whether the CEL solution is the only possible AF model with
the same field content and symmetries of \Eqref{eq:Sclassical}. More
specifically, can there be more AF solutions outside the family of
perturbatively renormalizable models? 
To address this possibility, we take inspiration
from the discovery that new AF trajectories
can be constructed in nonabelian Higgs models,
if functional RG equations are used
to explore the space of theories
including also couplings with negative mass dimension~\cite{Gies:2015lia,Gies:2016kkk}.
Therefore, as a first step of our investigation,
we now turn to the computation of such functional
RG equations for \HTQCD {} models.

%%%%%%%%%%%%%%%%%%%%%%%%%%%%%%%%%%%%%%%%
\section{Functional renormalization group} \label{sec:FRG}
%%%%%%%%%%%%%%%%%%%%%%%%%%%%%%%%%%%%%%%%

Since the work of Wilson, Wegner and Houghton, 
it is known that in a generic field theory one can construct functional RG
equations which are exact~\cite{Wilson:1973jj,Wegner:1972ih}.
For many purposes, the most useful form of these equations is the one, referring to the one-particle irreducible
effective action $\Gamma$,
which descends from adding a regularization kernel $R_k$ to the quadratic part of the bare action, in order to keep track of the successive inclusion of IR modes at a scale $k$.
Then the full (inverse) two-point function $\Gamma_k^{(2)}$ at this scale enters the one-loop computation, supplemented by the regulator $R_k$.
Differentiating with respect to the scale $k$ leads to the Wetterich equation~\cite{Wetterich:1992yh,Ellwanger:1993mw,Morris:1993qb,Bonini:1992vh}
\be
\de_t\Gamma_k[\Phi]=\frac{1}{2}\text{STr}\left\{\frac{\de_t R_k}{\Gamma^{(2)}_k[\Phi]+R_k}\right\}\, ,	\label{eq:Wetterich}
\ee
where $t=\log (k/k_{\mathrm{ref}})$ is the RG time with $k_{\mathrm{ref}}$ some reference scale.
Thanks to the derivative $\partial_t R_k$ in the numerator, all UV divergences are regulated as well.
The effective average action $\Gamma_k$ interpolates between a microscopic theory defined at some UV scale $\Lambda$, $\Gamma_{k=\Lambda}=S_\text{cl}$, and the effective action $\Gamma_{k=0}=\Gamma$, where all the quantum fluctuations are integrated out, see \cite{Berges:2000ew,Pawlowski:2005xe,Gies:2006wv,Delamotte:2007pf,Braun:2011pp} for reviews.

Equation~\eqref{eq:Wetterich} can be projected onto the RG flow of a specific coupling constant. 
 In addition, it is also well suited to study functional 
parametrizations of the dynamics, such as a general scalar effective potential.
These functional flow equations can then be used also outside the regime of small
field amplitudes, to address problems such as the existence of a nontrivial 
minimum or the global stability of the theory.

As we are interested in the properties of the beta functional of the scalar potential, we use
\begin{align}
	\Gamma_k&=\int_x \left[\frac{Z_\phi}{2}\de_\mu\phi\de^\mu\phi + U\left(\phi^{2}/2\right)+Z_\psi \bar\psi\I\slashed D\psi +\frac{\I \bar{h}}{\sqrt 2}\phi \bar\psi\psi\right.\notag\\
	&\quad \left. + \frac{Z_A}{4}F^i_{\mu\nu}F_i^{\mu\nu}+\frac{Z_A}{2{\alpha }}(\de_\mu A_i^\mu)^2+Z_\eta\bar\eta^i\de_\mu\nabla_{ij}^\mu\eta^j\right],
	\label{eq:Gammak}
\end{align}
as an approximation scheme for the effective average action. 
This derivative expansion has proven useful, especially in the analysis of the RG flow of the Higgs potential \cite{Gies:2013pma,Gies:2013fua,Gies:2014xha,Eichhorn:2014qka,Eichhorn:2015kea,Gies:2015lia,Jakovac:2015iqa,Jakovac:2015kka,Vacca:2015nta,Borchardt:2016xju,Gies:2016kkk,Jakovac:2017nsi,Gies:2017zwf,Gies:2017ajd,Sondenheimer:2017jin}.
The effective average potential $U$ which exhibits a discrete $\mathbb{Z}_{2}$ symmetry and the wave function renormalizations $Z_{\{\phi,\psi,A,\eta\}}$ are scale dependent, as well as the Yukawa coupling $h^2$ and the strong coupling $\gs$. 
Let us introduce a dimensionless renormalized scalar field in order to  fix the usual RG invariance of field rescalings
\be
	&2\rho = Z_\phi k^{2-d}\phi^2.	\label{eq:def_rho}
\ee
In a similar manner, also renormalized fields for the fermions and the gauge bosons might be introduced.
The dimensionless renormalized couplings read
\be
	h^2=\frac{\bar{h}^2 k^{d-4}}{Z_\phi Z_\psi^2},\quad \gs^2=\frac{\bar{g}_{\mathrm{s}}^2k^{d-4}}{Z_A}.
\ee
By plugging the ansatz for $\Gamma_k$ %\eqref{eq:Gammak}
into \Eqref{eq:Wetterich}, we can extract the flow equations for the dimensionless potential
\be
u(\rho)=k^{-d}U(Z_\phi^{-1}k^{d-2}\rho),
\ee
as well as the flow equation for the dimensionless renormalized Yukawa coupling, $\de_t h^2$. Similarly, we obtain the anomalous dimensions of the fields that are defined as
\be
	\eta_\phi=-\de_t\log Z_\phi, \quad\quad\eta_\psi=-\de_t\log Z_\psi,
\ee
encoding the running of the wave function renormalizations.

The functional flow equation for the full dimensionless renormalized potential is given by
\begin{align}
	\de_t u &= - du(\rho)+(d-2+\eta_\phi)\rho u^\prime(\rho) +2 v_d \, l_0^{(\rmB)d}(\omega,\eta_\phi) \notag\\
	&\quad - 2 v_d \Nc d_\gamma \, l_0^{(\rmF)d}(\omega_1,\eta_\psi),		
	\label{eq:betau}
\end{align}
where $v_d^{-1}=2^{d+1}\pi^{d/2}\Gamma(d/2)$ and
$\omega$ as well as $\omega_1$ are defined as
\be
\begin{split}
\omega=u^\prime(\rho)+2\rho u^{\prime\prime}(\rho), \quad
\omega_1=h^2\rho.
\label{eq:omega}
\end{split} \ee
Moreover, we have ignored field-independent contributions coming from a pure gluon or ghost loop which are irrelevant for the following investigations. 
The threshold functions $l_0^{(\rmB)d}$ and $l_0^{(\rmF)d}$ encode the nonuniversal regulator dependence of loop integrals and describe the decoupling of massive modes. 
Their general definitions as well as explicit representations for a convenient piece-wise linear regulator \cite{Litim:2000ci,Litim:2001up} to be used in the following, are listed, for instance, in Ref.~\cite{Gies:2017zwf}. 
Of course, it is straightforward to derive flow equations for particular scalar self-couplings up to an arbitrary order from this beta functional for the scalar potential. Additionally, it contains information beyond the RG evolution of polynomial approximations of the effective potential and keeps track of all relevant scales, the field amplitude as well as the RG scale. Thus, it allows to study global properties of the Higgs potential which we will discuss with regard to AF trajectories in the following.

The flow equation for the Yukawa coupling extracted from the Wetterich equation reads
\begin{align}
	\de_t h^2 &= (d-4+\eta_\phi+2\eta_\psi) h^2 + 4 v_d h^4 \, l_{11}^{(\rmF\rmB)d}(\omega_1,\omega;\eta_\psi,\eta_\phi) \notag\\
	& \quad - \frac{\Nc^2-1}{2N_c} (d+{\alpha }-1) 8 v_d \, h^2 \gs^2 \times \notag\\ 
	&\quad \times l_{11}^{(\rmF\rmB)d}(\omega_1,0;\eta_\psi,\eta_A) |_{\rho = \kappa}.	
	\label{eq:betahsquared}
\end{align}
Note, that this flow equation differs in the SSB regime from the one which was usually adopted in the literature for Yukawa models, e.g., \cite{Gies:2013fua,Gies:2014xha}. It has turned out that the running of $h$ extracted from a projection onto a field-dependent two-point function $\Gamma^{(2)}_{\bar{\psi}\psi}(\phi)$ shows better convergence upon the inclusion of higher-dimensional Yukawa interactions than the projection onto the three-point function $\Gamma^{(3)}_{\phi\bar{\psi}\psi}$ in case the system is in the SSB regime \cite{Pawlowski:2014zaa,Gies:2017zwf}.
The flow equation for the Yukawa coupling extracted from $\Gamma^{(3)}_{\phi\bar{\psi}\psi}$ can be obtained from Eq.~\eqref{eq:betahsquared} by taking a derivative with respect to $\rho$ before evaluating at $\rho = \kappa$ which coincides with flow equation $\de_t h^2$ derived in \cite{Gies:2013fua}.

Finally, the scalar and spinor anomalous dimensions read
\begin{align}
	\eta_\phi &= \frac{8 v_d\kappa}{d}  \big[3 u^{\prime\prime}(\rho)+2\kappa u^{\prime\prime\prime}(\rho)\big]^2 \, m_2^{(B)d}(\omega,\eta_\phi) \notag \\
	&\quad+\frac{4 v_d \Nc d_\gamma h^2}{d} \Big[ m_4^{(F)d}(\omega_1,\eta_\psi) \notag\\ 
	&\qquad\qquad\qquad\qquad  - h^2 \kappa \, m_2^{(F) d} (\omega_1,\eta_\psi) \Big]  \Big{|}_{\rho = \kappa},
	 \label{eq:etaphi}
\end{align}
and
\begin{align}
	\eta_\psi &= \frac{4 v_d h^2}{d} m_{12}^{(FB)d}(\omega_1,\omega;\eta_\psi,\eta_\phi)  \notag \\
	&\quad +\frac{8 v_d}{d} \frac{\Nc^2-1}{2 \Nc} \gs^2 \bigl[ (d-{\alpha }-1) m_{12}^{(FB)d}(\omega_1,0;\eta_\psi,\eta_A)  \notag  \\
	&\quad - (d-1)(1-{\alpha }) \tilde m_{11}^{(FB)d}(\omega_1,0;\eta_\psi,\eta_A) \bigr]\big{|}_{\rho = \kappa}, 
	\label{eq:etapsi}
\end{align}
with further threshold functions $m_{\dots}^{\dots}$ and $\tilde{m}_{\dots}^{\dots}$. Their arguments $\omega$ and $\omega_1$ in Eqs.~\eqref{eq:etaphi} and \eqref{eq:etapsi} are evaluated at the minimum of the potential $\kappa$, which means $\kappa=0$ in the symmetric regime and $u^\prime(\kappa)= 0$ in the SSB regime.
The precise definitions for all the threshold functions can be found in \cite{Gies:2017zwf}. For our quantitative analysis, we use the Landau gauge ${\alpha }\to0$, and a piece-wise linear regulator \cite{Litim:2000ci,Litim:2001up} for convenience.

In principle, functional flow equations can also be obtained for the gauge sector of the model.
Nevertheless as we are interested in the properties of the flow equations far above the QCD scale where $\gs$ is small, it is legitimate to treat the running of the gauge sector in a standard way. Therefore we will use the one-loop beta function for $\gs$ as shown in \Eqref{eq:gs-oneloop}.

As a matter of course, the universal one-loop coefficients of the beta function for the Yukawa as well as the quartic Higgs coupling and the one-loop expressions for the anomalous dimensions can be extracted from the flow Eqs.~\eqref{eq:betau}-\eqref{eq:etapsi}. For this purpose, 
one has to set all the 
anomalous dimensions occurring in the threshold functions to zero, but keep the anomalous dimensions entering the dimensional scaling of the renormalized couplings. The latter contribute to the perturbative one-loop flow equation via one-particle reducible graphs.
Furthermore, one has to take the limit toward the DER,
by setting
the mass parameter as well as the scalar vacuum expectation value to zero to neglect threshold effects.
Then, the anomalous dimension of the scalar field reduces to Eq.~\eqref{eq:etaphi_1loop}, and we obtain 
\begin{align}
 \eta_{\psi} = \frac{h^{2}}{32\pi^{2}} \equiv h^{2} \hat\eta_{\psi}, \label{eq:eta_psi_one-loop}
\end{align}
for the spinor anomalous dimension in the Landau gauge at one-loop order in $d=4$.
The flow equation for the Yukawa model reads in this limit 
\begin{align}
 \de_{t} h^{2} &= (\eta_{\phi} + 2\eta_{\psi})h^{2} + \frac{h^4}{8\pi^{2}} - \frac{3}{8\pi^{2}}\frac{\Nc^2-1}{N_c}  h^2 \gs^2.
\end{align}
Using the one-loop expressions for the anomalous dimensions, we obtain Eq.~\eqref{eq:hsquaredot}. 
In the rest of this paper we will drop the index $d$ from the threshold functions,
as we work in $d=4$ from now on.

The freedom to choose different regularization schemes is parametrized by the threshold functions $l,m,\dots$. This includes general mass-dependent schemes as well as mass-independent schemes as a particular limiting case. Using an EFT-like analysis, we investigate in the following whether the results in the more general mass-dependent schemes are sensitive to the assumption of working in the DER as a special case. It turns out below that the restriction to the DER is severe and legitimate only for the CEL solution. A more general class of asymptotically free solutions requires to take threshold effects into account.

%%%%%%%%%%%%%%%%%%%%%%%%%%%%%%%%%%%%%%%%
\section{Effective field theory analysis in the deep Euclidean region} \label{sec:effectiveFTdeepEuclidean}
%%%%%%%%%%%%%%%%%%%%%%%%%%%%%%%%%%%%%%%%

In the present section and \Secref{sec:effectiveFTthresholds}, we discuss a generalization of the construction outlined in \Secref{sec:perturbativeoneloop}, by including perturbatively nonrenormalizable interactions.
In adding higher-dimensional operators to \Eqref{eq:Sclassical}, we follow the EFT paradigm, but we do so only for momentum-independent scalar self-interactions. 
In fact, as will be explained in the next sections, a justification of the consistency of the new AF solutions we construct
 requires an infinite number of higher-dimensional operators, which cannot be 
 generally dealt with, unless further restrictions are imposed. 
The focus on point-like scalar self-interactions is one such additional specification, and it will be extensively discussed in the following.

Regardless of our choice to depart from a standard EFT setup, the AF solutions can be studied also within the latter.
The goal of the present section and of \Secref{sec:effectiveFTthresholds} is precisely to explain how to reveal these solutions and to properly account for some of their properties in a parameterization where a finite number of couplings with higher dimension is included.
These steps can be followed also when \emph{all} interactions up to some given dimensionality are included in the effective Lagrangian.
Still, the crucial ingredient in the construction is a treatment of the $\beta$ functions of these operators that slightly differs from the standard EFT one.
Namely, one has to treat the scale dependence of one coupling or Wilson coefficient in the EFT expansion as free.
Finally, we will show in the next sections that this additional freedom has to be present in any rigorous definition of the RG flow of the model, due to the infinite dimensionality of the theory space, and plays the role of a boundary condition in a functional representation of the quantum dynamics.

Let us start detailing the EFT-like analysis of the RG flow for the dimensionless potential.
To this end, we consider a systematic polynomial expansion of $u(\rho)$ around the actual scale-dependent flowing minimum $\kappa$, which can be either at vanishing field amplitude (SYM regime) or at some nontrivial value (SSB regime).
Assuming that the system is in the SSB regime, the potential is parametrized as 
\be
u(\rho)=\sum_{n=2}^{N_\mathrm{p}}
\frac{\lambda_n}{n!}(\rho-\kappa)^n. \label{eq:ExpansionAroundKappa}
\ee
Generically, we expect all couplings to be generated by fluctuations, i.e., $N_\mathrm{p}\to\infty$, whereas truncating the sum at some finite $N_\mathrm{p}$ corresponds to a polynomial approximation of the potential.

As we said above, in the present section we first study the DER where all mass parameters are neglected.
To implement this regime we restrict our analysis to the limit $\kappa\to 0$.
This ansatz is then plugged into \Eqref{eq:betau} such that, by setting the anomalous dimensions inside the threshold functions to zero, we recover the set of one-loop $\beta$ functions $\de_t\lambda_n$ for $n=2,\dots,N_\mathrm{p}$ in the DER.
As we are interested in constructing AF trajectories, we allow for any arbitrary scaling of the quartic coupling $\lambda_2$ with respect to the AF Yukawa coupling $h^2$, and introduce the finite ratio $\lresc_{2}$ defined in \Eqref{eq:L2def} for $\lambda=\lambda_{2}$.
Any QFP for $\lresc_{2}$ at a finite nonvanishing value of $h^2$ has the interpretation of an AF scaling solution for $\lambda_2$.
Similar arguments can be applied to the higher-order couplings $\lambda_n$, suggesting to define
\be
\lresc_{n}=\frac{\lambda_n}{h^{2P_n}}, 
\label{eq:lambdatoLn}
\ee
with $P_2=2P$, cf. \Eqref{eq:L2def}. 

Concerning the scaling of the Higgs coupling, namely the power $P$ of \Eqref{eq:L2def}, it will become clear soon that the only possibility in the DER is $P=1/2$.
In fact, since $\lresc_{3}$ and $\lresc_{4}$ contribute to the $\beta$ function of $\lresc_{2}$, $P$ cannot be fixed without fixing simultaneously all the other powers $P_n$ with $n>2$.
To simplify the discussion, we already start with the ansatz  $P=1/2$ and look for the corresponding values of $P_n$ and $\lresc_{n}$.
The flow equation for $\lresc_{3}$ then reads
\be
	\de_t \lresc_{3}=-\frac{81}{16 \pi ^2} \lresc_{2}^3 h^{2(3-P_3)} + \frac{9}{4 \pi ^2}  h^{2(3-P_3)}+2 \lresc_{3}+\mathcal{O}(h^2),	\notag
\ee
thus a QFP solution with finite $\lresc_{2}$ and $\lresc_{3}$ is possible only for $P_3=3$.
In the same way it is possible to fix the scaling of all the higher order couplings, and to conclude that
\begin{align}
	P=\frac{1}{2}\quad\text{and}\quad P_{n>2}=n.
	\label{eq:PnCEL}
\end{align}

The truncation of the polynomial expansion in \Eqref{eq:ExpansionAroundKappa} up to some integer value for $N_\mathrm{p}$ and for $\kappa=0$, provides a system of $N_\mathrm{p}$ equations in $N_\mathrm{p}$ variables when one looks at the QFP condition.
To give an example, the first four beta functions are shown here to leading order in $h^2$:
\begin{align}%N_\mathrm{p}
	\de_t \lresc_{2}&=h^2\left(\frac{9}{16\pi^2} \lresc_{2}^2+\frac{75}{16\pi^2} \lresc_{2}-\frac{3}{4\pi^2}\right)+\mathcal O(h^4)		\notag\\
	\de_t \lresc_{3}&=2  \lresc_{3}-\frac{81}{16 \pi ^2} \lresc_{2}^3+\frac{9}{4 \pi ^2}+\mathcal O(h^2)		\notag\\
	\de_t \lresc_{4}&=4 \lresc_{4}+\frac{243}{4\pi^2} \lresc_{2}^4-\frac{9}{\pi^2}+\mathcal O(h^2)		\notag\\
	\de_t \lresc_{5}&=6 \lresc_{5}-\frac{3645}{4\pi^2} \lresc_{2}^5+\frac{45}{\pi^2}+\mathcal O(h^2).		\label{SystemBetaFunctionsOneLoopDER}
\end{align}
By neglecting the subleading contributions, we have that the QFP solution for the scalar quartic coupling is $\lresc_{2}=\lresc_{2}^\pm$ as in \Eqref{eq:L2rootCEL}, and all the other higher-order couplings are functions of $\lresc_{2}$ only.
For the positive root $\lresc_{2}^+$  the sign of $\lresc_{n}$ with $n>2$ is alternating, 
whereas for the negative root $\lresc_{2}^-$ all the higher order couplings stay negative.
Furthermore, by solving numerically the system of QFP equations at the next-to-leading order in $h^2$, it is possible to see that only the positive root of $\lresc_{2}$ leads to a fully real solution for all $2<n\le N_\mathrm{p}$.

\begin{figure}[!t]
	\begin{center}
		\includegraphics[width=0.4\columnwidth]{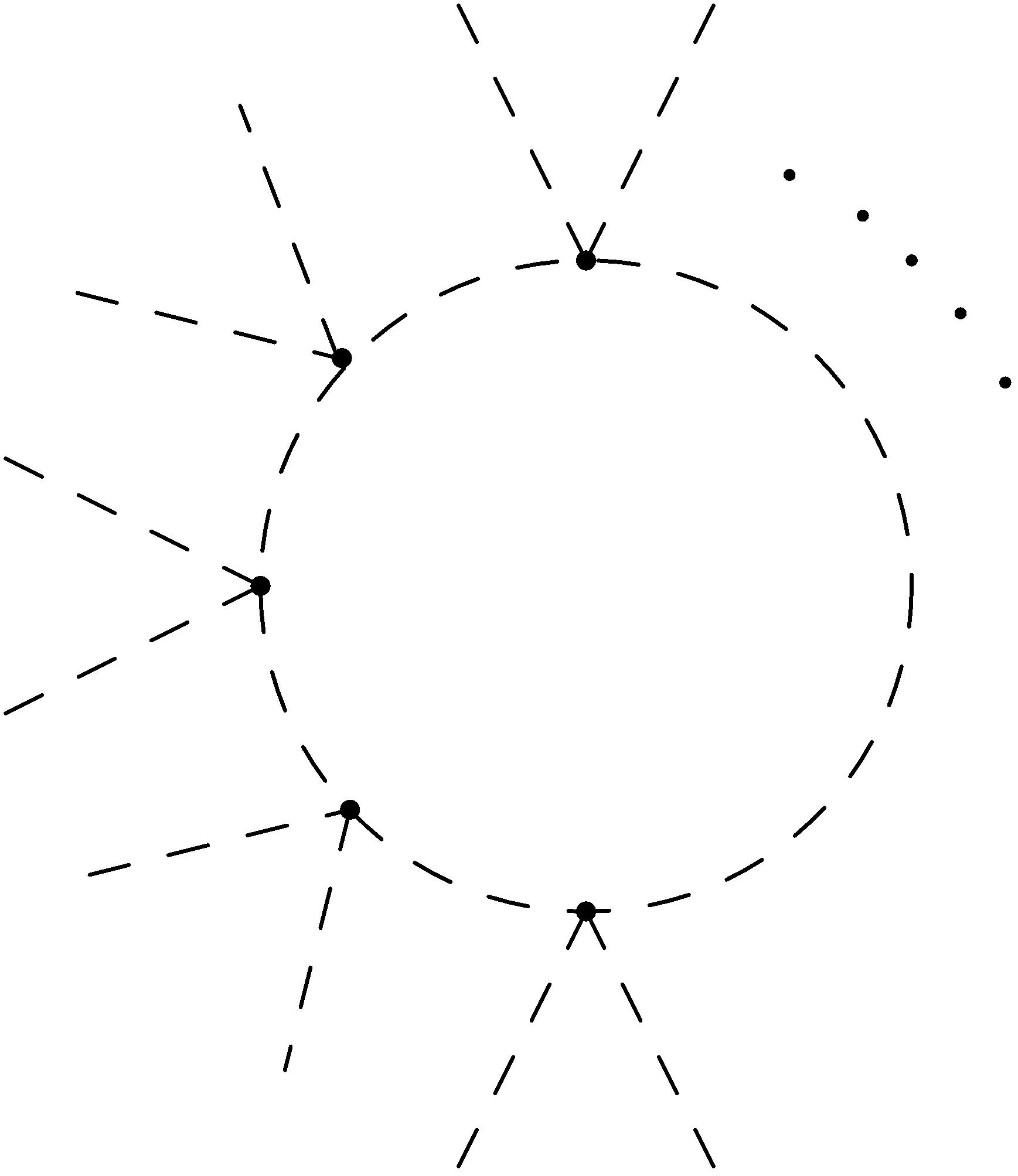}\hskip10mm
		\includegraphics[width=0.4\columnwidth]{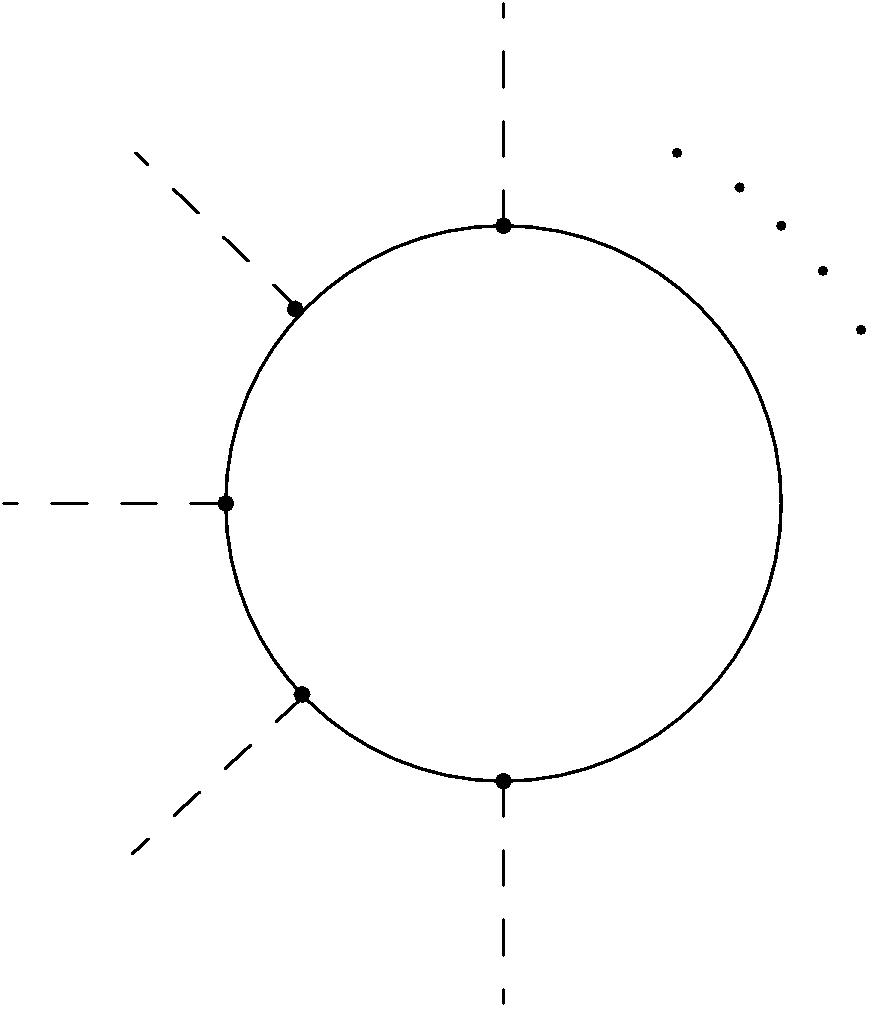}
	\end{center}
	\vskip-5mm
	\caption{Diagrammatic representation for the two contributions to the RG flow equation (\ref{SystemBetaFunctionsOneLoopDER}) of $\lresc_n$ in the DER and in the UV limit, where the rescaled Higgs quartic coupling $\lresc_2$ plays a dominant role. }
	\label{fig:scalarLoop}
\end{figure}

It is interesting to investigate the stability of the potential for the QFP solution $\lresc_{2}=\lresc_{2}^+$, once we sum the expansion in \Eqref{eq:ExpansionAroundKappa} for $N_\mathrm{p}\to\infty$.
To address this task let us consider the leading $h^2$ contribution for the $\beta_n$ function with $n\ge 2$. Its structure is
\be
	\beta_n= 2(n-2) \lresc_{n}+\frac{(-3)^n n!}{32\pi^2} \lresc_{2}^n+\frac{3 (-1)^{n+1} n!}{8\pi^2}
\ee
where $2(2-n)$ is the canonical dimension of $\lresc_{n}$, 
while the second and third terms are the contribution of a scalar loop with $n$ quartic self-interaction vertices 
and of a fermion loop with $2n$ Yukawa vertices, respectively.
This can be drawn diagrammatically as in \Figref{fig:scalarLoop}.
Thus, among all possible scalar self-interactions, the $\lresc_{2}$ coupling plays a dominant role in the UV.
This $\phi^4$-dominance regime can be studied by specifying a pure $\phi^4$ interaction in the bosonic threshold function that appears in the RG flow equation for $u(\rho)$.
This means
\be
\omega=3 \lambda_2 \rho	\label{eq:omega_phi^4_dominance}
\ee
in \Eqref{eq:betau}, where $\eta_{\phi}$ is given by \Eqref{eq:etaphi_1loop}.
Thanks to \Eqref{eq:PnCEL}, it is possible to encode all the rescalings from $\lambda_n$ for $n>2$ to $\lresc_{n}$  in a suitable redefinition of the field invariant $\rho$.
This can be achieved by defining
\begin{align}
z &= h^2\rho, \label{eq:z_def}\\
d_z &= 2+\eta_\phi+\eta_{h^2} \equiv 2+\eta_z. \label{eq:d_z_def}
\end{align}
By projecting the left-hand side of the above RG flow for $u(\rho)$, where \Eqref{eq:omega_phi^4_dominance} is substituted inside \Eqref{eq:betau}, onto the ansatz in \Eqref{eq:ExpansionAroundKappa} with $\kappa=0$ and $N_\mathrm{p}\to\infty$, it is possible to solve the QFP condition for all $\lresc_{n}$.
The solution is indeed
\begin{align}
	\lresc_{2}&=\lresc_{2}^\pm,\\
	\lresc_{n>2}&=\frac{(-1)^n n!}{32\pi^2}\,\frac{ (3\lresc_{2})^n - 12 }{4-n(2+\eta_z)},
\end{align}
and the resummation of the series has an analytic expression in terms of the hypergeometric function $_2F_1(a,b,c,z)$.
In fact the effective potential reads
\begin{align}
	u(\rho) &=\Bigg[\lresc_{2}\frac{z^2}{2h^2} \notag\\
	&\quad +\frac{1}{32\pi^2}\frac{(3\lresc_{2} z)^3}{2+3\eta_z} \, _2F_1\left(1,\frac{2+3\eta_z}{2+\eta_z},\frac{4+4\eta_z}{2+\eta_z},-3 \lresc_{2} z\right)\notag\\
	&\quad -\frac{3 }{8\pi^2}\frac{z^3}{ 2+3\eta_z }\times \notag \\
	&\quad \times _2F_1\left(1,\frac{2+3\eta_z}{2+\eta_z},\frac{4+4\eta_z}{2+\eta_z}, - z\right)\Bigg]_{z=h^2 \rho}
	\label{uhxDER}
\end{align}
which has the property
\be
	\lim_{z\to 0}u^{\prime\prime}(z)=\frac{\lresc_{2}}{h^2},
\ee
as it is clear from the chosen polynomial ansatz.

Since this solution is constructed by resummation of a local expansion for small field amplitudes, it might depart from the actual fixed-point potential at large values of $\rho$ due to nonanalytic terms.
We are interested mainly in the asymptotic region $\rho\to\infty$ in the UV where $h^2\to 0$.
However, because the QFP solution $u(\rho)$ is a function of both variables $\rho$ and $h^2$, 
there might be several such asymptotic regions, corresponding to different ways of taking the combined limit $\rho\to\infty$ and $h^2\to 0$.
To classify these possible limits, we address the dependence of loop effects on $h^2$ and $\rho$.
By inputting the asymptotic UV scaling of $\lambda_2$,
the threshold functions for the bosonic and fermionic loops in \Eqref{eq:betau} 
are functions of $\omega=3\lresc_{2} z$ and $\omega_1=z$ respectively.
Thus, the variable entering the threshold functions is $z$ as defined in \Eqref{eq:z_def}.
Therefore we can identify an outer region where $z\gg 1$ and an inner region where $z\ll 1$.
In \Appref{app:large fields} we address in more detail this combined limit and show that
it exists and is the same in both asymptotic regions,
such that \Eqref{uhxDER} does give a definite answer concerning the stability of the potential $u(\rho)$ for an arbitrarily small value of $h^2$.
In fact
\begin{align}
u(\rho)\widesim[2]{\rho\to \infty} \frac{1}{2}\lresc_{2} \frac{z^2}{h^2}. \label{eq:u(rho)_leading_behavior}
\end{align}
This proves that the CEL solution corresponds to a bounded potential in the DER.

%%%%%%%%%%%%%%%%%%%%%%%%%%%%%%%%%%%%%%%%
\section{Effective field theory analysis including thresholds} \label{sec:effectiveFTthresholds}
%%%%%%%%%%%%%%%%%%%%%%%%%%%%%%%%%%%%%%%%

In this section we relax the restriction adopted in \Secref{sec:effectiveFTdeepEuclidean} to the DER, and  we account for the running of the scalar mass term.
In other words, we include the possibility for a nontrivial minimum, by choosing a polynomial expansion of the scalar potential  around $\rho=\kappa\ne 0$ as in \Eqref{eq:ExpansionAroundKappa}.
By projecting the left-hand side of the \Eqref{eq:betau} onto this ansatz, we can derive the flow equations for the rescaled couplings $\lresc_{n}$ as defined in \Eqref{eq:L2def} and \Eqref{eq:lambdatoLn}.
Similarly, also the coupling $\kappa$ may scale asymptotically as a definite power of $h^2$.
We define
\be
\kap=h^{2Q}\kappa,
\label{eq:kappatoxikappa}
\ee
where the real power $Q$ is a priori arbitrary.

Let us denote by $\beta_n$ the beta function of $\lresc_{n}$, $\beta_n=\de_t\lresc_{n}$.
In order to construct polynomial solutions of the QFP equations for the couplings $\lresc_{n}$ and $\kap$, we set up the following recursive problem:
we solve the equation $\beta_{\kap}=0$ for $\lresc_{2}$, and $\beta_n=0$ for $\lresc_{n+1}$.
Upon truncating the series of equations at some $\beta_{N_\mathrm{p}}$, this can be achieved only if one more coupling $\lresc_{N_\mathrm{p}+1}$ is retained.
The result of this construction is a set of QFPs for $\lresc_{n}$ as functions of the couplings $h^2$ and $\kap$.
Also, some of the parameters $P$, $P_n$ and $Q$ might remain unconstrained.
A defining requirement for a viable QFP solution to represent an AF trajectory is that the couplings $\lresc_{n}$ and $\kap$ approach constants for $h^2\to 0$. 

Clearly, there is some freedom in the search for scaling solutions and particularly in the recursive procedure we have described.
Of course, it is likewise possible to treat another scalar coupling as a ``free'' parameter and to solve for $\kap$ in terms of some $\lresc_{n}$.
The question which coupling should meaningfully be treated as free parameter cannot be answered a priori and depends again on the precise details of the model.
We choose $\lresc_{N_\mathrm{p}+1}$ here to start with.
For definiteness, we concentrate in this work on solutions exhibiting the property that $\lresc_{2}\neq 0$ at the QFP (though this might be a scheme-dependent statement).

We now illustrate this process by considering $N_\mathrm{p}=2$; the analysis can straightforwardly be extended to any higher order.
Again we adopt the approximation of setting the anomalous dimension inside the threshold functions in Eqs.~\eqref{eq:betau}-\eqref{eq:etapsi} to zero.

\begin{figure*}[!t]
	\begin{center}
		\includegraphics[width=0.95\columnwidth]{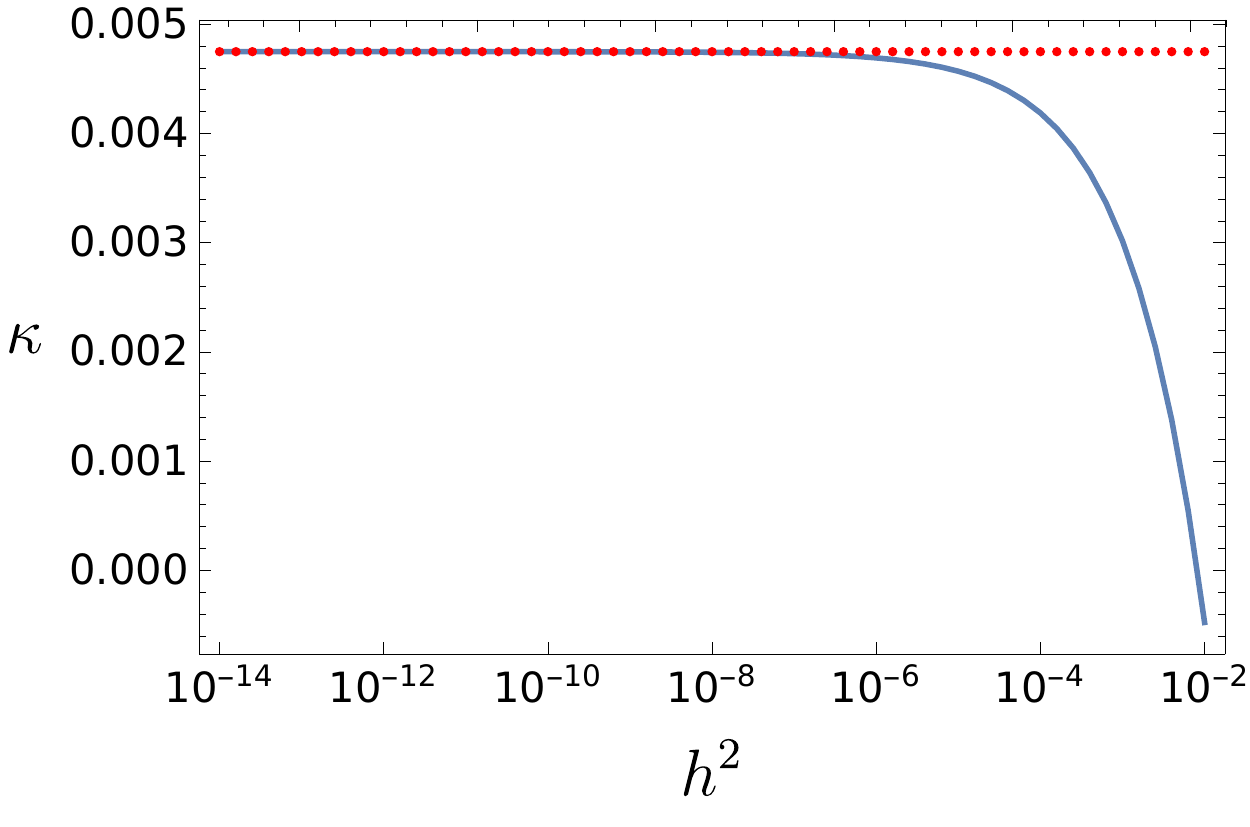}\hskip6mm
		\includegraphics[width=0.95\columnwidth]{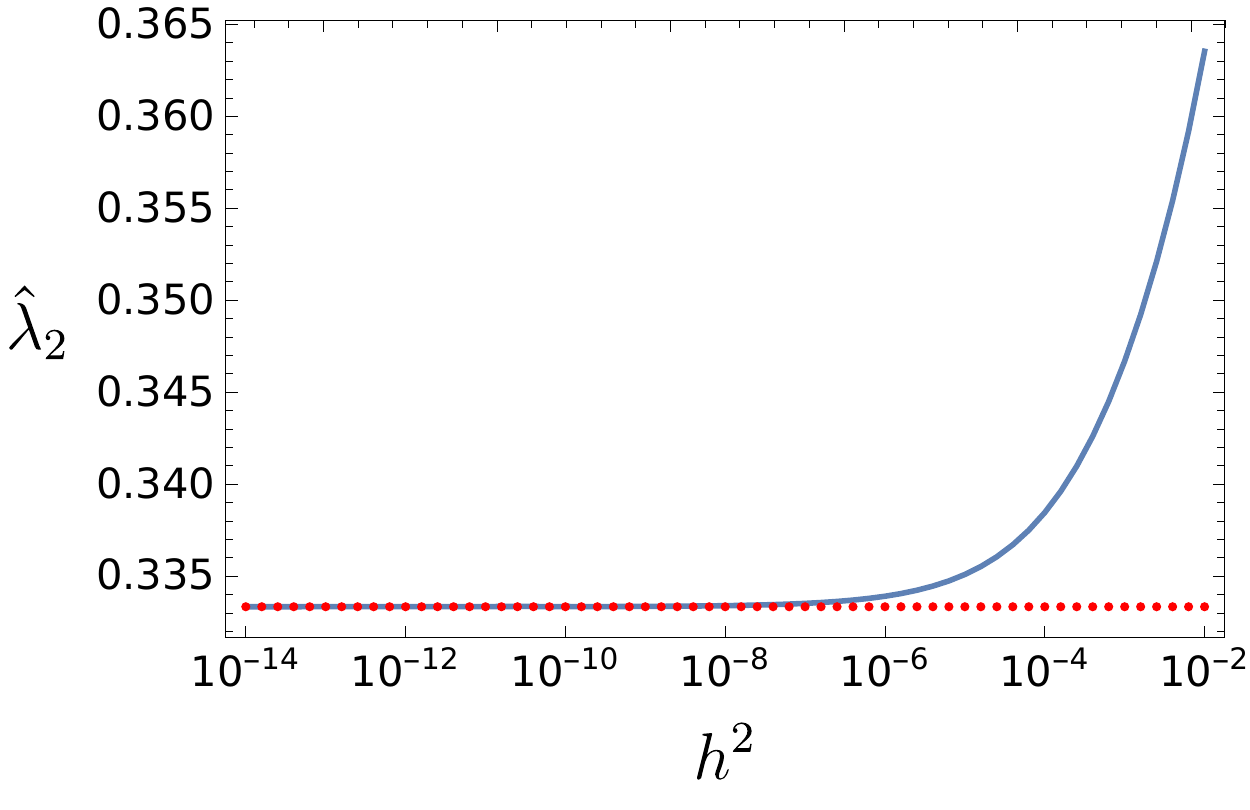}
	\end{center}
	\vskip-6mm
	\caption{Effective field theory analysis including thresholds for the case $P=1/4$.
		We plot the full numerical solutions  $\kappa\left(h^2\right)$ (upper panel) and $\lresc_{2}\left(h^2\right)$ (lower panel) 
		as a function of $h^2$ for the fixed value $\lresc_{3}=1$ as blue solid line.
		The red dashed lines correspond to the leading-order constant solutions highlighted in  \Eqref{eq:kappa_P<1/2} and \Eqref{eq:L2_P<1/2}.}
	\label{fig:EFTP_less_than_1over4}
\end{figure*}

%%%%%%%%%%%%%%
\subsection{$P\in \left(0,1/2\right)$}
%%%%%%%%%%%%%%

Because of the qualitative similarity between the flow equations of the present model and
those analyzed in Refs.~\cite{Gies:2015lia,Gies:2016kkk}, we know that the finite ratio $\kap$ defined in \Eqref{eq:kappatoxikappa} is actually $\kappa$
itself for $P$ being equal or smaller then $1/2$. Thus, we immediately make the ansatz $Q=0$, which turns out to be
the correct solution.
Indeed the leading orders in $h^2$ in 
the flow equations of the rescaled couplings are
\begin{align}
	\de_t \lresc_{2}&=-\frac{\lresc_{3}\,h^{2P_3-4P}}{16\pi^2} +\frac{9 \lresc_{2}^2\,h^{4P}}{16\pi^2}+\frac{\kappa\lresc_3^2\,h^{4P_3-8P}}{16\pi^2\lresc_2},\\
	\de_t\kappa&=-2\kappa+\frac{3}{32\pi^2}+\frac{\kappa \lresc_3\,h^{2 P_3-4 P}}{16\pi^2\lresc_2}.
	\label{eq:kappadot_P<1/2}
\end{align}
The QFP condition admits two solutions, each of them is a one parameter family of solutions.
One solution corresponds to the case where the contribution coming from $\lresc_3$ is subleading in \Eqref{eq:kappadot_P<1/2}, i.e., $P_3>2P$,	and it reads
\begin{alignat}{2}
	\kappa &=\frac{3}{64\pi^2},\quad\quad\quad &Q&=0, \label{eq:kappa_P<1/2}\\
	\lresc_{2}^2&=\frac{ \lresc_{3}}{9},\quad\quad\quad &P_3&=4P,\label{eq:L2_P<1/2}
\end{alignat}
thus $\lresc_{3}$ must be positive, but is otherwise arbitrary.
In \Figref{fig:EFTP_less_than_1over4} 
%On the left panel is shown the linear dependence between $\lresc_{2}^2$ and $ \lresc_{3}$ while on the central and right panels
it is shown how the numerical solutions for the full $h^2$-dependent flow equations 
(in the approximation detailed at the beginning of %\Secref{sec:effectiveFTthresholds}
the present section) are in agreement with the leading order approximation and approach the constant values in \Eqref{eq:kappa_P<1/2} and \Eqref{eq:L2_P<1/2} in the $h^2\to 0$ limit.

By contrast, the second solution corresponds to the case where the $\lresc_3$ term contributes to the flow equation for $\kappa$ in the UV limit, i.e., $P_{3}=2P$.
Indeed, we have that
\begin{alignat}{2}
	\kappa &=\frac{5}{64\pi^2}, \qquad &Q&=0,\label{eq:kappa_P<1/2_bis}\\
	\lresc_{2}&=\frac{5\lresc_3}{64\pi^2},\qquad &P_3&=2P,
	\label{eq:L2_P<1/2_bis}
\end{alignat}
where again the rescaled cubic scalar coupling remains a free parameter.

While the first class of solutions in Eqs.~(\ref{eq:kappa_P<1/2}) and
(\ref{eq:L2_P<1/2}) had already been discovered
in Refs.~\cite{Gies:2015lia,Gies:2016kkk},
the second one given by Eqs.~(\ref{eq:kappa_P<1/2_bis}) and
(\ref{eq:L2_P<1/2_bis}) is new.
These solutions were not
observed in Refs.~\cite{Gies:2015lia,Gies:2016kkk} because of
simplifying approximations in the analysis of the RG equations.
In particular, only linear insertions of the coupling
$\lambda_3$ into the beta functions of lower-dimensional 
parameters were considered.

%%%%%%%%%%%%%%
\subsection{$P=1/2$}\label{subsec:effectiveFTthresholdsP1half}
%%%%%%%%%%%%%%

For the following $P\ge 1/2$ cases we confine the discussion to analytical approximations
to leading order in the $h^2\to 0$ limit. Plots analogous to \Figref{fig:EFTP_less_than_1over4} with the numerical solutions capturing the full $h^2$ dependence
of the QFP solutions would show a similar agreement between the two descriptions.

\begin{figure*}[!t]
	\begin{center}
		\includegraphics[width=0.95\columnwidth]{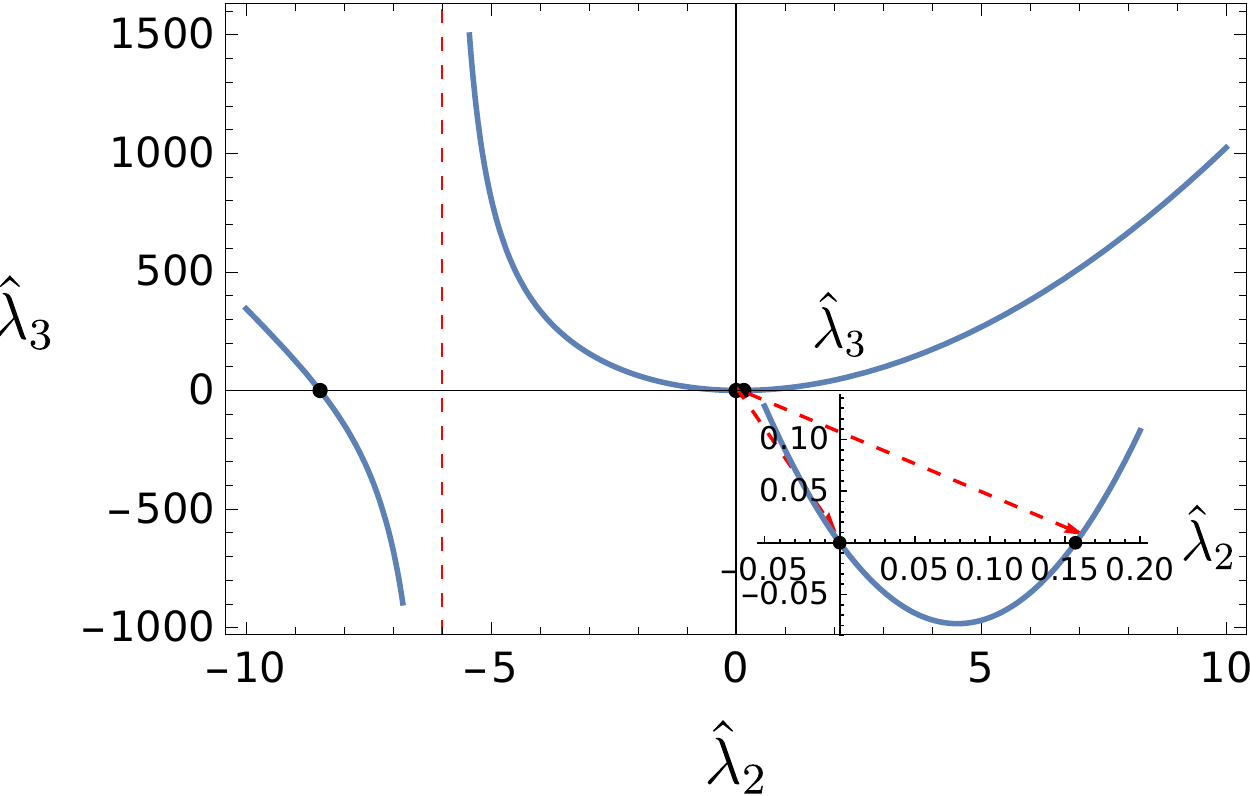}\hskip6mm
		\includegraphics[width=0.95\columnwidth]{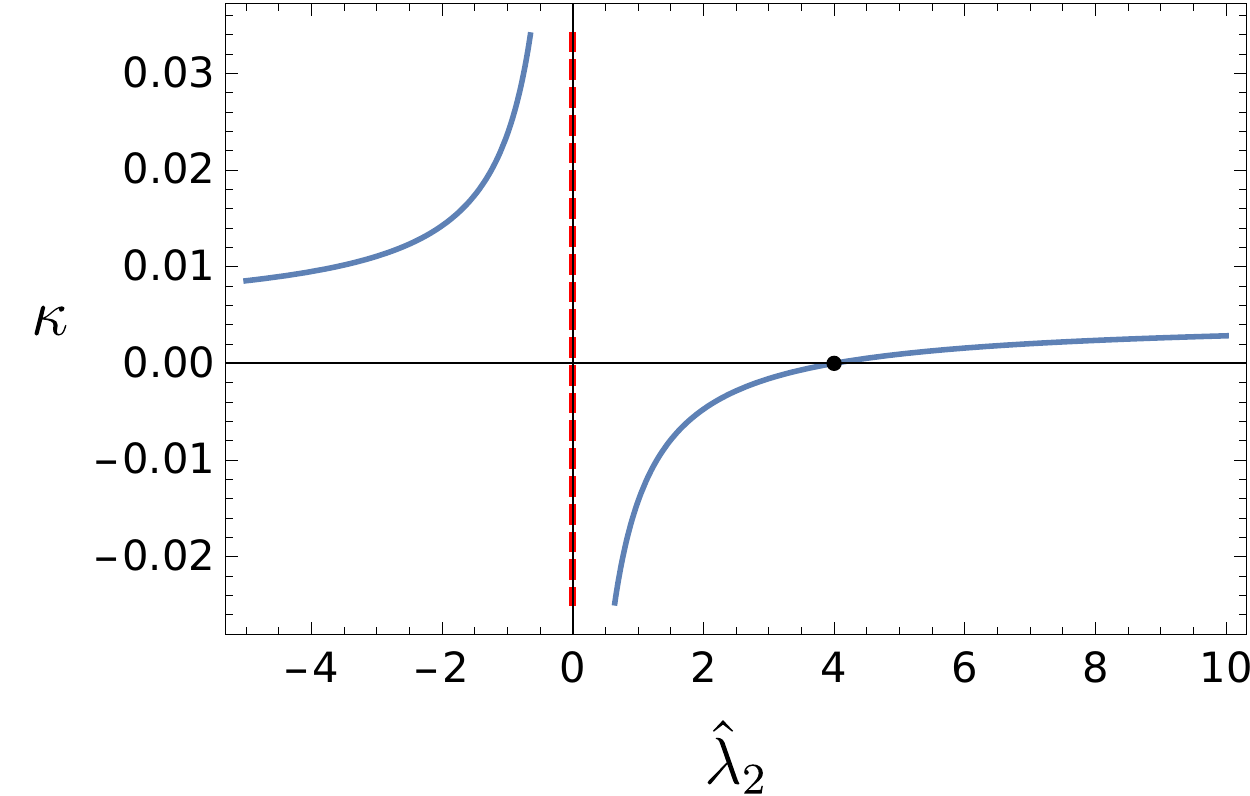}
	\end{center}
\vskip-5mm
	\caption{Effective field theory analysis including thresholds for $P=1/2$.
	We show the QFP solutions $\lresc_{3}(\lresc_{2})$ (left) and $\kappa(\lresc_{2})$ (right) for the case $P_3=2$ to leading order in the $h^2\to 0$ limit, as expressed in \Eqref{eq:L3_P3=2} and \Eqref{eq:kappa_P3=2}.
	The left panel illustrates that for every $\lresc_{2}\neq -6$ there is only one associated value for the free parameter $ \lresc_{3}$; the magnified inset near the origin highlights the CEL solution $\lresc_{2}=\lresc_{2}^+$ as in \Eqref{eq:L2rootCEL}.
	The right panel shows that there are solutions with a positive nontrivial minimum and scalar quartic coupling only for $\lresc_{2}>4$.
}
	\label{fig:EFTP_P1half_gamma2}
\end{figure*}

For $P=1/2$, the analysis is less straightforward and there are several possibilities.
The leading $h^2$-contributions to the RG flow of $\lresc_{2}$ and $\kappa$ are
\begin{align}
	\de_t \lresc_{2}&=h^2 \left(\frac{9\lresc_{2}^2}{16\pi^2}+\frac{75\lresc_{2}}{16\pi^2}-\frac{3}{4\pi^2}\right)+\frac{\kappa\lresc_3^2\,h^{4 P_3-4}}{16\pi^2 \lresc_2}\notag\\
				   &\quad -h^{2P_3-2}\left(\frac{\lresc_{3}}{16\pi^2}-\frac{3\lresc_{3}}{8\pi^2\lresc_2}\right), \label{dotxi2P1half}\\
	\de_t\kappa&=-2\kappa+\frac{3}{32\pi^2}-\frac{3}{8\pi^2 \lresc_{2}}+\frac{\kappa \lresc_{3}\,h^{2P_3-2}}{16\pi^2\lresc_{2}}.    \label{dotkappaP1half}
\end{align}
If $P_3>2$ the contributions due to $\lresc_{3}$ are negligible in the $h^2\to 0$ limit and we recover the CEL solution of \Eqref{eq:L2rootCEL}.
Moreover a positive (negative) solution for $\lresc_{2}$ leads to a negative (positive) solution for $\kappa$, suggesting that the stable CEL
potential possesses only the trivial minimum.

If $1<P_3<2$, the contribution coming from $\lresc_{3}$ plays the dominant role in the RG flow of $\lresc_{2}$ but is subleading for $\kappa$.
%The solution of the corresponding QFP equations has a negative unphysical result for the scalar self interaction which is $\lresc_{2}=-6$ and a positive nontrivial minimum that is $\kappa=5/(64\pi^2)$.
The solution of the corresponding QFP equations is $\kappa=5/(64\pi^2)$ and $\lresc_{2}=-6$, implying that the expansion point is a nontrivial maximum. 
As we have assumed in our analysis that the expansion point of the Taylor series is a minimum of the potential, we reject this solution albeit it might lead to further interesting solutions if an appropriate expansion scheme is used. 
Thus, the only two new solutions correspond to $P_3=1$ and $P_3=2$.

In the first case, $P_3=1$, the solution of $\de_t \lresc_{2}=0$ is determined only by the $h^{0}$-terms.
Together with \Eqref{dotkappaP1half}, this leads to a $\lresc_{2}$ which depends linearly on $\lresc_{3}$.
The solution is indeed
\begin{alignat}{2}
\kappa &=\frac{5}{64\pi^2}, \qquad &Q&=0, \\
\lresc_{2}&=\frac{5\lresc_3}{64 \pi^2} - 6, \qquad &P_3&=1.
\end{alignat}

In the second case where $P_3=2$, the contribution given by $\lresc_{3}$ in \Eqref{dotkappaP1half} is subleading and the corresponding QFP equation provides us $\kappa(\lresc_{2})$.
This solution can be substituted into \Eqref{dotxi2P1half} and the latter one can be solved in term of $\lresc_{3}(\lresc_{2})$.
The corresponding solution reads
\begin{alignat}{2}
\kappa &=\frac{3}{64\pi^2}\frac{\lresc_{2} - 4}{\lresc_{2}}, \qquad &Q&=0, \label{eq:kappa_P3=2}\\
\lresc_{3}&=3 \lresc_{2}\frac{3\lresc_{2}^2+25 \lresc_{2} - 4}{\lresc_{2}+6}, \qquad &P_3&=2. \label{eq:L3_P3=2}
\end{alignat}

We plot this solution for $P_3=2$ in \Figref{fig:EFTP_P1half_gamma2}.
The three black dots in the left panel highlight the three roots corresponding to $ \lresc_{3}=0$. 
For one of these roots, we find $\lresc_{2}=0$ which can be discarded as the QFP value for $\kappa$ is singular in this case. 
The other two roots are the $\lresc_{2}^\pm$ of \Eqref{eq:L2rootCEL}. 
Moreover, it is clear from \Eqref{eq:kappa_P3=2} that the condition $\lresc_{2}>4$ has to hold to obtain a positive nontrivial minimum and at the same time a positive quadratic scalar coupling. This can also be seen in the right panel of \Figref{fig:EFTP_P1half_gamma2}.

%%%%%%%%%%%%%%
\subsection{$P\in \left(1/2,1\right)$}
%%%%%%%%%%%%%%
By following again the gauged-Higgs model discussed in Refs.~\cite{Gies:2015lia,Gies:2016kkk} we can assume that for $P>1/2$ the nontrivial minimum goes to infinity according to some power of $h^2$ such that its scaling $Q$ is positive.
Choosing $Q=2 P-1$ as in the gauged-Higgs model turns out to be the correct scaling also for the present system. 
However, we prefer to be more general and consider $Q$ as an undetermined positive power in the first place.
It is possible to verify that, under the assumptions that $Q>0$, $P_3>0$, and $P>1/2$, the only terms that can contribute to the leading parts in the RG flow for $\lresc_{2}$ and $\kap$ are
\begin{align}
\de_t\lresc_{2} &= \frac{\lresc_{3}^2\kap\, h^{4 P_3-8P-2Q}}{16\pi^2\lresc_2 (1+2\lresc_2\kap h^{4P-2Q})^2} - \frac{3\,h^{4-4P}}{4\pi^2(1+h^{2-2 Q}\kap)^3}\notag\\
&\quad - \frac{3\lresc_3 h^{2-8P+2\gamma}}{8\pi^2\xi_2 (1+h^{2-2Q}\kap)^2} +\frac{\kap^2\lresc_{3}^2\,h^{4P_3-4Q-4P}}{4\pi^2(1+2h^{4 P-2 Q}\lresc_{2}\kap)^3}\notag\\
&\quad +\frac{\kap^3\lresc_3^2\,h^{4P_3-6Q}}{2\pi^2(1+2h^{4P-2Q}\lresc_2\kap)^4}\lresc_2(1-P),         \label{eq:lresc2_P>1/2}\\
\de_t\kap&=-2\kap+(Q-1)\frac{\kap^4\lresc_{3}^2\,h^{4P_3-6Q}}{4\pi^2(1+2h^{4P-2 Q}\lresc_{2}\kap)^4}     \notag\\
&\quad -\frac{3\,h^{2-4P+2Q}}{8\pi^2\lresc_{2}(1+h^{2-2 Q}\kap)^2}. \label{eq:kap_P>1/2}
\end{align}
By analyzing all the possible combinations among the three powers $Q$, $P_3$ and $P$, one has to take care that the two powers of $h^2$ in the denominators, i.e., $2P-Q$ and $1-Q$, give different contributions to the $\beta$ functions depending on whether they are positive or negative.
Moreover, we have to keep in mind that -- by definition of the finite ratios -- $\lresc_{2}$ and $\kap$ have to approach their QFP values in the UV limit up to subleading corrections in some positive power of $h^2$.
Among the set of all possible configurations there are only two QFP solutions.
One of these corresponds to the case where the contribution arising from $\lresc_3$ is subleading in \Eqref{eq:kap_P>1/2}:
\begin{alignat}{2}
\kap &= \frac{3}{8\pi^2\lresc_{3}}, \qquad &Q&=2P-1, \label{eq:kappa_1/2<P<1}\\
\lresc_{2} &= -\frac{\lresc_{3}}{2}, \qquad &P_3&=2P+1, \label{eq:L2_1/2<P<1}
\end{alignat}
where $\lresc_{3}$ is a free parameter. 

By contrast, the second solution is the one where $\lresc_3$ provides a leading contribution to the flow equation for $\kap$.
By solving the QFP condition in terms of the nontrivial minimum this solution reads
\begin{alignat}{2}
\lresc_2 &=\frac{-2\pm \sqrt{2P-1}}{2(5-2P)\kap}, \qquad &Q&=2P,\\
\lresc_{3}^2 &=\frac{8\pi^2(1+2\kap\lresc_2)^4}{(2P-1)\kap^3}, \quad &P_3&=3P.
\end{alignat}
We can therefore deduce that there are no reliable solutions that fulfill our assumptions for $P\in (1/2,1)$ because it is not possible to simultaneously satisfy the condition that both the Higgs quartic coupling and the nontrivial expansion point $\kappa$ are positive.

%%%%%%%%%%%%%%
\subsection{$P=1$}	\label{subsec:EFTP=1}
%%%%%%%%%%%%%%
Starting from \Eqref{eq:lresc2_P>1/2} and \Eqref{eq:kap_P>1/2}, it is possible to prove that for $P=1$ there are again two QFP solutions corresponding to different combinations for the two left powers $P_3$ and $Q$.
One solution is
\begin{alignat}{2}
	\lresc_{3}&=\frac{3}{8\pi^2\kap (1+\kap)^3}, \qquad &P_3&=3, \label{eq:L3_P=1}\\
	\lresc_{2}&=-\frac{3}{16\pi^2\kap(1+\kap)^2}, \qquad &Q&=1, \label{eq:L2_P=1}
\end{alignat}
whereas the second one reads
\begin{alignat}{2}
	\lresc_{3}^2 &=\frac{128\pi^2}{81\kap^3}, \qquad &P_3&=3, \\
	\lresc_{2} &= - \frac{1}{6\kap},\qquad &Q&=2.
\end{alignat}
We observe once more that there are no solutions with positive $\kappa$ and a positive scalar quartic coupling such that we expand the potential around a nonvanishing vacuum expectation value of the scalar field.

In \Appref{app:P>1} we complete the EFT analysis
of the present section, by discussing $P>1$.
Also in this case we conclude that all
the QFP solutions we observe have either
$\lresc_{2}$ or $\kap$ negative.

%%%%%%%%%%%%%%%%%%%%%%%%%%%%%%%%%%%%%%%%
\section{Full effective potential in the $\phi^4$-dominance approximation} \label{sec:EFT_for_f}
%%%%%%%%%%%%%%%%%%%%%%%%%%%%%%%%%%%%%%%%
So far, we have projected the RG flow of the potential onto a polynomial basis and studied only the running of the various coefficients.
Now, we investigate the functional RG flow of an arbitrary scalar potential which also includes nonpolynomial structures~\cite{Coleman:1973jx,Jackiw:1974cv}.
The latter is obtained by performing a one-loop computation with field-dependent thresholds.
The loop integrals are evaluated by using the piece-wise linear regulator \cite{Litim:2000ci,Litim:2001up}.
To simplify the discussion, we neglect the possible appearance of higher-dimensional couplings in the other $\beta$ functions and anomalous dimensions, and ignore contributions which would be present only in the SSB regime.
Thus in the following, we use \Eqref{eq:betau} together with Eqs.~\eqref{eq:gs-oneloop}, \eqref{eq:hsquaredot}, and the one-loop value for the anomalous dimension of the scalar field 
given in \Eqref{eq:etaphi_1loop}.

We pursue the identification of AF trajectories in the space of all flows described by integration of \Eqref{eq:betau} for generic boundary conditions.
We already know from the previous sections that AF solutions can in fact be constructed by
simply looking for QFPs of the flow of $h^2$-rescaled interactions.
To implement this condition in a functional set-up, we define a new field variable and its potential
\be
\begin{split}
	x&=h^{2P}\rho, \quad\quad
	f(x)=u(\rho). \label{eq:rhotox_utof}
\end{split}
\ee
We denote the minimum by $x_0$ and the couplings by $\xi_{n}$,
\be
\begin{split}
	f'(x_0)&=0, \qquad
	f^{(n)}(x_0)=\xi_n. \label{eq:xin_def}
\end{split}
\ee
The arbitrary rescaling power $P$ is chosen to be that of \Eqref{eq:L2def} so that $\xi_2=\lresc_{2}$, because we specifically look for QFPs where $\lresc_{2}\neq0$.
It might happen that at a QFP $x_0\neq\kap$, and $\xi_n\neq \lresc_{n}$ for $n>2$, such that solutions of the equation $\partial_t f(x)=0$ might differ from the actual scaling solutions.
Thus, the rescaling of \Eqref{eq:rhotox_utof} is expected to be useful as long as the quartic scalar coupling is the leading term in the approach of the scalar potential towards flatness.

As a first-level approximation, we consider an intermediate step between the polynomial and the functional approaches, which is based on the expectation that the marginal quartic coupling plays a dominant role in the UV.
Therefore, we assume that the contribution coming from the scalar fluctuations is dominated by a plain quartic interaction. 
More precisely, we use $\omega=3\lambda_2\rho$ on the right-hand side of \Eqref{eq:betau}, but we consider the scalar potential as an unknown arbitrary function in the scaling term and on the left-hand side of the flow equation itself.
This leads to the following flow equation
\begin{align}
	\de_t f(x)&= - 4 f(x) + d_x x f^\prime (x)+\frac{1}{32\pi^2}\,\frac{1}{1+3\xi_2 h^{2P} x}\notag\\
	&\quad-\frac{3}{8\pi^2}\,\frac{1}{1+h^{2-2P}x}.	\label{eq:beta_f_phi^4dominance}
\end{align}
where
\be
d_x=2+\eta_\phi+P\eta_{h^2} \equiv 2+\eta_x\ .
\label{eq:dx_def}
\ee
The anomalous dimension $\eta_x$ of the rescaled field invariant $x$ 
 includes also the introduced anomalous dimension of the Yukawa coupling $\eta_{h^2}$ defined in \Eqref{eq:defetah2}. %and  \Eqref{eq:etahath2}.

By setting the left-hand side to zero, we get a first-order linear ordinary differential equation that can be solved analytically for generic $P$ and its QFP solution is
\begin{align}
	f(x) &= \frac{1}{128\pi^2}\,_2F_1 \left(1,-\frac{4}{2+\eta_x},\frac{-2+\eta_x}{2+\eta_x},-3\xi_2 h^{2P} x\right) \notag\\
	            &\quad - \frac{3}{32 \pi^2}\,_2F_1\left(1,-\frac{4}{2+\eta_x},\frac{-2+\eta_x}{2+\eta_x}, -h^{2-2P}x\right)\notag \\
	            &\quad + C_\mathrm{f}\, x^\frac{4}{2+\eta_x}, 		
	            \label{eq:f_FP_phi^4dominance}
\end{align}
where the term proportional to the free integration constant $C_{\mathrm{f}}$ is the homogeneous solution of Eq.~\eqref{eq:beta_f_phi^4dominance} while the Gau\ss\ hypergeometric functions are particular solutions obtained by integrating the non-homogeneous part.

For $C_{\mathrm{f}}=0$ we can straightforwardly impose the consistency condition $f^{\prime\prime}(0)=\xi_2$.
Instead, for any nonvanishing $C_{\mathrm{f}}$, the QFP potential behaves as a nonrational power of $x$ at the origin.
Its second order derivative is not defined at the origin as long as $\eta_{x}>0$ which is generically the case for a potential in the symmetric regime.
This problem might be avoided if there is at least one nontrivial minimum $x_0$, in the spirit of the Coleman-Weinberg mechanism~\cite{Coleman:1973jx}. In fact, we can impose $f''(x_0)=\xi_2$ for this xcase.

As a first analysis, we want to understand the asymptotic properties of the full $h^2$-dependent solution $f(x)$. Specifically, we want to identify parameter ranges for $C_{\mathrm{f}}$ and $\xi_2$ for which the potential is bounded from below.
To this end, we focus on the asymptotic behavior of the solution, $x\to+\infty$. 
In particular, we are interested in the UV regime where $h^2\to 0$.
Since the QFP potential $f(x)$ for given $C_{\mathrm{f}}$, which might also depend on $h^2$, is a function of the two variables $x$ and $h^2$, 
we have to take the limit process with care to investigate the asymptotic behavior of $f$ in the deep UV.

In order to address the asymptotic behavior of the QFP potential in a systematic way, we analyze the flow for fixed arguments, 
\begin{align}
z_\rmB &= 3\xi_2h^{2P}x, \qquad 
z_\rmF = h^{2(1-P)}x, \label{eq:z_BF_def}
\end{align}
of the hypergeometric functions.
For small enough $h^2$ and $P\gtrless 1/2$, we have $z_\rmF\gtrless z_\rmB$. 
Thus, one can divide the interval $x \in [0,\infty)$ into three distinct domains. 
Suppose $z_\rmF < z_\rmB$, then we define the $h$-dependent boundary $x_{1}(h)$ of an inner interval $x \in [0,x_{1})$ by requiring $z_{\rmB}=1$ and the boundary $x_{2}(h)$ of an outer interval $(x_{2},\infty)$ by $z_{\rmF}=1$ for fixed $P$ and $\xi_{2}$. For $z_\rmF > z_\rmB$, the requirement $z_{\rmB}=1$ and $z_{\rmF}=1$ will define $x_{2}$ and $x_{1}$, respectively.
In case $P<1$, the two boundaries $x_{1}$ and $x_{2}$ grow towards larger values and always fulfill $x_{2}>x_{1}$ when we send $h \to 0$. 

Approximating the hypergeometric functions for small but fixed arguments $z_{\rmB/\rmF} \ll 1$, we obtain a valid approximation of the potential in the first interval as this also implies $x \ll x_{1}$. 
Thus, we are able to reliably check the asymptotic behavior by first performing the limit $h \to 0$ and afterwards $x \to \infty$ in this region. In case the hypergeometric functions shall be investigated for large arguments, we have to perform first the limit $x \to \infty$ before sending $h \to 0$ to investigate the asymptotic behavior such that one stays in the outer interval as only there the results can be trusted for the used approximations. Further details can be found in \Appref{app:largefield_f}. The rescaled potential $f(x)$ turns out to be stable in the deep UV for both regimes,
and the two asymptotic behaviors are in agreement.

%%%%%%%%%%%%%%%%%%%%%%%%%%%%%%%%%
\subsection{Large-field behavior} \label{subsec:large_field}
%%%%%%%%%%%%%%%%%%%%%%%%%%%%%%%%%

\begin{figure*}[!t]
	\begin{center}
		\includegraphics[width=0.95\columnwidth]{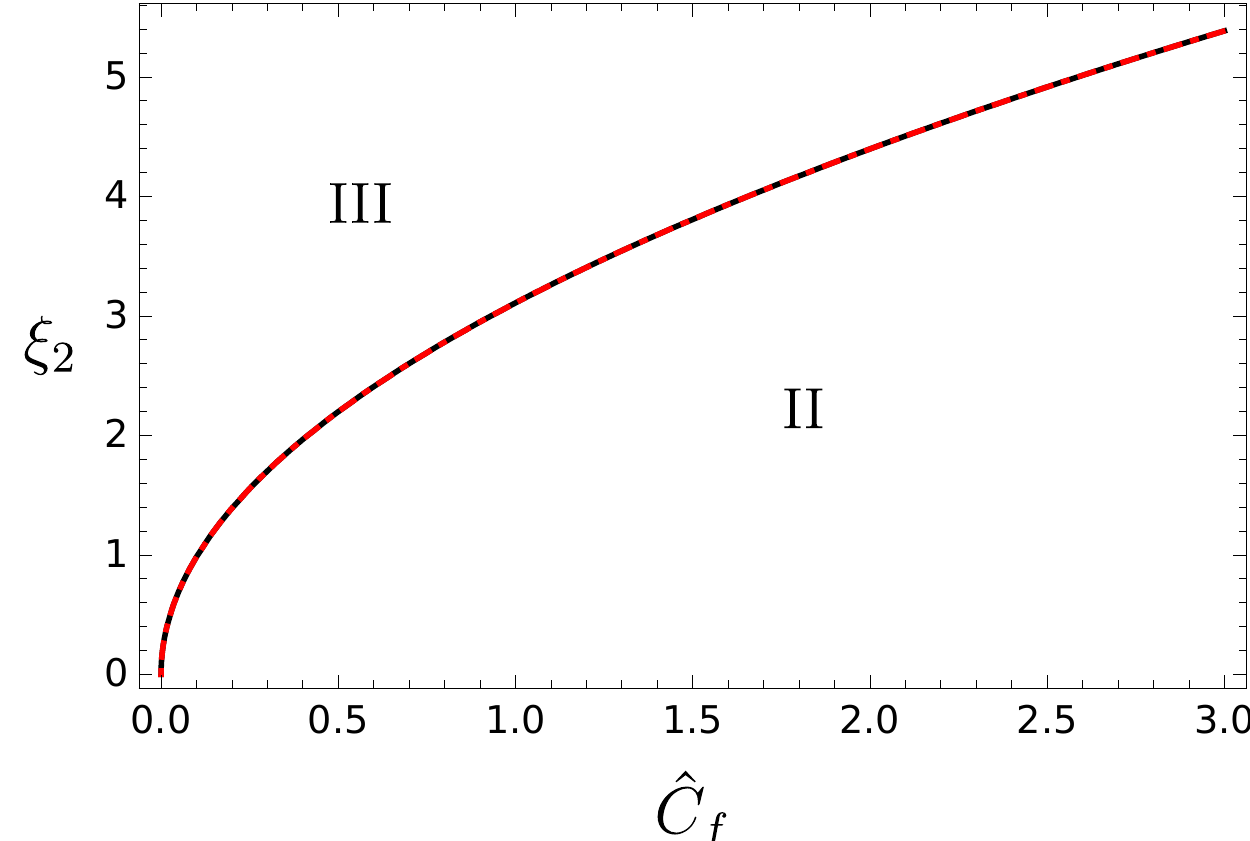}\hskip6mm
		\includegraphics[width=0.95\columnwidth]{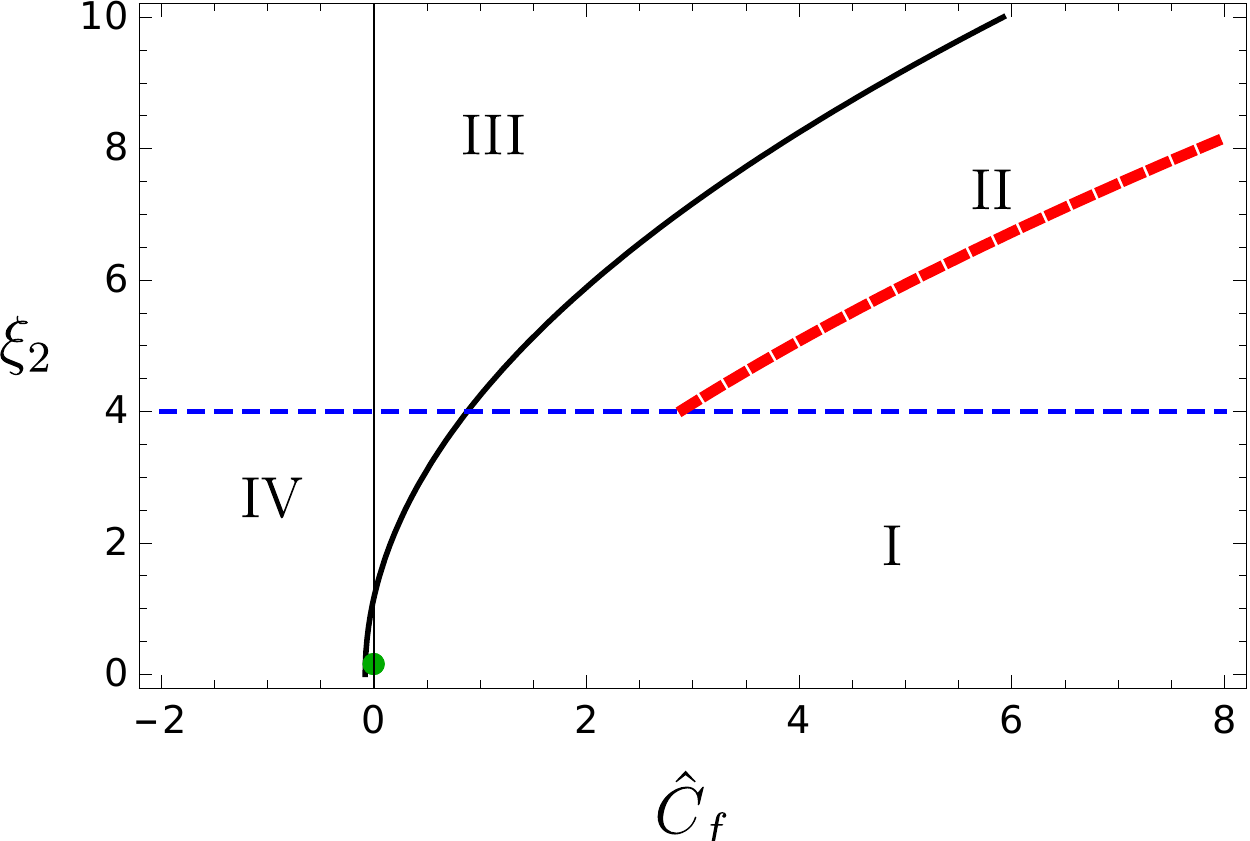}\vskip3mm
		I\hskip2mm\begin{minipage}{0.25\columnwidth} \includegraphics[width=\columnwidth]{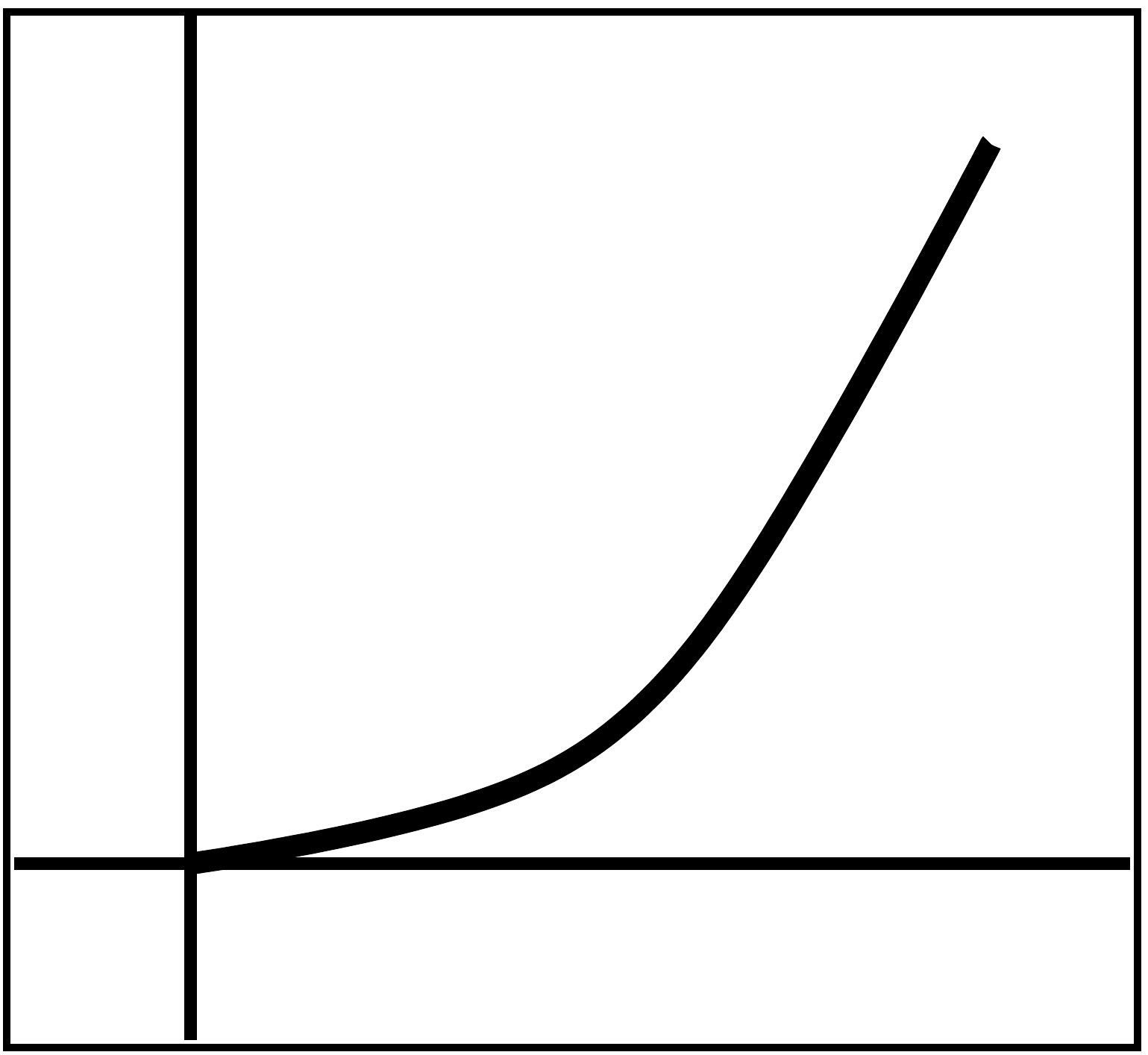}
		\end{minipage}
		\hskip10mm
		II\hskip2mm
		\begin{minipage}{0.25\columnwidth}
			\includegraphics[width=\columnwidth]{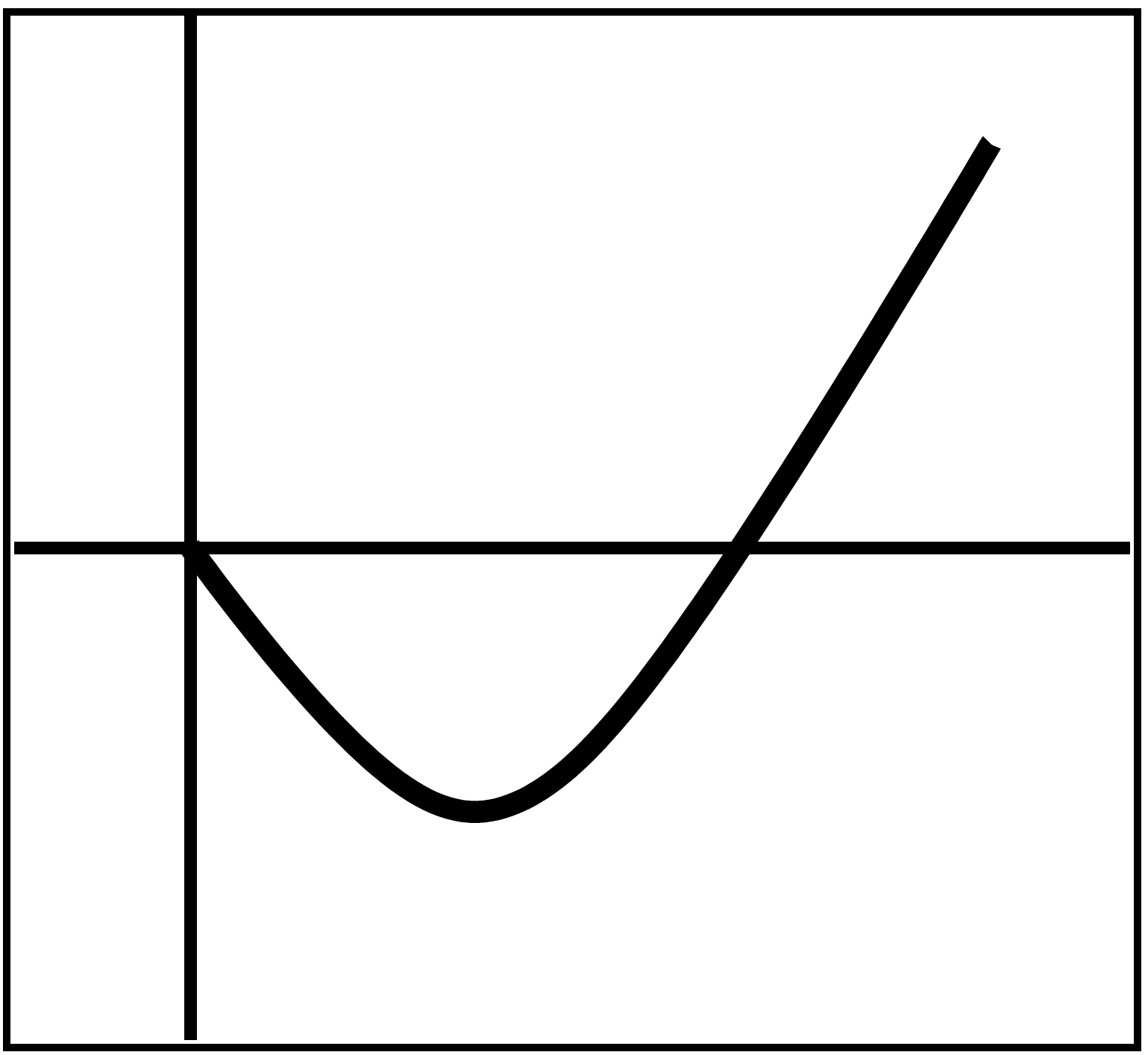}
		\end{minipage}			
		\hskip10mm
		III\hskip2mm\begin{minipage}{0.25\columnwidth}
			\includegraphics[width=\columnwidth]{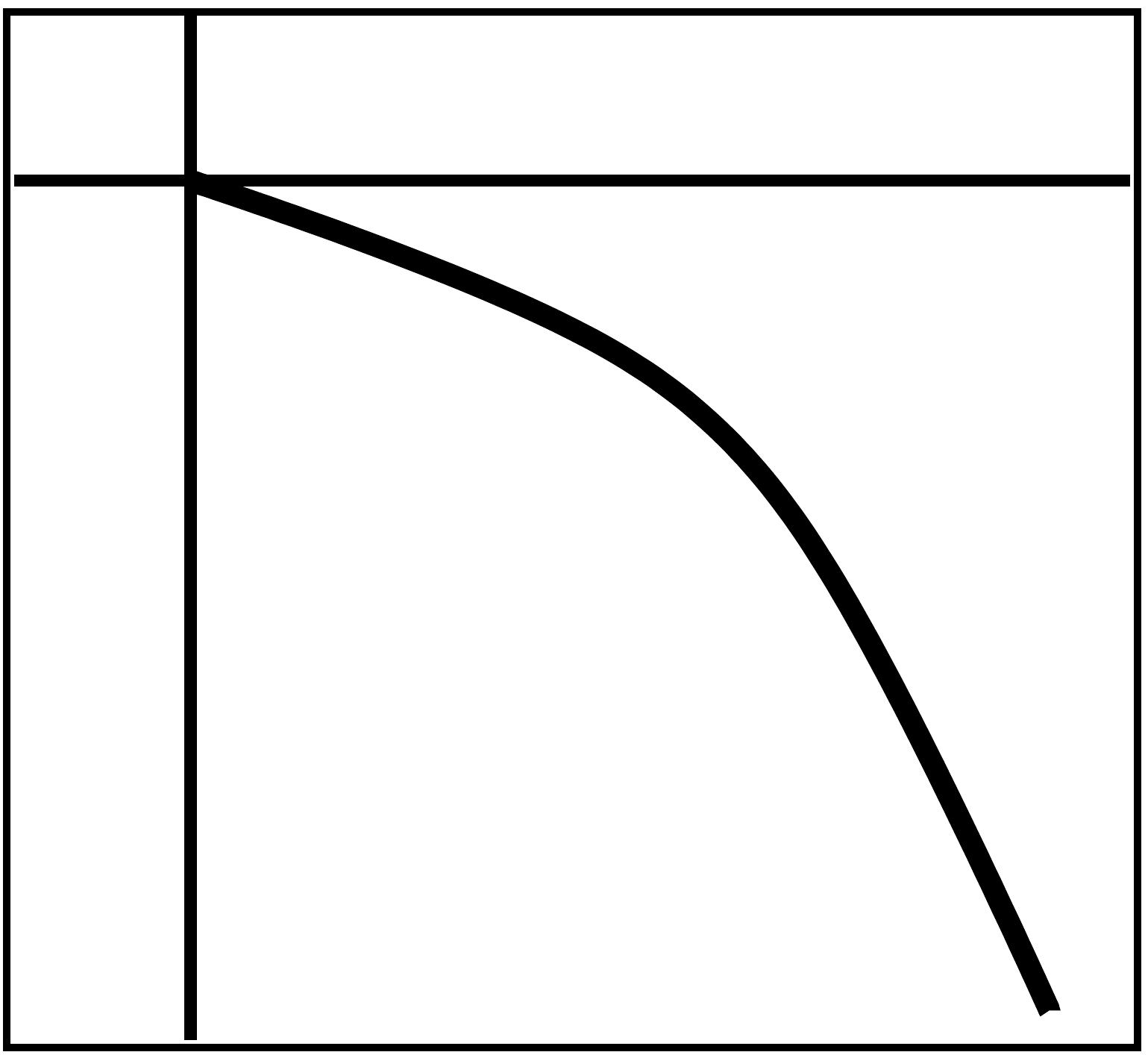}
		\end{minipage}			
		\hskip10mm
		IV\hskip2mm\begin{minipage}{0.25\columnwidth}
			\includegraphics[width=\columnwidth]{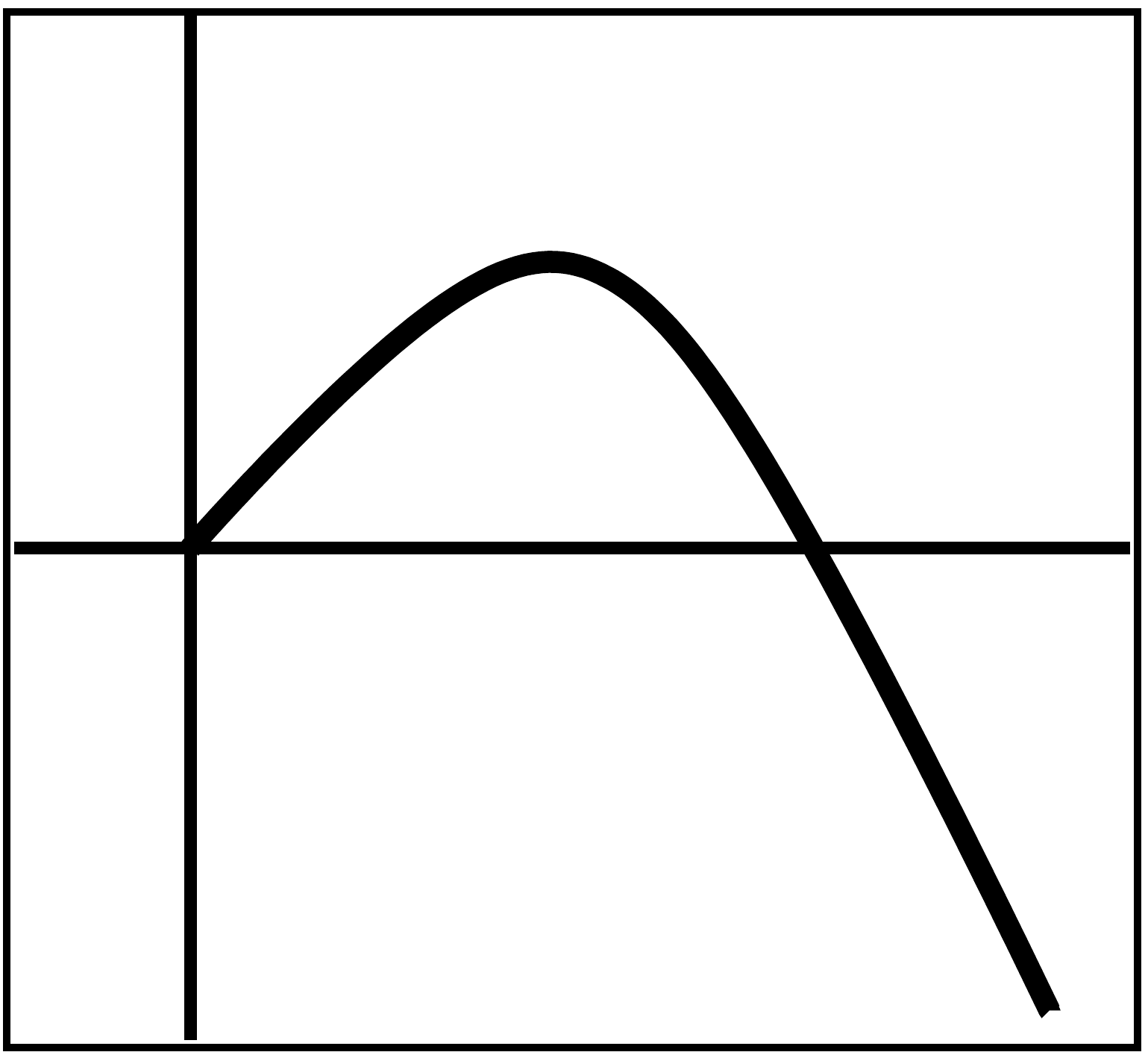}
		\end{minipage}		
	\vskip2mm
		\caption{
			Stability properties of the effective potential $f(x)$, see \Eqref{eq:f_FP_phi^4dominance}, for $P\in (0,1/2)$ (upper left) and $P=1/2$ (upper right).
			The two black lines separate the left-hand side regions where the potential is unbounded from below from the right-hand side regions where it is bounded, and in the $h^2\to 0$ limit. Their equations are obtained imposing the condition $\hat C_{\mathrm{f},\infty}=0$ in \Eqref{eq:barCf_infty}.
			Sketches of the potential
			shapes in the different regions are given in the lower panels.
			Upper right: the blue dashed line $\xi_2=4$ identifies the locus of points where $f^\prime(0)=0$.
			For $\xi_2<4$, the potential has an unstable minimum in region IV.  It is monotonically increasing to $+\infty$ in region I.
			For $\xi_2>4$,  $f(x)$ has a stable minimum in region II. It is monotonically decreasing to $-\infty$ in region III. 
			The green point $\{0,\lresc_{2}^+\}$ highlights the CEL solution, see \Eqref{eq:L2rootCEL}, being regular at $x=0$.
			The red dashed line in region II shows the one parameter family of new solutions satisfying the consistency condition $f^{\prime\prime}(x_0)=\xi_2$, as expressed in \Eqref{eq:Cf_P=1/2_new_solutions}.
			Upper left: only the regions of type II and III are present.
		}
		\label{fig:contourPlotXifLambda}
\end{center}\end{figure*}

For finite values of $h^2$, we can investigate the asymptotic behavior in the interval $(x_{2},\infty)$ by expanding the QFP potential in \Eqref{eq:f_FP_phi^4dominance} around $x=\infty$.
The analytic expansion yields
\begin{align}
	f(x)= C_{\mathrm{f},\infty}\, x^\frac{4}{2+\eta_x}+\mathcal{O}(x^{-1}),
	\label{eq:leading_largex_exp}
\end{align}
where the asymptotic coefficient in front of the scaling term depends on the different parameters characterizing the RG trajectory $C_{\mathrm{f},\infty}(C_{\mathrm{f}}, \xi_2, h^2, P)$.
The full expression is given in \Appref{app:largefield_f}, cf. \Eqref{eq:C_f,infty}. 
%Its $h^2$ dependence can be studied by expanding for small Yukawa coupling.
We investigate its $h^{2}$ dependence in the deep UV by an expansion at vanishing Yukawa coupling. 
This yields a scaling $C_{\mathrm{f},\infty}\, {\sim}\, h^{-2(1-2 P)}$ for $P\in (0,1/2)$ and $C_{\mathrm{f},\infty}\, {\sim}\, h^{-2(2 P-1)}$ for $P\in (1/2,1)$ for fixed $C_{\mathrm{f}}$.
%From this expansion it is possible to infer that at fixed $C_{\mathrm{f}}$ the coefficient $C_{\mathrm{f},\infty}$ scales as $h^{-2(1-2 P)}$ for $P\in (0,1/2)$, while it scales as $h^{-2(2 P-1)}$ for $P\in (1/2,1)$. 
We call $\hat C_{\mathrm{f},\infty}$
the corresponding finite ratio.
For the sake of clarity, it is therefore useful to define a new variable
\be
\hat C_{\mathrm{f}}=\left\{
\begin{aligned}
&h^{2(1-2P)}C_{\mathrm{f}}		& &\text{if}\quad P\in (0,1/2),\\
&C_{\mathrm{f}} 				& &\text{if}\quad P=1/2,\\
&h^{2(2P-1)}C_{\mathrm{f}}  	& &\text{if}\quad P\in (1/2,1).
\end{aligned}\right.\label{eq:barCf_P}
\ee
From this rescaling we obtain that the asymptotic coefficient has to be
\be
\hat C_{\mathrm{f},\infty}=\left\{
\begin{aligned}
&\hat C_{\mathrm{f}}-\frac{9}{64\pi^2\hat\eta_x}\xi_2^2  			& &\text{if}\quad P\in (0,1/2),\\
&C_{\mathrm{f}} - \frac{9\xi_2^2 - 12}{64\pi^2\hat\eta_x}	& &\text{if}\quad P=1/2,\\
&\hat C_{\mathrm{f}}+\frac{3}{16\pi^2\hat\eta_x}  	& &\text{if}\quad P\in (1/2,1),
\end{aligned}\right.\label{eq:barCf_infty}
\ee
in leading order in $h^2$ where $\hat\eta_x=\eta_x/h^2$.
The locus of points that satisfies the condition $\hat C_{\mathrm{f},\infty}=0$ for $P\leq 1/2$ are plotted in \Figref{fig:contourPlotXifLambda} by black lines. They characterize the transition from the region in the $(\hat C_{\mathrm{f}},\xi_2)$ plane where the potential is bounded from below (right side) to the region where the potential is unbounded (left side).

%%%%%%%%%%%%%%%%%%%%%%%%%%%%%%%%%
\subsection{Small-field behavior and the CEL solution} \label{subsec:small_field}
%%%%%%%%%%%%%%%%%%%%%%%%%%%%%%%%%

Next, we study the properties of the solution $f(x)$ for small arguments $x \ll 1$.
This is relevant to address both the $x\to 0$ limit at fixed $h^2$, and also to inspect the large field asymptotics for $P<1$ in the limit where $h^2\to 0$ and $x\to+\infty$ at $z_{\rmB/\rmF}\ll 1$.
For this purpose, we start from the expansion of the QFP potential $f(x)$ for small $x$, which can be found in \Appref{app:largefield_f}, cf. \Eqref{eq:f(x)_expansion_small_z_B/F}.
The Gau\ss\  hypergeometric functions are analytical for small $x$, but the scaling term is not, due to the nonrational power of $x$.
The first derivative at the origin is
\be
	f^\prime(0)=\frac{12 - 3 h^{2(2P-1)}\xi_2}{32\pi^2 (2-\eta_x)}h^{2(1-P)},
\ee
thus by keeping the leading order in $h^2$ we have
\be
f^\prime(0)=\left\{
\begin{aligned}
	& - \frac{3\,\xi_2}{64\pi^2} h^{2P} & &\text{if}\quad P\in (0,1/2), \\
	&\frac{12 - 3\xi_2}{64\pi^2}\,h & &\text{if}\quad P=1/2,\\
	&\frac{3 }{16\pi^2}\,h^{2(1-P)} & &\text{if}\quad P\in (1/2,1).
\end{aligned}\right.
\ee 
Thus, we observe that $f'(0)$ is negative for $P<1/2$ and
$\xi_2>0$ while it is always positive for $P>1/2$.

For $P=1/2$, the first derivative at the origin changes sign at $\xi_2=4$.
In this case, we find that the two lines $C_{\mathrm{f},\infty}=0$ and $\xi_2=4$ divide the ($C_{\mathrm{f},\infty},\xi_{2}$) plane in four regions with different qualitative behavior for $f(x)$,
as represented in the right panel of \Figref{fig:contourPlotXifLambda} with solid black line and dashed blue line respectively.
In region II the QFP potential is bounded from below and has a nontrivial stable minimum.
In region IV the potential has a nontrivial maximum but is unbounded from below.
Instead in regions I and III the function $f(x)$ is monotonically increasing towards $+\infty$ and decreasing to $-\infty$, respectively.
For $P<1/2$, there are only regions of type II and III.

In region I, where the potential is bounded from below and its minimum
is located at the origin,
we have to check as to whether it is possible to impose the consistency condition $f^{\prime\prime}(0)=\xi_2$.
The answer is positive if we remove the log-type singularity in the second derivative at the origin by requiring $C_{\mathrm{f}}=0$.
With this choice, we obtain
\begin{align}
	&\xi_2= 3 h^{2(1-2P)}\,\frac{4  - 3h^{4(2P-1)}\xi_2^2}{32\pi^2 \hat\eta_x}& & \text{if}\quad C_{\mathrm{f}}=0
\end{align}
where the rescaled quartic scalar coupling $\xi_2$, by definition, must be finite in the $h^2\to 0$ limit.
Therefore the only possible solution is
\begin{align}
	\xi_2=\lresc_{2}^\pm,\quad P=\frac{1}{2},
\end{align}
that is precisely the CEL solution described in \Secref{sec:perturbativeoneloop}.
The positive root $\lresc_{2}^+$ is highlighted by a a green dot in the right panel of \Figref{fig:contourPlotXifLambda}.

Having constructed a full effective potential for the
CEL solution, we can ask whether this is stable for large
field amplitudes and how it is related to the $u(\rho)$ of \Eqref{uhxDER}.
As shown in \Appref{app:u(rho)_and_f(x)_compared}, we have
\begin{align}
&\lim_{d_x\to d_z} f(x)=u(\rho)
+\frac{12 - 3\xi_2}{32\pi^2 (2-\eta_z)}hx,
\label{eq:f_reduces_to_u}
\end{align}
where an irrelevant additive constant has been neglected.
Therefore the full solution $f(x)$ includes all the information about $u(\rho)$ plus a linear term that was discarded in \Secref{sec:effectiveFTdeepEuclidean}
by the definition of the DER.
Furthermore, \Eqref{eq:leading_largex_exp} 
and \Eqref{eq:barCf_infty} apply to all values of $C_{\mathrm{f}}$,
thus by choosing $P=1/2$ and $C_{\mathrm{f}}=0$ in these equations,
and specifying the QFP value of $\xi_2$, we deduce
that the asymptotic behavior for the CEL potential is
\begin{align}
	f(x)\widesim[2]{x\to \infty}	\frac{\xi_2}{2} x^2.
\end{align}
Thus, we conclude that the CEL solution is stable for arbitrary small values of the Yukawa coupling.

%%%%%%%%%%%%%%%%%%%%%%%%%%%%%%%%%
\subsection{New solutions with a nontrivial minimum} \label{subsec:small_field_new_solution}
%%%%%%%%%%%%%%%%%%%%%%%%%%%%%%%%%

Within region II, the potential is stable and has a nontrivial minimum. Here, we demand the consistency condition to hold at the minimum, $f^{\prime\prime}(x_0)=\xi_2$.
To simplify the discussion we adopt the same small-field
expansion discussed above,
which corresponds to neglecting subleading powers of $x_0$,
for small values of the vacuum expectation value.
The defining condition for the minimum, $f^\prime(x_0)=0$, gprovides an expression for $C_{\mathrm{f}}$ as a function of $x_0$, $h^2$ and $\xi_2$ which is
\begin{align}
	C_{\mathrm{f}}=  x_0^{\frac{\eta_x-2}{2+\eta_x}} \frac{h^{2(1-P)}(2+\eta_x)}{128\pi^2 (2-\eta_x)}\left[3\xi_2h^{2(2P-1)} - 12\right].
	\label{eq:Cf_of_x0,h^2,xi2}
\end{align}
The second derivative of the potential in $x_0$ is thus
\begin{align}
	f^{\prime\prime}(x_0)= \frac{3\xi_2 h^{2(2P-1)} - 12}{32\pi^2 (2+\eta_x) x_0} h^{2(1-P)},
\end{align}
which, together with $f^{\prime\prime}(x_0)=\xi_2$, provides us with an expression for the nontrivial minimum as a function of $h^2$ and $\xi_2$
\begin{align}
	x_0=\frac{3\xi_2 h^{2(2P-1)} - 12}{32\pi^2 \xi_2 (2+\eta_x)}h^{2(1-P)}.
\end{align}
Different powers of $P$ lead to different leading behaviors in $h^2$ for the latter expression. These can be summarized in the following way
\be
x_0=\left\{
\begin{aligned}
&\frac{3}{64\pi^2}h^{2P} & &\text{if}\quad P\in (0,1/2), \\
&\frac{3\xi_2 - 12}{64\pi^2 \xi_2}h & &\text{if}\quad P=1/2,\\
&-\frac{3 }{16\pi^2 \xi_2}h^{2(1-P)} & &\text{if}\quad P\in (1/2,1).
\end{aligned}
\right.
\label{eq:x0_EFT_f(x)}
\ee
These results are in agreement with the EFT analysis including thresholds presented in \Secref{sec:effectiveFTthresholds}.
In fact Eqs.~\eqref{eq:kappa_P<1/2}, \eqref{eq:kappa_P3=2}, and \eqref{eq:kappa_1/2<P<1} are identical to those in \Eqref{eq:x0_EFT_f(x)}, recalling that $x_0=h^{2P}\kappa$.

Moreover, we can substitute the expression for the minimum $x_0 (\xi_2, h^2)$ inside the parametrization for $C_{\mathrm{f}}$ in \Eqref{eq:Cf_of_x0,h^2,xi2} for $P=1/2$. Considering the leading order in $h^2$, we find
\begin{align}
	C_{\mathrm{f}}&=\frac{9\xi_2^2 - 12}{64\pi^2 \hat\eta_x}+\frac{\xi_2}{2}\quad\text{for}\quad P=\frac{1}{2},
	\label{eq:Cf_P=1/2_new_solutions}
\end{align}
that describes a one-parameter family of QFP solutions satisfying the consistency condition at the nontrivial minimum, i.e., $f^{\prime\prime}(x_0)=\xi_2$.
These solutions are represented in the right panel of \Figref{fig:contourPlotXifLambda}  as a red dashed line laying in Reg.~II.
The asymptotic behavior for the latter solutions is obtained by plugging \Eqref{eq:Cf_P=1/2_new_solutions} into \Eqref{eq:barCf_infty}. 
It turns out that these solutions obey the same asymptotic behavior as the CEL solution which is given by a quadratic function in $x$,
\begin{align}
	f(x)\widesim[2]{x\to \infty}	\frac{\xi_2}{2} x^2\quad\text{for}\quad P=\frac{1}{2}.
\end{align}

Also for $P<1/2$ it is possible to find a parametrization $C_{\mathrm{f}}(\xi_2)$ for the QFP solutions with a nontrivial minimum satisfying the consistency condition in $x_0$.
Its leading order contribution in $h^2$ reads
\begin{align}
	\hat C_{\mathrm{f}}=\frac{9\,\xi_2^2}{64\pi^2 \,\hat\eta_x}\quad\text{for}\quad P\in (0,1/2),
\end{align}
and coincides exactly with the solution to the condition $\hat C_{\mathrm{f},\infty}=0$. 
%Indeed upon substitution into \Eqref{eq:barCf_infty}, 
Thus, we find the asymptotic behavior
\begin{align}
	f(x)\widesim[2]{x\to \infty}0\quad\text{for}\quad P\in(0,1/2).
\end{align}
Therefore, the QFP solutions for $P<1/2$ are asymptotically flat.

Along these two families of QFP solutions for $P\leq1/2$, it is interesting to evaluate the rescaled cubic coupling at $x_0$.
It is given by the third derivative of the homogenous scaling part with respect to $x$ which reads
\begin{align}
  f^{\prime\prime\prime}(x_0)=-C_{\mathrm{f}}\frac{8\eta_x (2-\eta_x)}{(2+\eta_x)^3}x_0^{-\frac{2+3\eta_x}{2+\eta_x}}.
\end{align}
By inserting $x_0(\xi_2,h^2)$ and $C_{\mathrm{f}}(\xi_2)$, the leading contribution in $h^2$ is given by
\be
\xi_3=\left\{
\begin{aligned}
&-6\xi_2^2h^{2P} & &\text{if}\quad P\in (0,1/2), \\
&-2\xi_2 h\, \frac{9\xi_2^2+32\pi^2\hat\eta_x\xi_2 - 12}{3\xi_2 - 12} & & \text{if}\quad P=1/2. 
\end{aligned}\right. 
\label{eq:xi3_f(x)}
\ee
From the definitions \eqref{eq:lambdatoLn} and \eqref{eq:xin_def}, we deduce that the transformation between the rescaled cubic coupling for $f(x)$ and the finite ratio $\lresc_{3}$ is
\be
	 \xi_3=\lresc_{3} h^{2(P_3-3P)}.
\ee
From \Eqref{eq:xi3_f(x)} we can conclude that $P_3=2$ for $P=1/2$ and $P_3=4P$ for $P\in(0,1/2)$.
This $h^2$-dependent behavior is in agreement with the EFT analysis including thresholds described in \Secref{sec:effectiveFTthresholds}.
However, the expression for the finite ratio $ \lresc_{3}$ is different, since we are treating the threshold functions in the $\phi^4$-dominance approximation in this section.

Finally, let us summarize once more the results of the fixed-point potential analysis for $f(x)$ and for general $P<1$.
Starting from a pure quartic scalar interaction for the potential given by $\lambda_2\rho^2/2=\xi_2 x^2/{2}$ with a trivial minimum at the origin, we obtain a QFP potential of the same type and with the required property $f^{\prime\prime}(0)=\xi_2$ only for the particular choice for the parameters $\{P,C_{\mathrm{f}},\xi_2\}=\{1/2,0,\lresc_{2}^\pm\}$.
This is the CEL solution. We argued that it is stable with a well defined asymptotic behavior in the combined limit $x\to\infty$ and $h^2\to 0$.
In addition for $P\leq1/2$, we discovered in the $(\hat C_{\mathrm{f}},\xi_2)$ plane the existence of a one-parameter family of \emph{new} solutions.
Despite the presence of a log-type singularity at the origin, these solutions have a nontrivial minimum $x_0$ which satisfies the consistency condition $f^{\prime\prime}(x_0)=\xi_2$.
For $P=1/2$ these new solutions are stable and present the same quadratic asymptotic behavior as for the CEL solution.
For $P<1/2$, the QFP potential becomes asymptotically flat in the combined limit $x\to\infty$ and $h^2\to 0$, because $\hat C_{\mathrm{f},\infty}=0$.

%%%%%%%%%%%%%%%%%%%%%%%%%%%%%%%%%%%%%%%%
\section{Full effective potential in the weak-coupling expansion} \label{sec:weakhexpansion}
%%%%%%%%%%%%%%%%%%%%%%%%%%%%%%%%%%%%%%%%

%
Let us discuss yet another analytic functional approximation, obtained by expanding the full functional equation for the rescaled potential $f(x)$ in powers of $h^2$.
The one-loop flow equation for $f(x)$ takes the form
\be
\partial_t f =-4f+d_x x f^\prime+\frac{1}{32\pi^2}\frac{1}{1+\omega_f}
-\frac{\Nc d_\gamma}{32\pi^2} \frac{1}{1+\omega_{1f}}, \qquad \label{eq:betafFull1Loop}
\ee
where we have chosen again the piecewise linear regulator for the
evaluation of the
threshold functions $l_0^{(\rmB/\rmF)}$, as in \Eqref{eq:betau}, which parametrize the result of the boson/fermion loop integrals.
Here, $d_x$ is the same as in \Eqref{eq:dx_def} and represents the quantum dimension of $x$.
The arguments $\omega_f$ and $\omega_{1f}$, defined as
\be
\begin{split}
\omega_f&=h^{2P}\left(f'+2xf''\right), \qquad
\omega_{1f}&=h^{2-2P}x,
\label{eq:omegaf}
\end{split}
\ee
are related to the scalar and Yukawa vertices, respectively.

The dimension of the rescaled field $d_x$ depends on $\eta_\phi$ and $\eta_{h^2}$ and thus is of order $h^{2}$, cf. \Eqref{eq:etaphi_1loop} and \Eqref{eq:etahath2}. 
Therefore, they can be neglected for $P<1$ and at leading order in $h^2$ the flow equation can be written as
%The contributions in $d_x$ due to $\eta_\phi$ and $\eta_{h^2}$ are of order $h^2$, cf. \Eqref{eq:etaphi_1loop} and \Eqref{eq:etahath2}, therefore for $P<1$ they can be neglected and at the leading order in $h^2$ the flow equation can be written as
\be
	\beta_f=\left[\beta_f\right]_{h^2\to 0}+\delta \beta_f,		\label{eq:beta_f_weak-h}
\ee
where the first term is just the $\beta$-function in the limit $h^2\to 0$ and the second one can be derived from the expansion of the bosonic and fermionic loops.
An $h^2$-independent contribution from the quantum fluctuations is
present only for $P=1$, and equals the fermion loop. Therefore
in $d=4$ one has
\be
	\left[\beta_f\right]_{h^2\to 0}=
\begin{cases}
- 4 f + 2 x f^\prime , \quad &\text{$P<1$,} \\	
 - 4 f + 2 x f^\prime - \frac{\Nc d_\gamma}{32\pi^2}
  \frac{1}{1+x}, \quad&\text{$P=1$.}
\end{cases} \qquad
\label{eq:zerothorder_betaf}
\ee
For $P<1$ the zeroth order in $h^2$ is trivial since no quantum fluctuations are retained.
On the other hand for $P=1$, the properties of the QFP solutions
depend on the current choice of the regulator.
Let us now discuss all interesting cases, for $P\leq 1$. 
For the case $P>1$, we demonstrate in \Appref{app:P>1} that no reliable solution can be constructed which is compatible with our assumptions and approximations.

%%%%%%%%%%%%%%
\subsection{$P\in(0,1/2)$}
%%%%%%%%%%%%%%

In this case only the scalar loop contributes to the first order correction to $\beta_f$. The scalar vertices scale like $h^{2P}$. 
Therefore $\delta\beta_f$ can be approximated by the linear term of a Taylor
expansion of the scalar threshold function  
at vanishing argument, reading
\be
	\delta \beta_f= -\frac{1}{32\pi^2} \, h^{2P} (f^\prime+2xf^{\prime\prime}).
\ee
Upon inclusion of this leading order correction, the flow equation now becomes a second order ODE that can be solved analytically. 
We find two linearly independent solutions. The first is given by the following polynomial 
\be
&f(x)=c\left(x^2-\frac{3}{32\pi^2} h^{2P}x\right).
\label{eq:fCf_weakh_Pless1/2}
\ee
where $c$ is an integration constant. 
The second solution grows exponentially for large field amplitudes.
However, we are only interested in solutions that obey power-like scaling for $x\to\infty$, since
in this case a scalar product can be defined on the space of eigenperturbations of these
solutions~\cite{Morris:1996nx,ODwyer:2007brp,Bridle:2016nsu}. 
Thus, we set the second integration constant to zero.

Clearly, the solution in \Eqref{eq:fCf_weakh_Pless1/2} is easily translated into the polynomial language of \Secref{sec:effectiveFTthresholds}, by identifying
\be\begin{aligned}
	\xi_2&=2 c, \\
	x_0&=\frac{3}{64\pi^2} h^{2P}.
	\end{aligned}
\ee
which agrees with \Eqref{eq:kappa_P<1/2} and \Eqref{eq:L2_P<1/2}.

%%%%%%%%%%%%%%
\subsection{$P=1/2$}
%%%%%%%%%%%%%%

For $P=1/2$, both the scalar and the fermion loops contribute to the first correction of $\beta_f$ that is
\be
	\delta \beta_f=-\frac{1}{32\pi^2} h (f^\prime+2xf^{\prime\prime})+\frac{1}{32\pi^2} \Nc d_\gamma h x.
\ee
The QFP equation is again a second order ODE whose analytic solution will have two integration constants. 
Again, we discard the solution which scales exponentially for large $x$ by setting the corresponding integration constant to zero.
The remaining solution is a quadratic polynomial that has a free quartic coupling $\xi_2$ and a minimum at
\be
	x_0=h\frac{3\xi_2 - \Nc d_\gamma }{64\pi^2\xi_2}.
\ee
By setting $\Nc = 3$ and working with an irreducible representation of the Clifford algebra in $d=4$, i.e., $d_\gamma=4$, one recovers the result of \Secref{subsec:effectiveFTthresholdsP1half} for $P_3>1$. As in that case, the nontrivial minimum only exists if $\xi_2>4$.
The straightforward generalization of this requirement reads
\be
\xi_2>\frac{\Nc d_\gamma }{3}
\ee
for a generic field content.

%%%%%%%%%%%%%%
\subsection{$P\in (1/2,1)$}
%%%%%%%%%%%%%%

In this case only the fermion loop contributes to the first correction of $\beta_f$ and is given by
\be
\delta \beta_f=\frac{1}{32\pi^2} \Nc d_\gamma h^{2(1-P)} x.
\ee
The differential equation remains a first order ODE and its analytical solution is
\be
f(x)=\frac{\xi_2}{2} x^2+\frac{1}{64\pi^2} \Nc d_\gamma h^{2(1-P)}x,
\ee
where $\xi_2$ is an arbitrary integration constant.
For any color number or representation of the Clifford algebra, the potential exhibits only the trivial minimum at vanishing field amplitude and thus the QFP solution is in the symmetric regime.
In fact the corresponding nontrivial minimum
\be
x_0=-\frac{\Nc d_\gamma}{64\pi^2\xi_2}h^{2(1-P)},
\ee
would be negative for any positive $\xi_2$.
This is again in agreement with the EFT analysis, cf. \Eqref{eq:kappa_1/2<P<1} and \Eqref{eq:L2_1/2<P<1}.

For all values of $P<1$ in the present approximation,
 we have obtained QFP solutions which are analytic in $x$.
 In \Secref{sec:effectiveFTthresholds}, this was implemented by construction,
since we have projected the functional flow equation onto a polynomial ansatz.
In the present analysis, this happens because the contributions to $\beta_f$ producing non-analyticities are accompanied by subleading powers of $h^2$ for $P<1$.
Indeed, both the anomalous dimension of $x$ and contributions from
the loops proportional to $x^2$ would produce a logarithmic
singularity of $f''(x)$ at $x=0$ for any $h^2\neq0$, as discussed in \Secref{subsec:small_field}, see also below. 
Knowing about the presence of this singularity for any $P$ for $h^2\neq 0$, 
we can accept the previous solutions
only if $x_0>0$, which appears to be impossible for $P\in (1/2,1)$.

%%%%%%%%%%%%%%%%%%
\subsection{$P=1$}   \label{sec:weak-h^2_P=1}
%%%%%%%%%%%%%%%%%%

As shown in \Eqref{eq:zerothorder_betaf},
already the zeroth order in $h^2$ accounts for nontrivial dynamical effects for $P=1$. The corresponding QFP solution 
for the piecewise linear regulator is
\be
f(x)=	c \,x^2+\frac{\Nc d_\gamma}{64\pi^2} \left(x+x^2\log\frac{x}{1+x}\right).
\label{eq:QFPf_Pone_weakh}
\ee
The second derivative of this potential has a log-type singularity at the origin.
We expect that this feature  survives also in the full $h^2$-dependent solution,
as addressed in the next section. 

The freedom in the choice of the parameter $c$ allows to
construct physical QFP solutions, that avoid the divergence at 
small fields by developing a nontrivial minimum.
The defining equation $f'(x_0)=0$ for this minimum,
where $f$ is given by \Eqref{eq:QFPf_Pone_weakh},
can 
straightforwardly be solved for $c$ in terms of $x_0$.
From the point of view where the latter is the free parameter
labeling the QFP solutions, the natural question then is
as to whether it can be chosen such that $f''(x_0)=\xi_2$ is positive and finite
for $h^2\to 0$.
The answer is negative, since the piecewise linear regulator gives
\be
\xi_2=-\frac{\Nc d_\gamma}{64\pi^2 x_0 (1+x_0)^2},
\ee
which is in agreement with \Eqref{eq:L2_P=1}.

%%%%%%%%%%%%%%%%%%%%%%%%%%%%%%%%%%%%%%%%
\section{Numerical solutions} \label{sec:numeric_sol}
%%%%%%%%%%%%%%%%%%%%%%%%%%%%%%%%%%%%%%%%

In this section, we test our previous analytical results by integrating numerically the full one-loop nonlinear flow equation for $f(x)$ as in \Eqref{eq:betafFull1Loop}, where we have computed the threshold functions $l_0^{\rmB/\rmF}$ in Eq.~\eqref{eq:betau} by choosing the piece-wise linear regulator.
We make a further approximation evaluating the anomalous dimensions $\eta_\phi$, $\eta_\psi$ and $\eta_{h^2}$ in the DER, leading to the expressions in Eqs.~\eqref{eq:etaphi_1loop}~,~\eqref{eq:etahath2} and \eqref{eq:eta_psi_one-loop}.
We are moreover interested in the $P=1/2$ case characterized by the existence of the CEL solution, regular at the origin, and a family of new QFP potentials, singular in $x=0$ but featuring  a nontrivial minimum $x_0\ne 0$.
To address this numerical issue we exploit two different methods.
First, we study the global behavior of the CEL solution using pseudo-spectral methods. And second, we corroborate the existence of the new QFP family of solutions using the shooting method.
s

\begin{figure}[!t]
	\begin{center}
		\includegraphics[width=0.95\columnwidth]{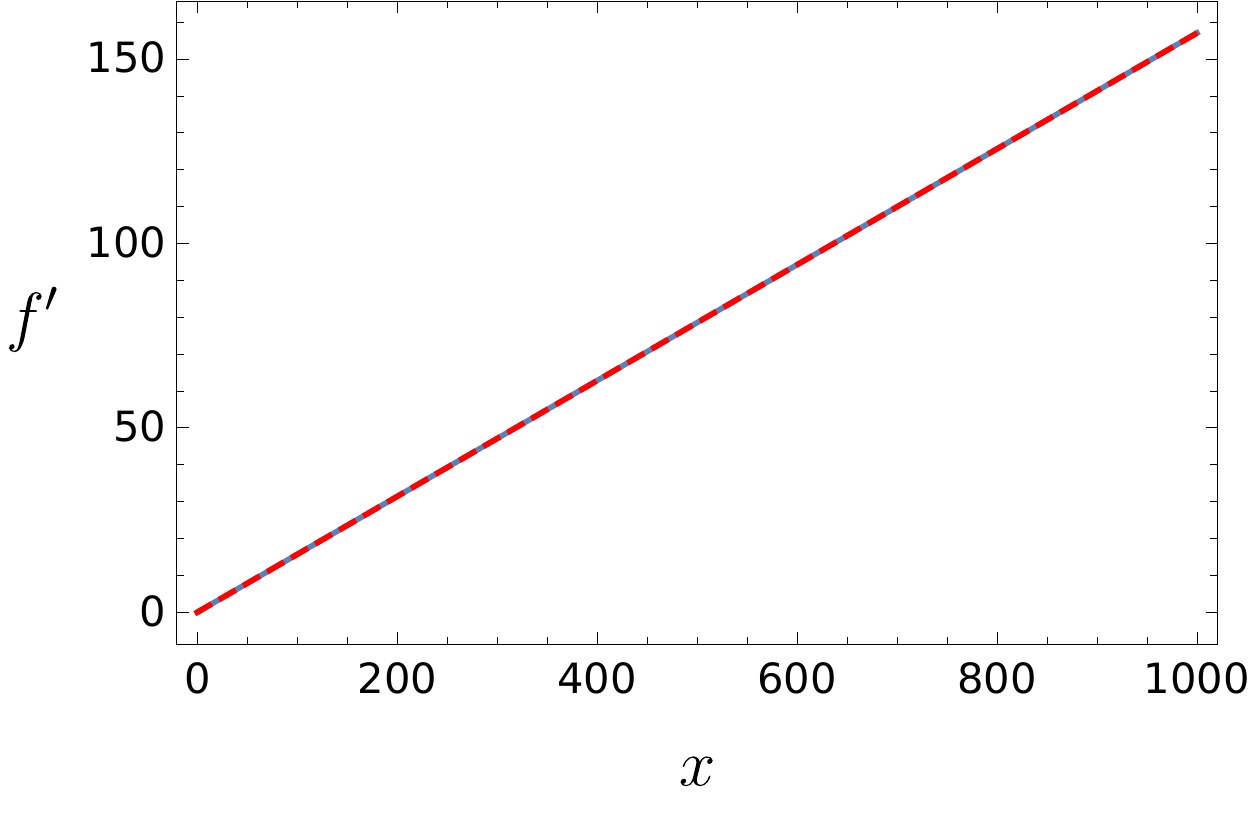}
	\end{center}
	\vskip-5mm
	\caption{CEL solution: first derivative of the potential $f^\prime(x)$ for $h^2=10^{-4}$,  $P=1/2$ and $\xi_2=\lresc_{2}^+$ as in \Eqref{eq:L2rootCEL}. The numerical solution of the full one-loop flow equation obtained from the pseudo-spectral method (solid blue line) lies exactly on top of the analytic solution in \Eqref{eq:f_FP_phi^4dominance} for $C_{\mathrm{f}}=0$ (red dashed line).}
	\label{fig:df_of_x_spectral_methods}
\end{figure}

%%%%%%%%%%%%%%%%%
\subsection{Pseudo-spectral methods}
%%%%%%%%%%%%%%%%%

Pesudo-spectral methods provide for a powerful tool to numerically
solve functional RG equations, provided the desired solution can be
spanned by a suitable set of basis functions. Here, we are interested
in a numerical construction of global properties of the QFP function
$f(x)$. We follow the method presented in \cite{Borchardt:2015rxa}, as
this approach has proven to be suited for this purpose, see
\cite{Borchardt:2016xju,Borchardt:2016pif,Borchardt:2016kco,Heilmann:2014iga}
for a variety of applications, and \cite{Fischer:2004uk} for earlier
FRG implementations; a more general account of pseudo-spectral methods can be found
in \cite{Boyd:ChebyFourier,Robson:1993,Ansorg:2003br,Macedo:2014bfa}.

In order to solve the differential equation given by \Eqref{eq:betafFull1Loop}
globally on $\mathbb R_+$, the strategy is to decompose the potential $f(x)$ into two series of Chebyshev polynomials.
The first series is defined over some domain $[0,x_\mathrm{M}]$ and is spanned in terms of Chebyshev polynomials of the first kind $T_i(z)$.
The second series is defined over the remaining infinite domain $[x_\mathrm{M},+\infty)$ and expressed in terms of rational Chebyshev polynomials $R_i(z)$.
Moreover, to capture the correct asymptotic behavior of $f(x)$, the latter series is multiplied by the leading asymptotic term $x^{d/d_x}$, which is in fact the solution of the homogeneous scaling part of \Eqref{eq:betafFull1Loop}.
Finally the ansatz reads
\be
	f(x)=\left\{\begin{aligned}
	&\sum_{i=0}^{N_a}\, a_i\, T_i\left(\frac{2x}{x_\mathrm{M}} - 1\right),&\quad x\le x_\mathrm{M}, \\
	&x^\frac{d}{d_x}\,\sum_{i=0}^{N_b}\, b_i\, R_i\bigl(x-x_\mathrm{M}),&\quad x\ge x_\mathrm{M}.
	\end{aligned}\right.
\ee

We thus convert the initial equation into an algebraic set of $N_a+N_b+2$ equations that can be solved applying the collocation method, for example by choosing the roots of  $T_{N_a+1}$ and $R_{N_b+1}$.
At the matching point $x_\mathrm{M}$, the continuity of $f(x)$ and $f^\prime(x)$ must be taken into account. The solutions presented in the following are obtained by choosing $x_\mathrm{M}=2$. We have further examined that the results do not change once $x_\mathrm{M}$ is varied.

In \Figref{fig:df_of_x_spectral_methods}, we compare the first derivative $f^\prime(x)$ obtained from this pseudo-spectral method and the analytical solution derived from the $\phi^4$-dominance EFT approximation, see \Eqref{eq:f_FP_phi^4dominance}, for a fixed value of $h^2=10^{-4}$ and $\xi_2=\lresc_{2}^+$. The two solutions lie perfectly on top of each other within the numerical error.
Moreover, the coefficients $a_i$ and $b_i$ exhibit an exponential decay with increasing $N_a$ and $N_b$ -- and thus indicate an exponentially small error of the numerical solution -- until the algorithm hits machine precision.

The pseudo-spectral method thus allows us to provide clear numerical evidence for the global existence of the CEL solution within the full non-linear flow equation in the one-loop approximation.
To our knowledge, this is the first time that results about global stability have been obtained for the scalar potential of this model.

We emphasize that the expansion around the origin in Chebyshev polynomials is an expansion over a set of basis functions that are in $C^\infty$.
Unfortunately, they do not form a suitable basis for the new QFP solutions parametrized by $C_{\mathrm{f}}(\xi_2)$ as in \Eqref{eq:Cf_P=1/2_new_solutions}, because of the presence of the log-type singularity at the origin.
Naively applying the same pseudo-spectral methods to this case does, in fact, not lead to numerically stable results. 

 \begin{figure}[!t]
 	\begin{center}
 		\includegraphics[width=0.95\columnwidth]{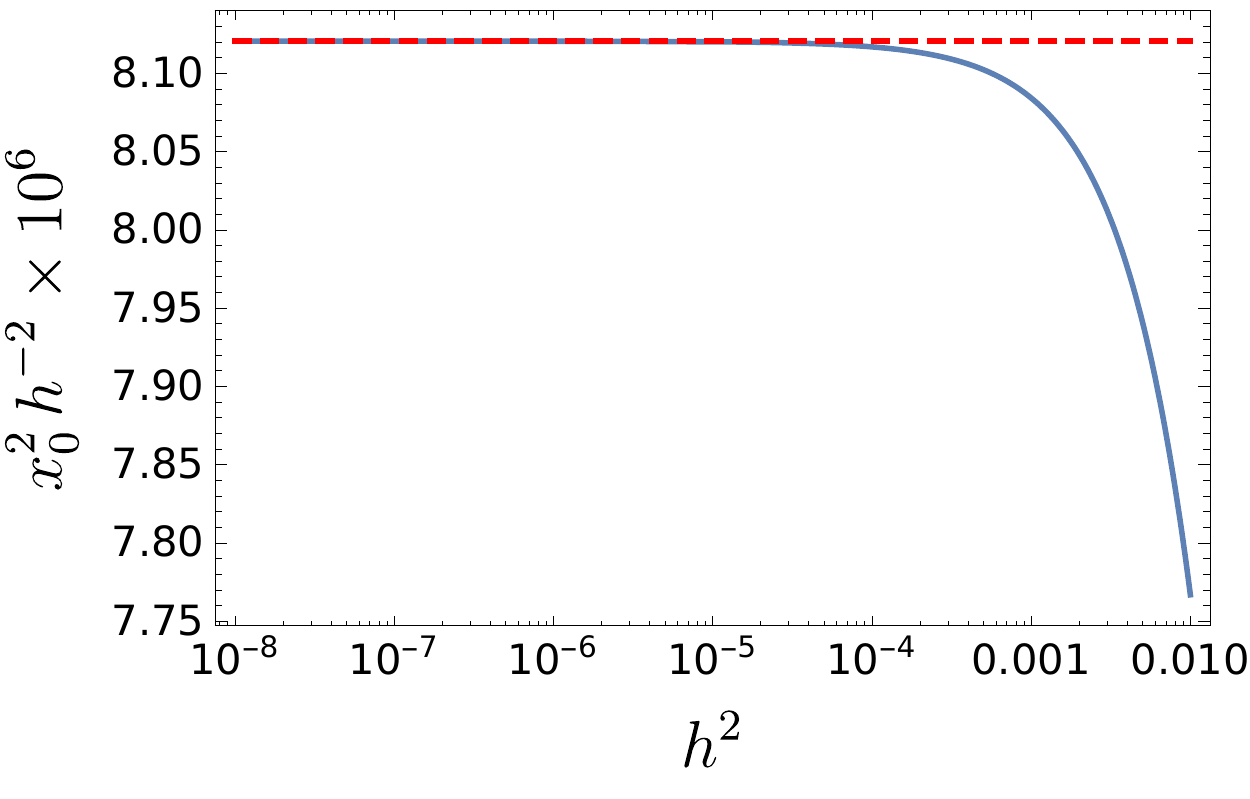}
 	\end{center}
 	\vskip-5mm
 	\caption{The ratio $x_0^2/h^2$ as a function of $h^2$ for $P=1/2$ and $\xi_2=10$. The solid line represents the numerical solution from the shooting from the minimum whereas the dashed line represents the analytic solution, which can be found in \Eqref{eq:x0_EFT_f(x)}.}
 	\label{fig:x0_over_h_of_h}
 \end{figure}

%%%%%%%%%%%%%%%%%
\subsection{Shooting method}
%%%%%%%%%%%%%%%%%

Let us therefore use the shooting method that allows to deal with the presence of the log-singularity to some extent. For this, we integrate \Eqref{eq:betafFull1Loop} starting from the minimum $x_0$ towards both the origin and infinity.
The boundary conditions that have to be fulfilled are
\be
	f^\prime(x_0)=0,\quad f^{\prime\prime}(x_0)= \xi_2,
\ee
which are just the definitions of the minimum and the quartic coupling.
The set of parameters is $x_0$, $\xi_2$, and $h^2$. For the present type of equations, it is well known that the integration outwards $x\to + \infty$ is spoiled by the presence of a movable singularity $x_{s+}$ \cite{Morris:1998da,Morris:1996nx,Morris:1994ki,Codello:2012sc,Codello:2012ec,Vacca:2015nta}. Here, the solutions from shooting develop a peak of maximum value of $x_{s+}$ only for a particular choice of initial parameters.
In the latter 3-dimemsional space, we therefore have a surface that can be parametrized, for example, by $x_0(\xi_2,h^2)$.
In the $\phi^4$-dominance EFT, we have seen that the leading contribution in $h^2$ to the nontrivial minimum $x_0$ is given by \Eqref{eq:x0_EFT_f(x)} for $P=1/2$.
\Figref{fig:x0_over_h_of_h} shows how the full numerical solution converges to the analytical one in the limit $h^2\to 0$ for the fixed value of $\xi_2=10$.
Repeating the numerical analysis for different values of $\xi_2>4$,  we find a similar agreement with the analytic solution in all studied cases. 

\begin{figure}[!t]
	\begin{center}
		\includegraphics[width=0.95\columnwidth]{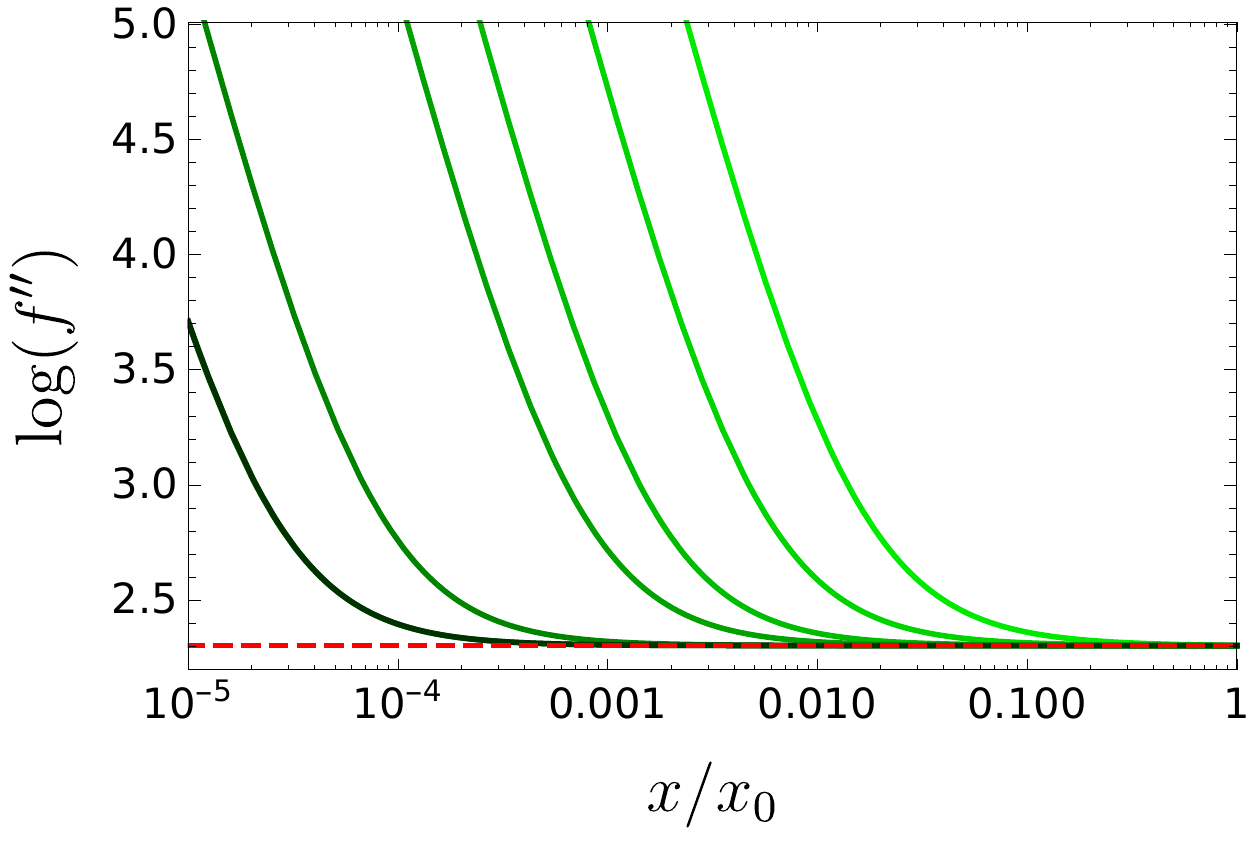}
	\end{center}
	\vskip-5mm
	\caption{$\text{Log}\, (f^{\prime\prime})$ as a function of $x/x_0$ for $P=1/2$ and fixed value of $\xi_2=10$. Comparison between the numerical solutions from the shooting from the minimum (solid lines) and the analytic ones (dashed lines) for different values of the Yukawa coupling: $h^2\in\{10^{-8},10^{-7},10^{-6},3\cdot10^{-5},10^{-4},6\cdot10^{-4},3\cdot 10^{-3}\}$ from the darker (left) to the lighter (right) curve.}
	\label{fig:ddf_of_x}
\end{figure}

\begin{figure}[!t]
	\begin{center}
		\includegraphics[width=0.95\columnwidth]{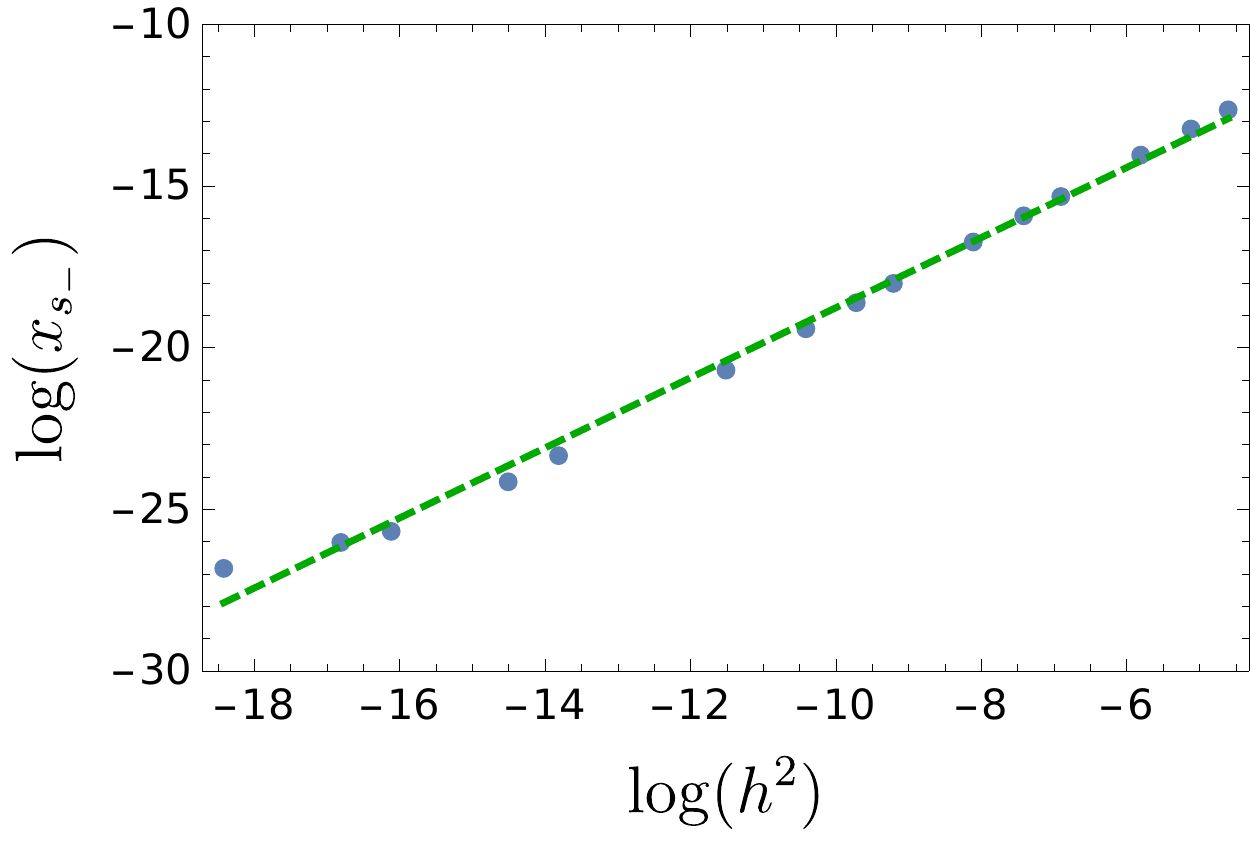}
\end{center}
\vskip-5mm
		\caption{Measure of the singular region due to the presence of the fixed singularity in $x=0$ as a function of $h^2$. The with is estimated by the criteria $f^{\prime\prime}(x_{s-})=4$. For fixed $\xi_2=10$ we have found that the numerical points are well approximated by the power law (dashed line) $x_{s-}\sim h^{2 a}$ where $a\simeq 1.084$.}
		\label{fig:xs-function_of_h}
\end{figure}

Additionally, we have also seen in the $\phi^4$-dominance EFT approximation that the family of solutions with a nontrivial minimum are singular at the origin from the second derivative on.
Very close to the origin this fixed singularity in $f^{\prime\prime}(x)$ may spoil standard integration algorithms and the numerical integration stops at some $x_{s-}$ value.
This kind of feature has been studied also in the non-abelian Higgs model \cite{Gies:2016kkk}.
In principle, these singularities in higher derivatives could contradict asymptotic freedom if they persisted in the $h^2\to 0$ limit.
To verify that this is not the case, we first analyze the behavior of $f^{\prime\prime}(x)$ close to the origin and compare it to the analytic one.
From \Eqref{eq:f_FP_phi^4dominance}, we know that the term responsible for the fixed singularity is the scaling term $C_{\mathrm{f}} x^{d/d_x}=C_{\mathrm{f}} x^{4/(2+\eta_x)}$. Indeed, taking the log of the second derivative gives
\be
	\log f^{\prime\prime}(x)\sim-\frac{2\eta_x}{2+\eta_x}\log x\quad\text{for $x\to 0$}.
\ee
In \Figref{fig:ddf_of_x}, we depict how the numerical solutions (green lines) deviate from this analytic one (dashed line) close to the origin and for different values of $h^2$ at fixed $\xi_2=10$.
This plot shows that the region of discrepancy progressively shrinks as $h^2$ gets smaller and smaller: indeed for smaller values of $h^2$ the point where the numerical solution deviates from the analytic one moves towards smaller values.
To measure this region, we have determined the onset of the singularity close to the origin as a function of $h^2$. Following the same idea as in \cite{Gies:2016kkk}, the criteria is to compute the position of $x_{s-}$ where $f^{\prime\prime}(x_{s-})$ assumes a sufficiently large value, let us say $\log f^{\prime\prime}(x_{s-})=4$.
An estimate of $x_{s-}$ is shown in \Figref{fig:xs-function_of_h} where a fit of the data confirms that the singular region shrinks to zero for $h^2\to 0$. In fact we have found numerically a power law $x_{s-}\sim h^{2a}$ with $a\simeq 1.084$ for the present model.

We conclude that the existence of the new solutions is confirmed with the shooting method. We find satisfactory qualitative agreement with the solutions identified in the $\phi^4$-dominance effective field theory approximation, which are singular at the origin and show a nontrivial minimum.

%%%%%%%%%%%%%%%%%%%%%%%%%%%%%%%%%%%%%%%%
\section{Conclusions} \label{sec:conclusions}
%%%%%%%%%%%%%%%%%%%%%%%%%%%%%%%%%%%%%%%%

Models that feature the existence of asymptotically free RG
trajectories represent ``perfect'' quantum field theories in the sense
that they could be valid and consistent models at any energy and
distance scale. Identifying such RG trajectories hence provides
information that can be crucial for our attempt at constructing
fundamental models of particle physics.  Based on the observation that
part of the standard model including the Higgs-top sector exhibits a
behavior reminiscent to an asymptotically free trajectory, we have
taken a fresh look at asymptotic freedom in a gauged-Yukawa model from a
perspective that supersedes conventional studies within standard
perturbation theory.

Gauged-Yukawa models form the backbone not only of the standard model,
but also of many models of new physics. Our study concentrates on the
simplest model, a \HTQCD\ model, that features asymptotically free
trajectories already in standard perturbation theory as first found in 
Ref.~\cite{Cheng:1973nv}. Using effective-field-theory methods as well as
various approximations based on the functional RG, we discover
additional asymptotically free trajectories. One key ingredient for
this discovery is a careful analysis of boundary conditions on the
correlation functions of the theory, manifested by the asymptotic
behavior of the Higgs potential in field space in our study. Whereas
standard perturbation theory corresponds to an implicit choice of
these boundary conditions, generalizing this choice explicitly yields
a further two-parameter family of asymptotically free trajectories. 

Our findings in this work generalize the strategy developed in
Refs.~\cite{Gies:2015lia,Gies:2016kkk} for gauged-Higgs models to systems
including a fermionic matter sector. The new solutions also share the
property that the quasi-fixed-point potentials, i.e., the solution to
the fixed-point equation for a given small value of the gauge
coupling, exhibit a logarithmic singularity at the origin in field
space. Nevertheless, standard criteria (polynomial boundedness of
perturbations, finiteness of the potential and its first derivative,
global stability) are still satisfied. Moreover, since the
quasi-fixed-point potential exhibits a nonzero minimum at any scale,
correlation functions defined in terms of derivatives at this minimum
remain well-defined to any order. Hence, we conclude that our
solutions satisfy all standard criteria 
that are known to be crucial for selecting physical solutions in
statistical-physics models
\cite{Morris:1996nx,ODwyer:2007brp,Bridle:2016nsu}.

The occurrence of a nontrivial minimum in the quasi-fixed-point
solutions also indicates that standard arguments based on {\em
  asymptotic symmetry} \cite{Lee:1974gua} are sidestepped:
conventional perturbation theory often focuses on the deep Euclidean
region (DER), thereby implicitly assuming the irrelevance of nonzero
minima or running masses for the RG analysis. In fact, all our new
solutions demonstrate that the inclusion of a nonzero minimum is
mandatory to reveal their existence. In this sense, the CEL solution
found in standard perturbation theory is just a special case that
features the additional property of asymptotic symmetry.

Our analysis is capable of extracting information about the global
shape of the quasi-fixed-point potential. In fact, the requirement of
global stability leads to constraints in the two-parameter family of
solutions. The scaling exponent is confined to the values $P\leq
1/2$. This constraint is new in the present model in comparison with
gauged-Higgs models \cite{Gies:2015lia,Gies:2016kkk}, and may be
indicative for the fact that further structures in the matter sector
may lead to further constraints. The CEL solution is a special
solution with $P=1/2$, such that our results provide direct evidence
for the first time that the CEL solution indeed features a globally
stable potential.

In our work, we so far concentrated on the flow of the effective
potential $u(\rho)$ (or $f(x)$). This does, of course, not exhaust all
possible structures that may be relevant for identifying
asymptotically free trajectories. A natural further step would be a
study of a full Yukawa coupling potential $h(\rho)$. This would
generalize the single Yukawa coupling $h$ which corresponds to the
coupling defined at the minimum  $h(\rho=\kappa)$. In fact, the functional RG methods
are readily available to also deal with this additional layer of
complexity
\cite{Zanusso:2009bs,Vacca:2010mj,Pawlowski:2014zaa,Braun:2014ata,Vacca:2015nta,Jakovac:2015kka,Knorr:2016sfs,Gies:2017zwf}.
As further boundary conditions have to be specified, it is an
interesting open question as to whether the set of asymptotically free
trajectories becomes more diverse or even more constrained.

In view of the standard model with its triviality problem arising from
the U(1) hypercharge sector, it also remains to be seen if our
construction principle can be applied to this U(1) sector. We believe
that the construction of UV complete quantum field theories with a U(1)
factor as part of the fundamental gauge-group structure should be a
valuable ingredient in contemporary model building.

%%%%%%%%%%%%%%%%%%%%%%%%%%%%%%%%%%%%%%%%
\section*{Acknowledgments}
%%%%%%%%%%%%%%%%%%%%%%%%%%%%%%%%%%%%%%%%

We thank J. Borchardt and B. Knorr for insightful discussions especially concerning the pseudo-spectral methods. Interesting discussion with C. Kohlf$\ddot{\text{u}}$rst and R. Martini are acknowledged. This work received funding
support by the DFG under Grants No. GRK1523/2 and
No. Gi328/9-1. RS and AU acknowledge support by the Carl-Zeiss foundation.

%%%%%%%%%%%%%%%%%%%%%%%%%%%%%%%%%%%%%%%%
\appendix
%%%%%%%%%%%%%%%%%%%%%%%%%%%%%%%%%%%%%%%%

%%%%%%%%%%%%%%%%%%%%%%%%%%%%%%%%%%%%%%%%%%%%%%%%%%%%%%%%%%%%%%%%%%%%%%%
\section{More perturbatively renormalizable asymptotically free solutions}
\label{app:moreAFYukawa}
%%%%%%%%%%%%%%%%%%%%%%%%%%%%%%%%%%%%%%%%%%%%%%%%%%%%%%%%%%%%%%%%%%%%%%%

In this Appendix, we complete the review of perturbatively renormalizable 
AF solutions allowed at one loop for the \HTQCD~model defined in \Eqref{eq:Sclassical}.
Our analysis is partly similar to that of Ref.~\cite{Giudice:2014tma}, but generalizes it with the notion of QFPs.
The flow in the $(\gs^2,h^2)$ plane, provided by \Eqref{eq:gs-oneloop}
and \Eqref{eq:hsquaredot}, is best understood by direct analytic integration of the RG equations,
and adopting $\gs^2$ as an (inverse) RG time. 
The solution of the flow reads
\be
h^2(\gs^2)= \frac{\gs^2}{c\,\gs^{2(1-\gamma)}+1/ \hresc_{*}^{2}},
\label{eq:h2ofg2}
\ee
where 
\be
\gamma=\frac{9}{11\Nc -2\Nf}\frac{\Nc^{2}-1}{\Nc}. \label{eq:defgamma}
\label{eq:defB}
\ee
The QFP $\hresc_{*}^{2}$ is defined in \Eqref{eq:xihFP} and $c$ is an integration constant.
Notice that $\gamma$ as defined in \Eqref{eq:defgamma} is
positive as long as $\gs^2$ is AF, according to \Eqref{eq:gs-oneloop}.
Also, the condition $\hresc_{*}^{2}>0$, which further restricts the viable field 
content as in \Eqref{eq:AFyukawawindow}, is equivalent to $\gamma>1$,
as follows from \Eqref{eq:defgamma}.
In fact, the standard-model case, $\Nc=3$ and $\Nf=6$, 
leads to $\gamma=8/7$.
If one initializes the flow at some arbitrary RG scale $\Lambda$,
with a gauge coupling $\gsL^2$ and a Yukawa coupling $ h_\Lambda^2$,
then $c$ is given by
\be
c\,\gsL^{2(1-\gamma)}=\left(\frac{\gs^2}{ h^2}\right)_{\!\!\Lambda}-1/ \hresc_{*}^{2}.
\label{eq:initializationh2}
\ee
There  is only one trajectory along which $h^2$ exhibits an asymptotic scaling proportional to $\gs^2$,
and it corresponds to $h^2_\Lambda= g^2_{\text{s}\Lambda}\, \hresc_{*}^{2}$ and $c=0$.
If the initial condition is chosen in this way,
the strong coupling drives the Yukawa coupling to zero in the UV. 
If instead the initial condition is different, then $c\neq0$
in \Eqref{eq:h2ofg2} and the fate of the system depends on the sign of $c$.
For $c<0$, which corresponds to $\hresc_{\Lambda}^{2}>\hresc_{*}^{2}$ according to \Eqref{eq:initializationh2},
either $h^2<0$ for all $\gs^2<1$, or $h_\Lambda^2>0$ and
the Yukawa coupling hits a Landau pole in the UV, i.e.,
it diverges at a finite RG time. 
For $c>0$, namely $\hresc_{\Lambda}^{2}<\hresc_{*}^{2}$, there is no Landau pole and
the trajectories are also AF, but with an asymptotic scaling
that differs from the one defined by \Eqref{eq:xihdef} and \Eqref{eq:xihFP}.
In fact, in this case
\be
h^2(\gs^2)\widesim[2]{\gs\to 0} \frac{1}{c}\, \gs^{2\gamma},
\label{eq:moreh2QFP}
\ee
for any $c\neq 0$, thanks to the assumption that \Eqref{eq:AFyukawawindow} holds,
such that $\gamma>1$.
Also this scaling solution should be amenable to an interpretation as a QFP
for the flow of a suitable ratio.
Indeed, we could define
\be
\hresc^{\prime\, 2}=\frac{h^2}{\gs^{2\gamma}}\label{eq:xihprimedef}.
\ee
For this ratio we would find the following $\beta$ function
\begin{align}
\de_t \hresc^{\prime\, 2} = \frac{9\gs^{2\gamma}}{16\pi^2}\hresc^{\prime\, 4}.	
\label{eq:betaXiprimeh}
\end{align}
Here the second term in \Eqref{eq:hsquaredot} has been canceled
by the contribution $-\gamma\eta_A\hresc^{\prime\, 2}$ coming from the
rescaling, due to the value of $\gamma$ given in \Eqref{eq:defgamma}.
While \Eqref{eq:betaXiprimeh} does not vanish for any finite value of the strong coupling constant $\gs^2\neq 0$,
the fact that the would-be-leading contribution proportional to 
$\gs^2$
vanishes for any $\hresc^{\prime\, 2}$ signals the presence of a QFP
with arbitrary value of $\hresc^{\prime\, 2}$. This is only approximately realized at finite $\gs^2\neq0$
and becomes exact in the $\gs^2\to 0$ limit.

Let us now address the stability properties of the AF trajectories
plotted in \Figref{fig:StreamPloth2g2}.
From the previous discussion it is clear that an infinitesimal perturbation of 
a trajectory characterized by $c\neq0$, along a direction which changes the
value of the Yukawa coupling, i.e., $c$ itself, results in a
new trajectory which is still a scaling solution. Thus, one
moves from a given $\hresc^{\prime\, 2}$ to another $\hresc^{\prime\, 2}+\delta \hresc^{\prime\, 2}$,
and the distance between the two trajectories stays constant in RG time
in the UV limit if measured 
in terms of the rescaled coupling $\hresc^{\prime\, 2}$. Hence, we can call this a marginal perturbation.
These QFP solutions are neither stable nor unstable.
Yet, as it is clear from the left panel of \Figref{fig:StreamPloth2g2},
quantification of the distance between trajectories in terms of 
the unrescaled $h^2$ would lead to a different conclusion, since
such a distance would decrease as $g^2\to 0$.
The unique trajectory with $c=0$ has a rather different behavior, 
as already discussed in \Secref{sec:perturbativeoneloop}.

The AF solutions of \Eqref{eq:moreh2QFP} in the Yukawa sector,
 translate into corresponding
AF trajectories in the Higgs sector.
As we did for the CEL solution, we inspect the running of the
finite ratio $\lresc_{2}$ defined in \Eqref{eq:L2def}, and $P$ still
to be determined.
We restrict $h^2$ such that 
the ratio in \Eqref{eq:xihprimedef} attains an arbitrary 
finite value in the UV.
In this case, the reduced anomalous dimension enters the $\beta$ function of $\lresc_{2}$ of \Eqref{eq:simplestbetaL_2},
\be
\eta_{h^2}'=\left[-\frac{\partial_t h^2}{h^2}\right]_{h^2=\gs^{2\gamma}\hresc^{\prime\, 2}}
\widesim[2]{\gs\to 0}\frac{3}{8\pi^2}\frac{\Nc^2-1}{\Nc}\gs^2,
\ee
where we have neglected a second contribution which is proportional
to $\gs^{2\gamma}$, since $\gamma>1$.
Inserting this into the flow equation for $\lresc_{2}$ and replacing
$\gs^2=(c\, h^2)^{1/\gamma}$,
where $c^{-1}>0$  is the QFP value of $\hresc^{\prime\, 2}$,
one gets four terms. Each of these terms scales with a different power of $h^2$.
In order to have a QFP solution with a positive $\lresc_{2}$, it 
is necessary that the contributions from the fermions
be the leading ones, which requires 
\be
P=1-\frac{1}{2\gamma},
\ee
and results in
\be
\lresc_{2}=\frac{\Nc^2}{3(\Nc^2-1)}\frac{1}{\big(1-\frac{1}{2\gamma}\big) c^{1/\gamma}}.
\label{eq:L2_at_gammaQFP}
\ee
Notice that these QFPs do not correspond to any trajectory
plotted in \Figref{fig:StreamPlotPerturbative1Loop},
because they lie on a different hypersurface in the $\{\gs^2,h^2,\lambda\}$ space,
with a scaling defined by \Eqref{eq:xihprimedef} rather then \Eqref{eq:xihdef}. 
Still, one can produce similar plots on the hypersurface corresponding to 
\Eqref{eq:xihprimedef}, and they would look very similar to those
shown in \Figref{fig:StreamPlotPerturbative1Loop}. In fact, due to the positive sign 
of $\eta_{h^2}$, also the QFP in \Eqref{eq:L2_at_gammaQFP} is UV repulsive,
meaning that for a chosen initialization value of $h^2$, i.e., one $c$,
there is only one AF trajectory for $\lambda_2$ approaching the 
Gau{\ss}ian fixed point from positive values, and it corresponds to 
\Eqref{eq:L2_at_gammaQFP}.
Larger values of $\lresc_{2}$ would result in a Landau pole,
while smaller values would lead to negative $\lambda_2$ at 
high energy.

%%%%%%%%%%%%%%%%%%%%%%%%%%%%%%%%%%%%%%%%%%%%%%%%%%%%%%%%%%%%%%%%%%%%%%%
\section{Large field behavior of the asymptotically free potentials}
\label{app:large fields}
%%%%%%%%%%%%%%%%%%%%%%%%%%%%%%%%%%%%%%%%%%%%%%%%%%%%%%%%%%%%%%%%%%%%%%%

\subsection{EFT resummation of the effective potential in the DER}
\label{app:largefield_u}

For a study of the stability of the potential $u(\rho)$ (see \Eqref{uhxDER}) in the UV, it is necessary to address the combined limit $\rho\to\infty$ and $h^2\to 0$.
However, a meaningful and consistent result requires to take these limites in an appropriate order while remaining in the outer or inner asymptotic region, defined respectively as the region where the variable $z=h^2 \rho$ appearing in the bosonic and fermionic threshold functions is $z \gg 1$ or $z\ll 1$ as introduced in Sec.~\ref{sec:EFT_for_f}.

Let us start with the outer region.
If $h^2$ is small and finite we can address the asymptotic behavior by expanding 
the potential of \Eqref{uhxDER}  for $z\to\infty$. 
This gives the following result
\begin{align}
u(\rho) &= \frac{\lresc_{2}z^2}{2h^2}+\Gamma\left(\frac{4+4\eta_z}{2+\eta_z}\right)\frac{1}{32\pi^2 (2+3\eta_z)}	\times\notag\\
&\quad \times\Bigg\{\frac{\Gamma\left(\frac{2 \eta_z}{2+\eta_z}\right)}{\Gamma\left(\frac{2+3\eta_z}{2+\eta_z}\right)^2} 3 z^2 (3\lresc_{2}^2 - 4)\notag\\
&\quad +\Gamma\left(\frac{-2\eta_z}{2+\eta_z}\right) z^\frac{4}{2+\eta_z}\bigl[ (3\lresc_{2})^\frac{4}{2+\eta_z} - 12\bigr]\Bigg\}\notag\\
&\quad +\mathcal{O}(z) \Big|_{z=h^{2}\rho} \notag \\
&\equiv \left[ \frac{\lresc_{2}}{2h^{2}} +c_1(h^2)\right] z^2 + c_2(h^2)\, z^\frac{4}{2+\eta_z}+\mathcal O (z)\Big|_{z=h^{2}\rho}
\label{u_large_x}
\end{align}
where $\eta_z$ is the anomalous dimension of $z$ as given in \Eqref{eq:d_z_def}.
Now we can safely perform the limit $h^2\to 0$ in the outer asymptotic region where $z\gg 1$. We find that the leading order behavior is given only by the first term,
as in \Eqref{eq:u(rho)_leading_behavior}.
Indeed the noninteger power scaling $z^{4/(2+\eta_z)}$ behaves as $z^2$ for $h^2\to 0$  and the two coefficients $c_1$ and $c_2$ have simple poles in $h^2$ that cancel each other in the limit $h^2\to 0$.

Let us further consider the inner interval. Expanding the potential 
of \Eqref{uhxDER}  for small $h^2$, we obtain the following expression:
\begin{align}
u(\rho) &= \frac{\lresc_{2}}{h^2} \frac{z^2}{h^2}+\frac{3 z^2}{32\pi^2 (2+3\eta_z)}\Bigl[3\lresc_{2}^2\log(1+3 \lresc_{2} z)\notag\\
&\quad - 4 \log (1+z)\Bigr]+b_2 (z) \frac{\eta_z z^2}{2+3\eta_z}+\mathcal{O}(h^4 z^2).\label{u_small_epsilon}
\end{align}
The function $b_2(z)$ goes to zero with $z$, as is also true for the second term.
This expansion is valid only for those values of $\rho$ such that $h^2\rho\ll 1$.
Therefore for addressing the limit $\rho\to\infty$, it is necessary to take the limit $h^2\to 0$ while keeping $z\ll 1$, in such a way that the expansion in \Eqref{u_small_epsilon} still holds.
Doing so, we find that the leading term for the potential is again the one in \Eqref{eq:u(rho)_leading_behavior} also in the inner asymptotic region.
The same conclusion can be deduced by expanding the potential for small $z$ and keeping $h^2$ fixed.
Indeed the function $u(\rho)$ is analytic in $z$ and its expansion reads
\begin{align}
	u(\rho)=\frac{\lresc_{2}}{2}\frac{z^2}{h^2}+\frac{27 \lresc_{2}^3 - 12 }{32\pi^2 (2+3\eta_z)} z^3+ \mathcal O (z^4),
\end{align}
where again the $h^2\to 0$ limit with fixed $z\ll 1$ gives us the result in \Eqref{eq:u(rho)_leading_behavior}.

Let us emphasize once more that a consistent answer about the full stability of the potential $u(\rho)$ requires to take the two limits $\rho\to\infty$ and $h^2\to 0$  in such a way that the variable $z=h^2\rho$ entering the bosonic and fermionic loops controllably satisfies $z\gg 1$ (outer region) or $z\ll 1$ (inner region).
In these two asymptotic regions, the potential has the same positive asymptotic coefficient in front of the leading quadratic term. Therefore, we conclude that it is stable for arbitrarily small values of the Yukawa coupling.

%%%%%%%%%%%%%%%%%%%%%%%%%%%%%%%%%%
\subsection{$f(x)$ in the $\phi^4$-dominance approximation}
\label{app:largefield_f}

In \Secref{sec:EFT_for_f}, the same reasoning for taking the asymptotic limits as in the preceding section applies to the two loop-variables $z_\rmB$ and $z_\rmF$,
defined in \Eqref{eq:z_BF_def}.
Let us start by inspecting the potential $f(x)$ in the $\phi^4$-dominance approximation first in the outer region.
For finite values of $h^2$, we can assume that the loop contributions are negligible for large field amplitudes 
and therefore expand the scalar and fermionic loops in powers of $x^{-1}$.
Setting the left-hand side of \Eqref{eq:beta_f_phi^4dominance} to zero,
the QFP potential can then be expressed in terms of an infinite series
\begin{align}
f_\text{as}(x) &= - \frac{1}{32\pi^2}\sum_{n=1}^\infty (-)^n \frac{1 - 12 \,(3h^{2(2P-1)}\xi_2)^n}{\left(3\xi_2 h^{2P} x\right)^n(4+2n+n\eta_x)} \notag\\
&\quad + C_\text{as}\, x^\frac{4}{2+\eta_x},
\label{eq:f_as_series}
\end{align}
which can be resummed analytically
\begin{align}
&f_\text{as}(x)=C_\text{as} x^\frac{4}{2+\eta_x}+\frac{1}{32\pi^2 (6+\eta_x)}\times      \notag\\
&\times\Biggl[\frac{1}{3\xi_2 h^{2P} x}\,_2F_1\left(1,\frac{6+\eta_x}{2+\eta_x},\frac{8+2\eta_x}{2+\eta_x},-\frac{1}{3\xi_2 h^{2P} x}\right)\notag\\
&-\frac{12}{h^{2(1-P)}x}\, _2F_1\left(1,\frac{6+\eta_x}{2+\eta_x},\frac{8+2\eta_x}{2+\eta_x},-\frac{1}{h^{2(1-P)}x}\right)\Biggr].
\label{fasymptotic}
\end{align}
Using the following linear transformation among the hypergeometric functions
\begin{align}
&\frac{\sin \left(\pi (b-a)\right)}{\pi\Gamma(c)}\,_2F_1(a,b,c,z)=\notag\\
&=\frac{1}{(-z)^a}\frac{_2F_1 \left(a,a-c+1,a-b+1,z^{-1}\right)}{\Gamma(b)\Gamma(c-a)\Gamma(a-b+1)}		\notag\\
&-\frac{1}{(-z)^b}\frac{_2F_1 \left(b,b-c+1,b-a+1,z^{-1}\right)}{\Gamma(a)\Gamma(c-b)\Gamma(b-a+1)},
\label{sin_2F1}
\end{align}
it is possible to rewrite the solution $f(x)$ into $f_\text{as}(x)$.
Indeed, this becomes clear from the relation between the two integration constants $C_{\mathrm{f}}$ and $C_\text{as}$ which is
\begin{align}
C_\text{as}&=C_{\mathrm{f}}+\frac{\pi}{2+\eta_x}\left[\sin \left(\frac{4 \pi}{2+\eta_x}\right)\right]^{-1}h^\frac{8(1-P)}{2+\eta_x}\times\notag\\
&\times\frac{1}{32\pi^2} \left[ \left(3\xi_2 h^{2(2P-1)}\right)^\frac{4}{2+\eta_x} - 12 \right].   	\label{eq:C_as_of_C_f}
\end{align}
This mapping from $f_\text{as}(x)$ to $f(x)$ tells us that
the asymptotic behavior of the QFP solution is determined in the outer asymptotic region, where $z_{\rmB/\rmF}\gg 1$, only by the scaling terms in $\de_t f(x)=0$.
In fact, this property can be inferred also by expanding the solution $f(x)$, instead of its beta function, for large $z_{\rmB/\rmF}$
\begin{align}
	f(x) &= C_{\mathrm{f},\infty}\, x^\frac{4}{2+\eta_x}+\frac{\Gamma\left(-\frac{6+\eta_x}	{2+\eta_x}\right)\Gamma\left(\frac{-2+\eta_x}{2+\eta_x}\right)}{128\pi^2\, \Gamma\left(-\frac{4}{2+\eta_x}\right)^2}\times\notag\\
	&\quad \times\Biggl[-\frac{12}{z_\rmF}+\frac{1}{z_\rmB}+\mathcal O\left(z_\rmB^{-2}\right)+\mathcal O\left(z_\rmF^{-2}\right)\Biggr].	\label{eq:f_expansion_zB/F>>1}
\end{align}
The coefficient in front of the scaling term is a function of $C_{\mathrm{f}}$, $\xi_2$, $h^2$, and $P$
\begin{align}
C_{\mathrm{f},\infty}&=C_{\mathrm{f}}+\frac{1}{128\pi^2}\Gamma\left(\frac{-2+\eta_x}{2+\eta_x}\right)\Gamma\left(\frac{6+\eta_x}{2+\eta_x}\right)\times\notag\\
&\quad \times\left[\left(3\xi_2 h^{2P}\right)^\frac{4}{2+\eta_x} - 12 \,h^\frac{8 (1-P)}{2+\eta_x} \right].	\label{eq:C_f,infty}
\end{align}
It is not surprising that this scaling factor is exactly the asymptotic coefficient $C_\text{as}$.
By using one of the defining properties of the Gamma function, $\Gamma(1+z)=z\Gamma(z)$ as well as the following identity
\begin{align}
	\Gamma(z)\Gamma(1-z)=\frac{\pi}{\sin(\pi z)},		\label{eq:Gamma_sin_relation}
\end{align}
we recover precisely the expression in \Eqref{eq:C_as_of_C_f}, therefore
\be
C_{\mathrm{f},\infty}=C_\text{as}.
\ee
As we are interested in the asymptotic behavior in the UV, it is convenient to expand \Eqref{eq:C_f,infty} for small $h^2$ and keep only the leading terms,
\begin{align}
	&C_{\mathrm{f},\infty}\widesim[2]{h^2\to 0} C_{\mathrm{f}} - \frac{9 \xi_2^2 h^{2(2P - 1)} - 12\, h^{2(1-2P)}}{64\pi^2 \hat\eta_x}.\label{eq:C_f,infty_small_h2}
\end{align}
Moreover, all the subleading terms in \Eqref{eq:f_expansion_zB/F>>1} of order $\mathcal O(z_\rmB^{-1})$ and $\mathcal O(z_\rmF^{-1})$ are regular in the $h^2\to 0$ limit.
We can thus conclude that the asymptotic property of the QFP potential is correctly described by \Eqref{eq:C_f,infty_small_h2} both in the outer region, i.e., for large field amplitudes, and in the UV limit.

Let us address now the situation in the inner region, where we can expand the potential $f(x)$ either for $z_{\rmB/\rmF}\ll 1$ or for $x\ll 1$ while keeping $h^2$ finite.
In both cases the result is the same, and reads
\begin{align}
	f(x)&=-\frac{11}{128\pi^2} +C_{\mathrm{f}} x^\frac{4}{2+\eta_x} - \frac{z_\rmB - 12 z_\rmF}{32\pi^2(2-\eta_x)}\notag\\
	&\quad -\frac{z_\rmB^2 - 12 z_\rmF^2}{64\pi^2 h^2 \hat\eta_x}+\mathcal O(z_\rmB^3)+\mathcal O(z_\rmF^3). \label{eq:f(x)_expansion_small_z_B/F}
\end{align}
In the UV limit, the inner region increases and thus allows to address the asymptotic behavior of the potential.
Indeed this combined limit can be taken as long as $z_{\rmB/\rmF}\ll 1$ holds. From \Eqref{eq:f(x)_expansion_small_z_B/F}, we can deduce that
\begin{align}
	f(x)\widesim[2]{x\to\infty} \left[C_{\mathrm{f}}- \frac{9 \xi_2^2 h^{2(2P - 1)} - 12 h^{2(1-2P)}}{64\pi^2 \hat\eta_x}\right] x^2,\label{eq:f(x)_small_z_B/F_and_large_x}
\end{align}
where the coefficient in front of the quadratic term coincides with \Eqref{eq:C_f,infty_small_h2}.
The same information is obtained by expanding the hypergeometric functions in \Eqref{eq:f_FP_phi^4dominance} for small Yukawa coupling,
\begin{align}
	f(x)=C_{\mathrm{f}} x^\frac{4}{2+\eta_x}-\frac{z_\rmB^2 - 12\, z_\rmF^2}{64\pi^2 h^2 \hat\eta_x}+\mathcal O(h^0).
\end{align}
This is in agreement with \Eqref{eq:f(x)_small_z_B/F_and_large_x} for the asymptotic behavior within the inner region.

We finally conclude that it is possible to simultaneously take the $x\to\infty$ limit and the $h^2\to 0$ limit in both the inner and outer asymptotic regions.
This combined limit gives the same result in both regions, and can be summarized as
\begin{align}
	f(x)\widesim[3]{\substack{x\to\infty\\h^2\to 0}} h^{\pm 2(2P-1)}\,\,\hat C_{\mathrm{f},\infty}\,x^2,
\end{align}
where the $\pm$ sign is for $P\lessgtr1/2$. The expression for $\hat C_{\mathrm{f},\infty}$ can be found in the main text in \Eqref{eq:barCf_infty}.

%%%%%%%%%%%%%%%%%%%%%%%%%%%%%%%%%%%%
\subsection{Comparison between $u(\rho)$ and $f(x)$}
\label{app:u(rho)_and_f(x)_compared}
%%%%%%%%%%%%%%%%%%%%%%%%%%%%%%%%%%%%

The potential $u(\rho)$ in \Eqref{uhxDER}, obtained within the DER,
and the effective potential $f(x)$ in \Eqref{eq:f_FP_phi^4dominance}, derived in the $\phi^4$-dominance approximation, can be related to each other by exploiting a general identity among the Gau\ss\  hypergeometric functions
\begin{align}
&\frac{zb}{c}\,_2F_1(a+1,b+1,c+1,z)\notag\\
&=\,_2F_1(a+1,b,c,z)-\,_2F_1(a,b,c,z).
\end{align}
By setting $a=0$ in the latter expression, we get the following relationship
\begin{align}
_2F_1(1,b,c,z)=1+\frac{zb}{c}\,_2F_1(1,b+1,c+1,z). \label{2F1=1+2F1}
\end{align}
Iterating this twice, we can rewrite the solution $f(x)$ as
\begin{align}
f(x) &= \frac{1}{32\pi^2}\Biggl[-\frac{11}{4}+\frac{4 - \xi_2}{2-\eta_x}3hx+\frac{4 - \xi_2^2}{2\eta_x}3h^2x^2\notag\\
&\quad +\frac{(3\xi_2 h x)^3}{2+3\eta_x}\,_2F_1\left(1,\frac{2+3\eta_x}{2+\eta_x},\frac{4+4\eta_x}{2+\eta_x},-3\xi_2 hx\right)	\notag\\
&\quad -\frac{12 (h x)^3}{2+3\eta_x}\,_2F_1\left(1,\frac{2+3\eta_x}{2+\eta_x},\frac{4+4\eta_x}{2+\eta_x},- hx\right)\Biggr]\notag \\
&\quad + C_{\mathrm{f}}\, x^\frac{4}{2+\eta_x}. 
\label{eq:f(x)_compared_to_u(rho)}
\end{align}
Let us consider the CEL potential corresponding to $P=1/2$, $C_{\mathrm{f}}=0$, and $\xi_2=\lresc_{2}^+$.
Working in the limit $h^2\to 0$ where we can take  $\eta_x\to\eta_z$, cf. \Eqref{eq:dx_def} and \Eqref{eq:d_z_def}, \Eqref{eq:f(x)_compared_to_u(rho)} becomes
\begin{align}
	\lim_{\eta_x\to\eta_z} f(x)= u(\rho)\Big{|}_{x=h\rho}-\frac{11}{128\pi^2}+\frac{3 h x (4 - \xi_2)}{32\pi^2(2-\eta_z)}.
\end{align}
We conclude that the solution $f(x)$ becomes the solution $u(\rho)$ in the UV limit -- apart from a linear term in the field variable that is, in fact, discarded by the definition of the deep euclidean approximation.

%%%%%%%%%%%%%%%%%%%%%%%%%%%%%%%%%%%%%%%%%%%%%%%%%%%%%%%%%%%%%%%%%%%%%%%%%%%%
\section{The unphysical $P>1$ case}
\label{app:P>1}
%%%%%%%%%%%%%%%%%%%%%%%%%%%%%%%%%%%%%%%%%%%%%%%%%%%%%%%%%%%%%%%%%%%%%%%%%%%%

\subsection{Effective-field-theory analysis}
\label{subsec:EFTP>1}

In \Secref{sec:effectiveFTthresholds}, we have encountered an example in the RG equations for $\lresc_{2}$ and $\kap$ (cf. \Eqref{eq:lresc2_P>1/2} and \Eqref{eq:kap_P>1/2}) that for $P>1/2$ there are only few terms which may contribute to the leading part in the $h^2\to 0$ limit.
This is a consequence of the remaining legitimate configuration for the scaling powers $P$, $Q$, and $P_3$. The situation is very similar also for $P>1$, where there are only two consistent solutions,
\begin{alignat}{2}
	\kap&=\mp\left(\frac{3}{8\pi^2 \lresc_{3}}\right)^\frac{1}{4},\quad\quad\quad &Q&=\frac{2P+1}{3},\label{eq:kappa_P>1}\\
	\lresc_{2}&=\pm \frac{1}{2}\left(\frac{3 \lresc_{3}^3}{8\pi^2}\right)^\frac{1}{4},\quad\quad\quad &P_3&=\frac{8P+1}{3},\label{eq:L2_P>1}
\end{alignat}
and
\begin{alignat}{2}
\lresc_2 &=\frac{-2\pm \sqrt{2P-1}}{2(5-2P)\kap},\quad\quad\quad &Q&=2P, \\
\lresc_{3}^2&=\frac{8\pi^2(1+2\kap\lresc_2)^4}{(2P-1)\kap^3},\quad\quad\quad &P_3&=3P,
\end{alignat}
where $\lresc_{3}$ is a free parameter.
Moreover we notice that the scaling powers $Q$ and $P_3$ in \Eqref{eq:kappa_P>1} and \Eqref{eq:L2_P>1} are the same as in the model discussed in Refs.~\cite{Gies:2015lia,Gies:2016kkk}.
These analytical solutions are however unphysical due to the fact that positive $\kap$ correspond to negative $\lresc_{2}$ and \emph{vice versa}.

%%%%%%%%%%%%%%%%%%
\subsection{Weak-$h^2$ expansion}
%%%%%%%%%%%%%%%%%%

For $P>1$ the argument of the fermionic loop $\omega_{1f}$, defined in \Eqref{eq:omegaf},  diverges in the limit $h^2\to 0$ at fixed $x$.
Therefore, in order to capture the correct UV behavior, one has to Taylor expand the threshold function $l_0^{(\rmF)}(\omega_{1f})$ in powers of $\omega_{1f}^{-1}$.
Let us start the investigation for $1<P<2$.
In this case the leading $h^2$-dependent contribution to $\beta_f$ is
\begin{align}
\delta\beta_f&= - \frac{\Nc d_\gamma}{32\pi^2x}
h^{2(P-1)}& & \text{for $1<P<2$},
\end{align}
and the integration of the QFP condition $\beta_f=0$ gives us the potential
\begin{align}
f(x)=c\, x^2 - \frac{\Nc d_\gamma}{192\pi^2 x}h^{2(P-1)},
\end{align}
where $c$ is the integration constant of the first order ODE.
The defining equation $f^{\prime}(x_0)=0$ for the minimum fixes an expression for this integration constant $c(h^2,x_0)$, while the consistency condition $f^{\prime\prime}(x_0)=\xi_2$ provides how $x_0$ is related to the parameters $\xi_2$ and $h^2$. Indeed
\begin{align}
x_0^3= - \frac{ \Nc d_\gamma}{64\pi^2\xi_2}h^{2(P-1)}. 	\label{eq:x0^3_general_scheme_1<P<2weak-h}
\end{align}
By  setting $\Nc=3$ and $d_\gamma=4$, the latter equation becomes
\begin{align}
x_0^3= - \frac{ 3}{16\pi^2\xi_2}h^{2(P-1)},	\label{eq:x0^3_1<P<2_weak-h}
\end{align}
showing an agreement with the EFT analysis including thresholds.
Indeed, by taking Eqs.~\eqref{eq:kappa_P>1},\eqref{eq:L2_P>1}  and recalling that $x_0=h^{2 P}\kappa=h^{2(P-1)/3}\kap$, one gets exactly \Eqref{eq:x0^3_1<P<2_weak-h}.
Thus, there are no solutions with a nontrivial minimum for $\xi_2>0$.

It is worth to point out that the potential and all its derivatives are singular at the origin due to the Taylor expansion of $l_0^{(\rmF)}(\omega_{1f})$ for small $\omega_{1f}^{-1}$, producing a term in $\delta\beta_f$ proportional to $x^{-1}$.
This expansion is valid only for $x\gg h^{2(P-1)}$,
a condition which is not fulfilled in the $x\to 0$ limit at
fixed $h^2$.
Yet, the fermionic loop is finite at the origin and this suggests to retain its whole $x$-dependence.

Accounting for the full fermionic loop still allows for
an analytic solution of the QFP,
which leads,
for the piecewise linear cutoff regulator, to a  Coleman-Weinberg-like potential
\begin{align}
f(x)&=c\, x^2+\frac{\Nc d_\gamma}{64\pi^2} \Biggl[\frac{x}{h^{2(P-1)}}\notag\\
&\quad+\frac{x^2}{h^{4(P-1)}}\log\Bigl(\frac{x}{x+h^{2(P-1)}}\Bigr)\Biggr].
\end{align}
The corresponding quadratic rescaled scalar coupling, as a function of $h^2$ and $x_0$, reads
\begin{align}
\xi_2= - \frac{ \Nc d_\gamma}{64\pi^2} \frac{h^{2(P-1)}}{x_0 \left(x_0+h^{2(P-1)}\right)^2},	\label{eq:xi2_weak-h^2_Litim_1<P<2}
\end{align}
which yields \Eqref{eq:x0^3_1<P<2_weak-h} in the $h^2\to 0$ limit for $\Nc=3$ and $d_\gamma=4$.

\begin{figure}[!t]
	\begin{center}
		\includegraphics[width=0.95\columnwidth]{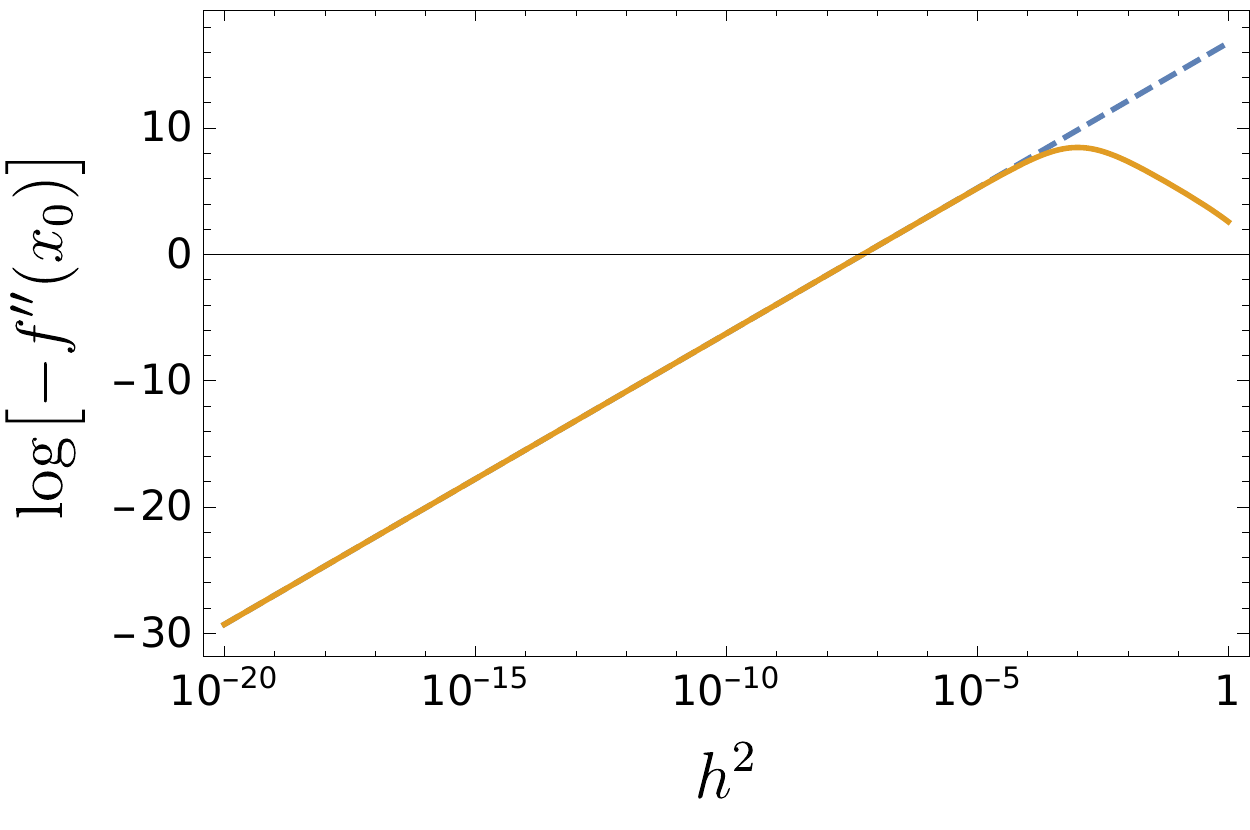}
	\end{center}
	\vskip-5mm
	\caption{
		The rescaled quadratic coupling $\xi_2$ as a function of $h^2$ for $P=2$ and fixed value for the nontrivial minimum $x_0=10^{-3}$.
		Dashed line: EFT approximation including thresholds, see Eqs.~\eqref{eq:kappa_P>1},\eqref{eq:L2_P>1}. Solid line: weak-$h^2$ expansion, see \Eqref{eq:xi2_weak-h^2_full_loop_Litim_P>2}.
	}
	\label{fig:weak_h_xi2_of_h_P2}
\end{figure}

The situation is slightly different for $P\ge 2$ due to the contribution coming from the anomalous dimensions $\eta_x$ in the scaling part of $\beta_f$.
Let us first expand the threshold function $l_0^{(\rmF)}(\omega_{1f})$ in powers of $\omega_{1f}^{-1}$ such that the leading correction to the beta function is
\begin{align}
\delta\beta_f&=\hat\eta_x h^2 x f^\prime - \frac{\Nc d_\gamma}{32\pi^2x}  h^{2(P-1)},& &\text{for $P\ge 2$}.
\end{align}
Due to the presence of a singular term at the origin, we expect that this pole survives also in the corresponding QFP solution which is in fact
\begin{align}
f(x)=c\, x^\frac{4}{2+\eta_x} - \frac{ \Nc d_\gamma}{(6+\eta_x)}\frac{h^{2(P-1)}}{32\pi^2 x}.
\end{align}
Additionally there is also a log-type singularity in the second derivative. Indeed by Taylor expanding the scaling term for small $h^2$, we get a term proportional to $x^2\log x$.
This potential has a nontrivial minimum $x_0$ whose analytical expression in terms of $h^2$ and $\xi_2$ is
\begin{align}
x_0^3= - \frac{ \Nc d_\gamma}{32 \pi^2 \xi_2 (2+\eta_x)}h^{2(P-1)}.	\label{eq:x0^3_weak-h^2_P>2}
\end{align}

If we instead consider the full fermionic loop,
 the general QFP solution reads
\begin{align}
f(x) &= - \frac{\Nc d_\gamma}{128\pi^2}\,_2F_1\left(1,-\frac{4}{2+\eta_x},\frac{-2+\eta_x}{2+\eta_x},-\frac{x}{h^{2(P-1)}}\right)\notag \\
&\quad + c\,x^\frac{4}{2+\eta_x}
\end{align}
where the Gau\ss\ hypergeometric function comes from the integration of the fermionic threshold function and it is analytic around the origin.
The corresponding quadratic rescaled scalar coupling reads
\begin{align}
\xi_2 = - \frac{ \Nc d_\gamma}{32\pi^2 (2+\eta_x)} \frac{h^{2(P-1)}}{x_0 \left(x_0+h^{2(P-1)}\right)^2},		\label{eq:xi2_weak-h^2_full_loop_Litim_P>2}
\end{align}
which is a generalization of \Eqref{eq:xi2_weak-h^2_Litim_1<P<2} due to the anomalous dimension.

For all of these cases computed within the piecewise linear regulator, we can thus conclude that the weak-$h^2$ expansion for $P>1$ agrees with the EFT approximation including thresholds analyzed in \Secref{subsec:EFTP>1}.
In fact Eqs.~\eqref{eq:x0^3_1<P<2_weak-h}, \eqref{eq:xi2_weak-h^2_Litim_1<P<2}, \eqref{eq:x0^3_weak-h^2_P>2}, and \eqref{eq:xi2_weak-h^2_full_loop_Litim_P>2} are all in agreement with Eqs.~\eqref{eq:kappa_P>1} and \eqref{eq:L2_P>1} in the $h^2\to 0$ limit and by fixing $\Nc=3$ and $d_\gamma=4$.
In \Figref{fig:weak_h_xi2_of_h_P2} we show indeed the rescaled quartic coupling $\xi_2$, i.e., the curvature of the potential at the nontrivial minimum $x_0$, as a function of the Yukawa coupling for fixed value of $x_0=10^{-3}$.
The dashed line represents the EFT analysis, see Eqs.~\eqref{eq:kappa_P>1} and \eqref{eq:L2_P>1}, whereas the solid line represents the weak-$h^2$ approximation in the case where the anomalous dimension $\eta_x$ and the full fermionic loop are taken into account, see \Eqref{eq:xi2_weak-h^2_full_loop_Litim_P>2}.

So we summarize this section by emphasizing once more that for $P>1$ it is not possible to have a solution with a nontrivial minimum and at the same time a positive $\xi_2$.

%\cleardoublepage
%%%%%%%%%%%%%%%%%%%%%%%%%%%%%%%%%%%%%%%%
%%%%%%%%%%%%%%%%%%%%%%%%%%%%%%%%%%%%%%%%

\bibliography{bibliography}

\begin{thebibliography}{100}%
\makeatletter
\providecommand \@ifxundefined [1]{%
 \@ifx{#1\undefined}
}%
\providecommand \@ifnum [1]{%
 \ifnum #1\expandafter \@firstoftwo
 \else \expandafter \@secondoftwo
 \fi
}%
\providecommand \@ifx [1]{%
 \ifx #1\expandafter \@firstoftwo
 \else \expandafter \@secondoftwo
 \fi
}%
\providecommand \natexlab [1]{#1}%
\providecommand \enquote  [1]{``#1''}%
\providecommand \bibnamefont  [1]{#1}%
\providecommand \bibfnamefont [1]{#1}%
\providecommand \citenamefont [1]{#1}%
\providecommand \href@noop [0]{\@secondoftwo}%
\providecommand \href [0]{\begingroup \@sanitize@url \@href}%
\providecommand \@href[1]{\@@startlink{#1}\@@href}%
\providecommand \@@href[1]{\endgroup#1\@@endlink}%
\providecommand \@sanitize@url [0]{\catcode `\\12\catcode `\$12\catcode
  `\&12\catcode `\#12\catcode `\^12\catcode `\_12\catcode `\%12\relax}%
\providecommand \@@startlink[1]{}%
\providecommand \@@endlink[0]{}%
\providecommand \url  [0]{\begingroup\@sanitize@url \@url }%
\providecommand \@url [1]{\endgroup\@href {#1}{\urlprefix }}%
\providecommand \urlprefix  [0]{URL }%
\providecommand \Eprint [0]{\href }%
\providecommand \doibase [0]{http://dx.doi.org/}%
\providecommand \selectlanguage [0]{\@gobble}%
\providecommand \bibinfo  [0]{\@secondoftwo}%
\providecommand \bibfield  [0]{\@secondoftwo}%
\providecommand \translation [1]{[#1]}%
\providecommand \BibitemOpen [0]{}%
\providecommand \bibitemStop [0]{}%
\providecommand \bibitemNoStop [0]{.\EOS\space}%
\providecommand \EOS [0]{\spacefactor3000\relax}%
\providecommand \BibitemShut  [1]{\csname bibitem#1\endcsname}%
\let\auto@bib@innerbib\@empty
%</preamble>
\bibitem [{\citenamefont {Gross}\ and\ \citenamefont
  {Wilczek}(1973{\natexlab{a}})}]{Gross:1973id}%
  \BibitemOpen
  \bibfield  {author} {\bibinfo {author} {\bibfnamefont {D.~J.}\ \bibnamefont
  {Gross}}\ and\ \bibinfo {author} {\bibfnamefont {F.}~\bibnamefont
  {Wilczek}},\ }\href {\doibase 10.1103/PhysRevLett.30.1343} {\bibfield
  {journal} {\bibinfo  {journal} {Phys. Rev. Lett.}\ }\textbf {\bibinfo
  {volume} {30}},\ \bibinfo {pages} {1343} (\bibinfo {year}
  {1973}{\natexlab{a}})}\BibitemShut {NoStop}%
%%CITATION = PRLTA,30,1343;%%
\bibitem [{\citenamefont {Politzer}(1973)}]{Politzer:1973fx}%
  \BibitemOpen
  \bibfield  {author} {\bibinfo {author} {\bibfnamefont {H.~D.}\ \bibnamefont
  {Politzer}},\ }\href {\doibase 10.1103/PhysRevLett.30.1346} {\bibfield
  {journal} {\bibinfo  {journal} {Phys. Rev. Lett.}\ }\textbf {\bibinfo
  {volume} {30}},\ \bibinfo {pages} {1346} (\bibinfo {year}
  {1973})}\BibitemShut {NoStop}%
%%CITATION = PRLTA,30,1346;%%
\bibitem [{\citenamefont {Gross}\ and\ \citenamefont
  {Wilczek}(1973{\natexlab{b}})}]{Gross:1973ju}%
  \BibitemOpen
  \bibfield  {author} {\bibinfo {author} {\bibfnamefont {D.~J.}\ \bibnamefont
  {Gross}}\ and\ \bibinfo {author} {\bibfnamefont {F.}~\bibnamefont
  {Wilczek}},\ }\href {\doibase 10.1103/PhysRevD.8.3633} {\bibfield  {journal}
  {\bibinfo  {journal} {Phys. Rev.}\ }\textbf {\bibinfo {volume} {D8}},\
  \bibinfo {pages} {3633} (\bibinfo {year} {1973}{\natexlab{b}})}\BibitemShut
  {NoStop}%
%%CITATION = PHRVA,D8,3633;%%
\bibitem [{\citenamefont {Cheng}\ \emph {et~al.}(1974)\citenamefont {Cheng},
  \citenamefont {Eichten},\ and\ \citenamefont {Li}}]{Cheng:1973nv}%
  \BibitemOpen
  \bibfield  {author} {\bibinfo {author} {\bibfnamefont {T.~P.}\ \bibnamefont
  {Cheng}}, \bibinfo {author} {\bibfnamefont {E.}~\bibnamefont {Eichten}}, \
  and\ \bibinfo {author} {\bibfnamefont {L.-F.}\ \bibnamefont {Li}},\ }\href
  {\doibase 10.1103/PhysRevD.9.2259} {\bibfield  {journal} {\bibinfo  {journal}
  {Phys. Rev.}\ }\textbf {\bibinfo {volume} {D9}},\ \bibinfo {pages} {2259}
  (\bibinfo {year} {1974})}\BibitemShut {NoStop}%
%%CITATION = PHRVA,D9,2259;%%
\bibitem [{\citenamefont {Gross}\ and\ \citenamefont
  {Wilczek}(1974)}]{Gross:1974cs}%
  \BibitemOpen
  \bibfield  {author} {\bibinfo {author} {\bibfnamefont {D.~J.}\ \bibnamefont
  {Gross}}\ and\ \bibinfo {author} {\bibfnamefont {F.}~\bibnamefont
  {Wilczek}},\ }\href {\doibase 10.1103/PhysRevD.9.980} {\bibfield  {journal}
  {\bibinfo  {journal} {Phys. Rev.}\ }\textbf {\bibinfo {volume} {D9}},\
  \bibinfo {pages} {980} (\bibinfo {year} {1974})}\BibitemShut {NoStop}%
%%CITATION = PHRVA,D9,980;%%
\bibitem [{\citenamefont {Politzer}(1974)}]{Politzer:1974fr}%
  \BibitemOpen
  \bibfield  {author} {\bibinfo {author} {\bibfnamefont {H.~D.}\ \bibnamefont
  {Politzer}},\ }\href {\doibase 10.1016/0370-1573(74)90014-3} {\bibfield
  {journal} {\bibinfo  {journal} {Phys. Rept.}\ }\textbf {\bibinfo {volume}
  {14}},\ \bibinfo {pages} {129} (\bibinfo {year} {1974})}\BibitemShut
  {NoStop}%
%%CITATION = PRPLC,14,129;%%
\bibitem [{\citenamefont {Chang}(1974)}]{Chang:1974bv}%
  \BibitemOpen
  \bibfield  {author} {\bibinfo {author} {\bibfnamefont {N.-P.}\ \bibnamefont
  {Chang}},\ }\href {\doibase 10.1103/PhysRevD.10.2706} {\bibfield  {journal}
  {\bibinfo  {journal} {Phys. Rev.}\ }\textbf {\bibinfo {volume} {D10}},\
  \bibinfo {pages} {2706} (\bibinfo {year} {1974})}\BibitemShut {NoStop}%
%%CITATION = PHRVA,D10,2706;%%
\bibitem [{\citenamefont {Chang}\ and\ \citenamefont
  {Perez-Mercader}(1978)}]{Chang:1978nu}%
  \BibitemOpen
  \bibfield  {author} {\bibinfo {author} {\bibfnamefont {N.-P.}\ \bibnamefont
  {Chang}}\ and\ \bibinfo {author} {\bibfnamefont {J.}~\bibnamefont
  {Perez-Mercader}},\ }\href {\doibase 10.1103/PhysRevD.18.4721,
  10.1103/PhysRevD.19.2515} {\bibfield  {journal} {\bibinfo  {journal} {Phys.
  Rev.}\ }\textbf {\bibinfo {volume} {D18}},\ \bibinfo {pages} {4721} (\bibinfo
  {year} {1978})},\ \bibinfo {note} {[Erratum: Phys.
  Rev.D19,2515(1979)]}\BibitemShut {NoStop}%
%%CITATION = PHRVA,D18,4721;%%
\bibitem [{\citenamefont {Fradkin}\ and\ \citenamefont
  {Kalashnikov}(1975)}]{Fradkin:1975yt}%
  \BibitemOpen
  \bibfield  {author} {\bibinfo {author} {\bibfnamefont {E.~S.}\ \bibnamefont
  {Fradkin}}\ and\ \bibinfo {author} {\bibfnamefont {O.~K.}\ \bibnamefont
  {Kalashnikov}},\ }\href {\doibase 10.1088/0305-4470/8/11/017} {\bibfield
  {journal} {\bibinfo  {journal} {J. Phys.}\ }\textbf {\bibinfo {volume}
  {A8}},\ \bibinfo {pages} {1814} (\bibinfo {year} {1975})}\BibitemShut
  {NoStop}%
%%CITATION = JPAGA,A8,1814;%%
\bibitem [{\citenamefont {Salam}\ and\ \citenamefont
  {Strathdee}(1978)}]{Salam:1978dk}%
  \BibitemOpen
  \bibfield  {author} {\bibinfo {author} {\bibfnamefont {A.}~\bibnamefont
  {Salam}}\ and\ \bibinfo {author} {\bibfnamefont {J.~A.}\ \bibnamefont
  {Strathdee}},\ }\href {\doibase 10.1103/PhysRevD.18.4713} {\bibfield
  {journal} {\bibinfo  {journal} {Phys. Rev.}\ }\textbf {\bibinfo {volume}
  {D18}},\ \bibinfo {pages} {4713} (\bibinfo {year} {1978})}\BibitemShut
  {NoStop}%
%%CITATION = PHRVA,D18,4713;%%
\bibitem [{\citenamefont {Bais}\ and\ \citenamefont
  {Weldon}(1978)}]{Bais:1978fv}%
  \BibitemOpen
  \bibfield  {author} {\bibinfo {author} {\bibfnamefont {F.~A.}\ \bibnamefont
  {Bais}}\ and\ \bibinfo {author} {\bibfnamefont {H.~A.}\ \bibnamefont
  {Weldon}},\ }\href {\doibase 10.1103/PhysRevD.18.1199} {\bibfield  {journal}
  {\bibinfo  {journal} {Phys. Rev.}\ }\textbf {\bibinfo {volume} {D18}},\
  \bibinfo {pages} {1199} (\bibinfo {year} {1978})}\BibitemShut {NoStop}%
%%CITATION = PHRVA,D18,1199;%%
\bibitem [{\citenamefont {Salam}\ and\ \citenamefont
  {Elias}(1980)}]{Salam:1980ss}%
  \BibitemOpen
  \bibfield  {author} {\bibinfo {author} {\bibfnamefont {A.}~\bibnamefont
  {Salam}}\ and\ \bibinfo {author} {\bibfnamefont {V.}~\bibnamefont {Elias}},\
  }\href {\doibase 10.1103/PhysRevD.22.1469} {\bibfield  {journal} {\bibinfo
  {journal} {Phys. Rev.}\ }\textbf {\bibinfo {volume} {D22}},\ \bibinfo {pages}
  {1469} (\bibinfo {year} {1980})}\BibitemShut {NoStop}%
%%CITATION = PHRVA,D22,1469;%%
\bibitem [{\citenamefont {Callaway}(1988)}]{Callaway:1988ya}%
  \BibitemOpen
  \bibfield  {author} {\bibinfo {author} {\bibfnamefont {D.~J.~E.}\
  \bibnamefont {Callaway}},\ }\href {\doibase 10.1016/0370-1573(88)90008-7}
  {\bibfield  {journal} {\bibinfo  {journal} {Phys. Rept.}\ }\textbf {\bibinfo
  {volume} {167}},\ \bibinfo {pages} {241} (\bibinfo {year}
  {1988})}\BibitemShut {NoStop}%
%%CITATION = PRPLC,167,241;%%
\bibitem [{\citenamefont {Giudice}\ \emph {et~al.}(2015)\citenamefont
  {Giudice}, \citenamefont {Isidori}, \citenamefont {Salvio},\ and\
  \citenamefont {Strumia}}]{Giudice:2014tma}%
  \BibitemOpen
  \bibfield  {author} {\bibinfo {author} {\bibfnamefont {G.~F.}\ \bibnamefont
  {Giudice}}, \bibinfo {author} {\bibfnamefont {G.}~\bibnamefont {Isidori}},
  \bibinfo {author} {\bibfnamefont {A.}~\bibnamefont {Salvio}}, \ and\ \bibinfo
  {author} {\bibfnamefont {A.}~\bibnamefont {Strumia}},\ }\href {\doibase
  10.1007/JHEP02(2015)137} {\bibfield  {journal} {\bibinfo  {journal} {JHEP}\
  }\textbf {\bibinfo {volume} {02}},\ \bibinfo {pages} {137} (\bibinfo {year}
  {2015})},\ \Eprint {http://arxiv.org/abs/1412.2769} {arXiv:1412.2769
  [hep-ph]} \BibitemShut {NoStop}%
%%CITATION = ARXIV:1412.2769;%%
\bibitem [{\citenamefont {Holdom}\ \emph {et~al.}(2015)\citenamefont {Holdom},
  \citenamefont {Ren},\ and\ \citenamefont {Zhang}}]{Holdom:2014hla}%
  \BibitemOpen
  \bibfield  {author} {\bibinfo {author} {\bibfnamefont {B.}~\bibnamefont
  {Holdom}}, \bibinfo {author} {\bibfnamefont {J.}~\bibnamefont {Ren}}, \ and\
  \bibinfo {author} {\bibfnamefont {C.}~\bibnamefont {Zhang}},\ }\href
  {\doibase 10.1007/JHEP03(2015)028} {\bibfield  {journal} {\bibinfo  {journal}
  {JHEP}\ }\textbf {\bibinfo {volume} {03}},\ \bibinfo {pages} {028} (\bibinfo
  {year} {2015})},\ \Eprint {http://arxiv.org/abs/1412.5540} {arXiv:1412.5540
  [hep-ph]} \BibitemShut {NoStop}%
%%CITATION = ARXIV:1412.5540;%%
\bibitem [{\citenamefont {Hansen}\ \emph {et~al.}(2017)\citenamefont {Hansen},
  \citenamefont {Janowski}, \citenamefont {Langaeble}, \citenamefont {Mann},
  \citenamefont {Sannino}, \citenamefont {Steele},\ and\ \citenamefont
  {Wang}}]{Hansen:2017pwe}%
  \BibitemOpen
  \bibfield  {author} {\bibinfo {author} {\bibfnamefont {F.~F.}\ \bibnamefont
  {Hansen}}, \bibinfo {author} {\bibfnamefont {T.}~\bibnamefont {Janowski}},
  \bibinfo {author} {\bibfnamefont {K.}~\bibnamefont {Langaeble}}, \bibinfo
  {author} {\bibfnamefont {R.~B.}\ \bibnamefont {Mann}}, \bibinfo {author}
  {\bibfnamefont {F.}~\bibnamefont {Sannino}}, \bibinfo {author} {\bibfnamefont
  {T.~G.}\ \bibnamefont {Steele}}, \ and\ \bibinfo {author} {\bibfnamefont
  {Z.-W.}\ \bibnamefont {Wang}},\ }\href@noop {} {\  (\bibinfo {year}
  {2017})},\ \Eprint {http://arxiv.org/abs/1706.06402} {arXiv:1706.06402
  [hep-ph]} \BibitemShut {NoStop}%
%%CITATION = ARXIV:1706.06402;%%
\bibitem [{\citenamefont {Hetzel}\ and\ \citenamefont
  {Stech}(2015)}]{Hetzel:2015bla}%
  \BibitemOpen
  \bibfield  {author} {\bibinfo {author} {\bibfnamefont {J.}~\bibnamefont
  {Hetzel}}\ and\ \bibinfo {author} {\bibfnamefont {B.}~\bibnamefont {Stech}},\
  }\href {\doibase 10.1103/PhysRevD.91.055026} {\bibfield  {journal} {\bibinfo
  {journal} {Phys. Rev.}\ }\textbf {\bibinfo {volume} {D91}},\ \bibinfo {pages}
  {055026} (\bibinfo {year} {2015})},\ \Eprint
  {http://arxiv.org/abs/1502.00919} {arXiv:1502.00919 [hep-ph]} \BibitemShut
  {NoStop}%
%%CITATION = ARXIV:1502.00919;%%
\bibitem [{\citenamefont {Pelaggi}\ \emph {et~al.}(2015)\citenamefont
  {Pelaggi}, \citenamefont {Strumia},\ and\ \citenamefont
  {Vignali}}]{Pelaggi:2015kna}%
  \BibitemOpen
  \bibfield  {author} {\bibinfo {author} {\bibfnamefont {G.~M.}\ \bibnamefont
  {Pelaggi}}, \bibinfo {author} {\bibfnamefont {A.}~\bibnamefont {Strumia}}, \
  and\ \bibinfo {author} {\bibfnamefont {S.}~\bibnamefont {Vignali}},\ }\href
  {\doibase 10.1007/JHEP08(2015)130} {\bibfield  {journal} {\bibinfo  {journal}
  {JHEP}\ }\textbf {\bibinfo {volume} {08}},\ \bibinfo {pages} {130} (\bibinfo
  {year} {2015})},\ \Eprint {http://arxiv.org/abs/1507.06848} {arXiv:1507.06848
  [hep-ph]} \BibitemShut {NoStop}%
%%CITATION = ARXIV:1507.06848;%%
\bibitem [{\citenamefont {Pica}\ \emph {et~al.}(2016)\citenamefont {Pica},
  \citenamefont {Ryttov},\ and\ \citenamefont {Sannino}}]{Pica:2016krb}%
  \BibitemOpen
  \bibfield  {author} {\bibinfo {author} {\bibfnamefont {C.}~\bibnamefont
  {Pica}}, \bibinfo {author} {\bibfnamefont {T.~A.}\ \bibnamefont {Ryttov}}, \
  and\ \bibinfo {author} {\bibfnamefont {F.}~\bibnamefont {Sannino}},\
  }\href@noop {} {\  (\bibinfo {year} {2016})},\ \Eprint
  {http://arxiv.org/abs/1605.04712} {arXiv:1605.04712 [hep-th]} \BibitemShut
  {NoStop}%
%%CITATION = ARXIV:1605.04712;%%
\bibitem [{\citenamefont {Molgaard}\ and\ \citenamefont
  {Sannino}(2016)}]{Molgaard:2016bqf}%
  \BibitemOpen
  \bibfield  {author} {\bibinfo {author} {\bibfnamefont {E.}~\bibnamefont
  {Molgaard}}\ and\ \bibinfo {author} {\bibfnamefont {F.}~\bibnamefont
  {Sannino}},\ }\href@noop {} {\  (\bibinfo {year} {2016})},\ \Eprint
  {http://arxiv.org/abs/1610.03130} {arXiv:1610.03130 [hep-ph]} \BibitemShut
  {NoStop}%
%%CITATION = ARXIV:1610.03130;%%
\bibitem [{\citenamefont {Heikinheimo}\ \emph {et~al.}(2017)\citenamefont
  {Heikinheimo}, \citenamefont {Kannike}, \citenamefont {Lyonnet},
  \citenamefont {Raidal}, \citenamefont {Tuominen},\ and\ \citenamefont
  {Veermäe}}]{Heikinheimo:2017nth}%
  \BibitemOpen
  \bibfield  {author} {\bibinfo {author} {\bibfnamefont {M.}~\bibnamefont
  {Heikinheimo}}, \bibinfo {author} {\bibfnamefont {K.}~\bibnamefont
  {Kannike}}, \bibinfo {author} {\bibfnamefont {F.}~\bibnamefont {Lyonnet}},
  \bibinfo {author} {\bibfnamefont {M.}~\bibnamefont {Raidal}}, \bibinfo
  {author} {\bibfnamefont {K.}~\bibnamefont {Tuominen}}, \ and\ \bibinfo
  {author} {\bibfnamefont {H.}~\bibnamefont {Veermäe}},\ }\href {\doibase
  10.1007/JHEP10(2017)014} {\bibfield  {journal} {\bibinfo  {journal} {JHEP}\
  }\textbf {\bibinfo {volume} {10}},\ \bibinfo {pages} {014} (\bibinfo {year}
  {2017})},\ \Eprint {http://arxiv.org/abs/1707.08980} {arXiv:1707.08980
  [hep-ph]} \BibitemShut {NoStop}%
%%CITATION = ARXIV:1707.08980;%%
\bibitem [{\citenamefont {Buttazzo}\ \emph {et~al.}(2013)\citenamefont
  {Buttazzo}, \citenamefont {Degrassi}, \citenamefont {Giardino}, \citenamefont
  {Giudice}, \citenamefont {Sala}, \citenamefont {Salvio},\ and\ \citenamefont
  {Strumia}}]{Buttazzo:2013uya}%
  \BibitemOpen
  \bibfield  {author} {\bibinfo {author} {\bibfnamefont {D.}~\bibnamefont
  {Buttazzo}}, \bibinfo {author} {\bibfnamefont {G.}~\bibnamefont {Degrassi}},
  \bibinfo {author} {\bibfnamefont {P.~P.}\ \bibnamefont {Giardino}}, \bibinfo
  {author} {\bibfnamefont {G.~F.}\ \bibnamefont {Giudice}}, \bibinfo {author}
  {\bibfnamefont {F.}~\bibnamefont {Sala}}, \bibinfo {author} {\bibfnamefont
  {A.}~\bibnamefont {Salvio}}, \ and\ \bibinfo {author} {\bibfnamefont
  {A.}~\bibnamefont {Strumia}},\ }\href {\doibase 10.1007/JHEP12(2013)089}
  {\bibfield  {journal} {\bibinfo  {journal} {JHEP}\ }\textbf {\bibinfo
  {volume} {12}},\ \bibinfo {pages} {089} (\bibinfo {year} {2013})},\ \Eprint
  {http://arxiv.org/abs/1307.3536} {arXiv:1307.3536 [hep-ph]} \BibitemShut
  {NoStop}%
%%CITATION = ARXIV:1307.3536;%%
\bibitem [{\citenamefont {Bednyakov}\ \emph {et~al.}(2015)\citenamefont
  {Bednyakov}, \citenamefont {Kniehl}, \citenamefont {Pikelner},\ and\
  \citenamefont {Veretin}}]{Bednyakov:2015sca}%
  \BibitemOpen
  \bibfield  {author} {\bibinfo {author} {\bibfnamefont {A.~V.}\ \bibnamefont
  {Bednyakov}}, \bibinfo {author} {\bibfnamefont {B.~A.}\ \bibnamefont
  {Kniehl}}, \bibinfo {author} {\bibfnamefont {A.~F.}\ \bibnamefont
  {Pikelner}}, \ and\ \bibinfo {author} {\bibfnamefont {O.~L.}\ \bibnamefont
  {Veretin}},\ }\href {\doibase 10.1103/PhysRevLett.115.201802} {\bibfield
  {journal} {\bibinfo  {journal} {Phys. Rev. Lett.}\ }\textbf {\bibinfo
  {volume} {115}},\ \bibinfo {pages} {201802} (\bibinfo {year} {2015})},\
  \Eprint {http://arxiv.org/abs/1507.08833} {arXiv:1507.08833 [hep-ph]}
  \BibitemShut {NoStop}%
%%CITATION = ARXIV:1507.08833;%%
\bibitem [{\citenamefont {Di~Luzio}\ \emph {et~al.}(2015)\citenamefont
  {Di~Luzio}, \citenamefont {Isidori},\ and\ \citenamefont
  {Ridolfi}}]{DiLuzio:2015iua}%
  \BibitemOpen
  \bibfield  {author} {\bibinfo {author} {\bibfnamefont {L.}~\bibnamefont
  {Di~Luzio}}, \bibinfo {author} {\bibfnamefont {G.}~\bibnamefont {Isidori}}, \
  and\ \bibinfo {author} {\bibfnamefont {G.}~\bibnamefont {Ridolfi}},\
  }\href@noop {} {\  (\bibinfo {year} {2015})},\ \Eprint
  {http://arxiv.org/abs/1509.05028} {arXiv:1509.05028 [hep-ph]} \BibitemShut
  {NoStop}%
%%CITATION = ARXIV:1509.05028;%%
\bibitem [{\citenamefont {Andreassen}\ \emph {et~al.}(2017)\citenamefont
  {Andreassen}, \citenamefont {Frost},\ and\ \citenamefont
  {Schwartz}}]{Andreassen:2017rzq}%
  \BibitemOpen
  \bibfield  {author} {\bibinfo {author} {\bibfnamefont {A.}~\bibnamefont
  {Andreassen}}, \bibinfo {author} {\bibfnamefont {W.}~\bibnamefont {Frost}}, \
  and\ \bibinfo {author} {\bibfnamefont {M.~D.}\ \bibnamefont {Schwartz}},\
  }\href@noop {} {\  (\bibinfo {year} {2017})},\ \Eprint
  {http://arxiv.org/abs/1707.08124} {arXiv:1707.08124 [hep-ph]} \BibitemShut
  {NoStop}%
%%CITATION = ARXIV:1707.08124;%%
\bibitem [{\citenamefont {Alekhin}\ \emph {et~al.}(2012)\citenamefont
  {Alekhin}, \citenamefont {Djouadi},\ and\ \citenamefont
  {Moch}}]{Alekhin:2012py}%
  \BibitemOpen
  \bibfield  {author} {\bibinfo {author} {\bibfnamefont {S.}~\bibnamefont
  {Alekhin}}, \bibinfo {author} {\bibfnamefont {A.}~\bibnamefont {Djouadi}}, \
  and\ \bibinfo {author} {\bibfnamefont {S.}~\bibnamefont {Moch}},\ }\href
  {\doibase 10.1016/j.physletb.2012.08.024} {\bibfield  {journal} {\bibinfo
  {journal} {Phys. Lett.}\ }\textbf {\bibinfo {volume} {B716}},\ \bibinfo
  {pages} {214} (\bibinfo {year} {2012})},\ \Eprint
  {http://arxiv.org/abs/1207.0980} {arXiv:1207.0980 [hep-ph]} \BibitemShut
  {NoStop}%
%%CITATION = ARXIV:1207.0980;%%
\bibitem [{\citenamefont {Bezrukov}\ and\ \citenamefont
  {Shaposhnikov}(2015)}]{Bezrukov:2014ina}%
  \BibitemOpen
  \bibfield  {author} {\bibinfo {author} {\bibfnamefont {F.}~\bibnamefont
  {Bezrukov}}\ and\ \bibinfo {author} {\bibfnamefont {M.}~\bibnamefont
  {Shaposhnikov}},\ }\href {\doibase 10.1134/S1063776115030152} {\bibfield
  {journal} {\bibinfo  {journal} {J. Exp. Theor. Phys.}\ }\textbf {\bibinfo
  {volume} {120}},\ \bibinfo {pages} {335} (\bibinfo {year} {2015})},\ \Eprint
  {http://arxiv.org/abs/1411.1923} {arXiv:1411.1923 [hep-ph]} \BibitemShut
  {NoStop}%
%%CITATION = ARXIV:1411.1923;%%
\bibitem [{\citenamefont {Gies}\ \emph {et~al.}(2014)\citenamefont {Gies},
  \citenamefont {Gneiting},\ and\ \citenamefont {Sondenheimer}}]{Gies:2013fua}%
  \BibitemOpen
  \bibfield  {author} {\bibinfo {author} {\bibfnamefont {H.}~\bibnamefont
  {Gies}}, \bibinfo {author} {\bibfnamefont {C.}~\bibnamefont {Gneiting}}, \
  and\ \bibinfo {author} {\bibfnamefont {R.}~\bibnamefont {Sondenheimer}},\
  }\href {\doibase 10.1103/PhysRevD.89.045012} {\bibfield  {journal} {\bibinfo
  {journal} {Phys. Rev.}\ }\textbf {\bibinfo {volume} {D89}},\ \bibinfo {pages}
  {045012} (\bibinfo {year} {2014})},\ \Eprint {http://arxiv.org/abs/1308.5075}
  {arXiv:1308.5075 [hep-ph]} \BibitemShut {NoStop}%
%%CITATION = ARXIV:1308.5075;%%
\bibitem [{\citenamefont {Branchina}\ and\ \citenamefont
  {Messina}(2013)}]{Branchina:2013jra}%
  \BibitemOpen
  \bibfield  {author} {\bibinfo {author} {\bibfnamefont {V.}~\bibnamefont
  {Branchina}}\ and\ \bibinfo {author} {\bibfnamefont {E.}~\bibnamefont
  {Messina}},\ }\href {\doibase 10.1103/PhysRevLett.111.241801} {\bibfield
  {journal} {\bibinfo  {journal} {Phys. Rev. Lett.}\ }\textbf {\bibinfo
  {volume} {111}},\ \bibinfo {pages} {241801} (\bibinfo {year} {2013})},\
  \Eprint {http://arxiv.org/abs/1307.5193} {arXiv:1307.5193 [hep-ph]}
  \BibitemShut {NoStop}%
%%CITATION = ARXIV:1307.5193;%%
\bibitem [{\citenamefont {Hegde}\ \emph {et~al.}(2014)\citenamefont {Hegde},
  \citenamefont {Jansen}, \citenamefont {Lin},\ and\ \citenamefont
  {Nagy}}]{Hegde:2013mks}%
  \BibitemOpen
  \bibfield  {author} {\bibinfo {author} {\bibfnamefont {P.}~\bibnamefont
  {Hegde}}, \bibinfo {author} {\bibfnamefont {K.}~\bibnamefont {Jansen}},
  \bibinfo {author} {\bibfnamefont {C.~J.~D.}\ \bibnamefont {Lin}}, \ and\
  \bibinfo {author} {\bibfnamefont {A.}~\bibnamefont {Nagy}},\ }\bibfield
  {booktitle} {\emph {\bibinfo {booktitle} {{Proceedings, 31st International
  Symposium on Lattice Field Theory (Lattice 2013)}}},\ }\href@noop {}
  {\bibfield  {journal} {\bibinfo  {journal} {PoS}\ }\textbf {\bibinfo {volume}
  {LATTICE2013}},\ \bibinfo {pages} {058} (\bibinfo {year} {2014})},\ \Eprint
  {http://arxiv.org/abs/1310.6260} {arXiv:1310.6260 [hep-lat]} \BibitemShut
  {NoStop}%
%%CITATION = ARXIV:1310.6260;%%
\bibitem [{\citenamefont {Gies}\ and\ \citenamefont
  {Sondenheimer}(2015)}]{Gies:2014xha}%
  \BibitemOpen
  \bibfield  {author} {\bibinfo {author} {\bibfnamefont {H.}~\bibnamefont
  {Gies}}\ and\ \bibinfo {author} {\bibfnamefont {R.}~\bibnamefont
  {Sondenheimer}},\ }\href {\doibase 10.1140/epjc/s10052-015-3284-1} {\bibfield
   {journal} {\bibinfo  {journal} {Eur. Phys. J.}\ }\textbf {\bibinfo {volume}
  {C75}},\ \bibinfo {pages} {68} (\bibinfo {year} {2015})},\ \Eprint
  {http://arxiv.org/abs/1407.8124} {arXiv:1407.8124 [hep-ph]} \BibitemShut
  {NoStop}%
%%CITATION = ARXIV:1407.8124;%%
\bibitem [{\citenamefont {Eichhorn}\ \emph {et~al.}(2015)\citenamefont
  {Eichhorn}, \citenamefont {Gies}, \citenamefont {Jaeckel}, \citenamefont
  {Plehn}, \citenamefont {Scherer},\ and\ \citenamefont
  {Sondenheimer}}]{Eichhorn:2015kea}%
  \BibitemOpen
  \bibfield  {author} {\bibinfo {author} {\bibfnamefont {A.}~\bibnamefont
  {Eichhorn}}, \bibinfo {author} {\bibfnamefont {H.}~\bibnamefont {Gies}},
  \bibinfo {author} {\bibfnamefont {J.}~\bibnamefont {Jaeckel}}, \bibinfo
  {author} {\bibfnamefont {T.}~\bibnamefont {Plehn}}, \bibinfo {author}
  {\bibfnamefont {M.~M.}\ \bibnamefont {Scherer}}, \ and\ \bibinfo {author}
  {\bibfnamefont {R.}~\bibnamefont {Sondenheimer}},\ }\href {\doibase
  10.1007/JHEP04(2015)022} {\bibfield  {journal} {\bibinfo  {journal} {JHEP}\
  }\textbf {\bibinfo {volume} {04}},\ \bibinfo {pages} {022} (\bibinfo {year}
  {2015})},\ \Eprint {http://arxiv.org/abs/1501.02812} {arXiv:1501.02812
  [hep-ph]} \BibitemShut {NoStop}%
%%CITATION = ARXIV:1501.02812;%%
\bibitem [{\citenamefont {Chu}\ \emph {et~al.}(2015)\citenamefont {Chu},
  \citenamefont {Jansen}, \citenamefont {Knippschild}, \citenamefont {Lin},\
  and\ \citenamefont {Nagy}}]{Chu:2015nha}%
  \BibitemOpen
  \bibfield  {author} {\bibinfo {author} {\bibfnamefont {D.~Y.~J.}\
  \bibnamefont {Chu}}, \bibinfo {author} {\bibfnamefont {K.}~\bibnamefont
  {Jansen}}, \bibinfo {author} {\bibfnamefont {B.}~\bibnamefont {Knippschild}},
  \bibinfo {author} {\bibfnamefont {C.~J.~D.}\ \bibnamefont {Lin}}, \ and\
  \bibinfo {author} {\bibfnamefont {A.}~\bibnamefont {Nagy}},\ }\href {\doibase
  10.1016/j.physletb.2015.03.050} {\bibfield  {journal} {\bibinfo  {journal}
  {Phys. Lett.}\ }\textbf {\bibinfo {volume} {B744}},\ \bibinfo {pages} {146}
  (\bibinfo {year} {2015})},\ \Eprint {http://arxiv.org/abs/1501.05440}
  {arXiv:1501.05440 [hep-lat]} \BibitemShut {NoStop}%
%%CITATION = ARXIV:1501.05440;%%
\bibitem [{\citenamefont {Chu}\ \emph {et~al.}(2014)\citenamefont {Chu},
  \citenamefont {Jansen}, \citenamefont {Knippschild}, \citenamefont {Lin},
  \citenamefont {Nagai},\ and\ \citenamefont {Nagy}}]{Chu:2015ula}%
  \BibitemOpen
  \bibfield  {author} {\bibinfo {author} {\bibfnamefont {D.~Y.~J.}\
  \bibnamefont {Chu}}, \bibinfo {author} {\bibfnamefont {K.}~\bibnamefont
  {Jansen}}, \bibinfo {author} {\bibfnamefont {B.}~\bibnamefont {Knippschild}},
  \bibinfo {author} {\bibfnamefont {C.~J.~D.}\ \bibnamefont {Lin}}, \bibinfo
  {author} {\bibfnamefont {K.-I.}\ \bibnamefont {Nagai}}, \ and\ \bibinfo
  {author} {\bibfnamefont {A.}~\bibnamefont {Nagy}},\ }\bibfield  {booktitle}
  {\emph {\bibinfo {booktitle} {{Proceedings, 32nd International Symposium on
  Lattice Field Theory (Lattice 2014)}}},\ }\href@noop {} {\bibfield  {journal}
  {\bibinfo  {journal} {PoS}\ }\textbf {\bibinfo {volume} {LATTICE2014}},\
  \bibinfo {pages} {278} (\bibinfo {year} {2014})},\ \Eprint
  {http://arxiv.org/abs/1501.00306} {arXiv:1501.00306 [hep-lat]} \BibitemShut
  {NoStop}%
%%CITATION = ARXIV:1501.00306;%%
\bibitem [{\citenamefont {Akerlund}\ and\ \citenamefont {\protect{de
  Forcrand}}(2016)}]{Akerlund:2015fya}%
  \BibitemOpen
  \bibfield  {author} {\bibinfo {author} {\bibfnamefont {O.}~\bibnamefont
  {Akerlund}}\ and\ \bibinfo {author} {\bibfnamefont {P.}~\bibnamefont
  {\protect{de Forcrand}}},\ }\href {\doibase 10.1103/PhysRevD.93.035015}
  {\bibfield  {journal} {\bibinfo  {journal} {{Phys. Rev.}}\ }\textbf {\bibinfo
  {volume} {D93}},\ \bibinfo {pages} {035015} (\bibinfo {year} {2016})},\
  \Eprint {http://arxiv.org/abs/1508.07959} {arXiv:1508.07959 [hep-lat]}
  \BibitemShut {NoStop}%
%%CITATION = ARXIV:1508.07959;%%
\bibitem [{\citenamefont {Sondenheimer}(2017)}]{Sondenheimer:2017jin}%
  \BibitemOpen
  \bibfield  {author} {\bibinfo {author} {\bibfnamefont {R.}~\bibnamefont
  {Sondenheimer}},\ }\href@noop {} {\  (\bibinfo {year} {2017})},\ \Eprint
  {http://arxiv.org/abs/1711.00065} {arXiv:1711.00065 [hep-ph]} \BibitemShut
  {NoStop}%
%%CITATION = ARXIV:1711.00065;%%
\bibitem [{\citenamefont {Holthausen}\ \emph {et~al.}(2013)\citenamefont
  {Holthausen}, \citenamefont {Kubo}, \citenamefont {Lim},\ and\ \citenamefont
  {Lindner}}]{Holthausen:2013ota}%
  \BibitemOpen
  \bibfield  {author} {\bibinfo {author} {\bibfnamefont {M.}~\bibnamefont
  {Holthausen}}, \bibinfo {author} {\bibfnamefont {J.}~\bibnamefont {Kubo}},
  \bibinfo {author} {\bibfnamefont {K.~S.}\ \bibnamefont {Lim}}, \ and\
  \bibinfo {author} {\bibfnamefont {M.}~\bibnamefont {Lindner}},\ }\href
  {\doibase 10.1007/JHEP12(2013)076} {\bibfield  {journal} {\bibinfo  {journal}
  {JHEP}\ }\textbf {\bibinfo {volume} {12}},\ \bibinfo {pages} {076} (\bibinfo
  {year} {2013})},\ \Eprint {http://arxiv.org/abs/1310.4423} {arXiv:1310.4423
  [hep-ph]} \BibitemShut {NoStop}%
%%CITATION = ARXIV:1310.4423;%%
\bibitem [{\citenamefont {Helmboldt}\ \emph {et~al.}(2016)\citenamefont
  {Helmboldt}, \citenamefont {Humbert}, \citenamefont {Lindner},\ and\
  \citenamefont {Smirnov}}]{Helmboldt:2016mpi}%
  \BibitemOpen
  \bibfield  {author} {\bibinfo {author} {\bibfnamefont {A.~J.}\ \bibnamefont
  {Helmboldt}}, \bibinfo {author} {\bibfnamefont {P.}~\bibnamefont {Humbert}},
  \bibinfo {author} {\bibfnamefont {M.}~\bibnamefont {Lindner}}, \ and\
  \bibinfo {author} {\bibfnamefont {J.}~\bibnamefont {Smirnov}},\ }\href@noop
  {} {\  (\bibinfo {year} {2016})},\ \Eprint {http://arxiv.org/abs/1603.03603}
  {arXiv:1603.03603 [hep-ph]} \BibitemShut {NoStop}%
%%CITATION = ARXIV:1603.03603;%%
\bibitem [{\citenamefont {Ahriche}\ \emph {et~al.}(2016)\citenamefont
  {Ahriche}, \citenamefont {Manning}, \citenamefont {McDonald},\ and\
  \citenamefont {Nasri}}]{Ahriche:2016ixu}%
  \BibitemOpen
  \bibfield  {author} {\bibinfo {author} {\bibfnamefont {A.}~\bibnamefont
  {Ahriche}}, \bibinfo {author} {\bibfnamefont {A.}~\bibnamefont {Manning}},
  \bibinfo {author} {\bibfnamefont {K.~L.}\ \bibnamefont {McDonald}}, \ and\
  \bibinfo {author} {\bibfnamefont {S.}~\bibnamefont {Nasri}},\ }\href
  {\doibase 10.1103/PhysRevD.94.053005} {\bibfield  {journal} {\bibinfo
  {journal} {Phys. Rev.}\ }\textbf {\bibinfo {volume} {D94}},\ \bibinfo {pages}
  {053005} (\bibinfo {year} {2016})},\ \Eprint
  {http://arxiv.org/abs/1604.05995} {arXiv:1604.05995 [hep-ph]} \BibitemShut
  {NoStop}%
%%CITATION = ARXIV:1604.05995;%%
\bibitem [{\citenamefont {Shaposhnikov}\ and\ \citenamefont
  {Shkerin}(2018)}]{Shaposhnikov:2018xkv}%
  \BibitemOpen
  \bibfield  {author} {\bibinfo {author} {\bibfnamefont {M.}~\bibnamefont
  {Shaposhnikov}}\ and\ \bibinfo {author} {\bibfnamefont {A.}~\bibnamefont
  {Shkerin}},\ }\href@noop {} {\  (\bibinfo {year} {2018})},\ \Eprint
  {http://arxiv.org/abs/1803.08907} {arXiv:1803.08907 [hep-th]} \BibitemShut
  {NoStop}%
%%CITATION = ARXIV:1803.08907;%%
\bibitem [{\citenamefont {Gies}\ and\ \citenamefont
  {Zambelli}(2015)}]{Gies:2015lia}%
  \BibitemOpen
  \bibfield  {author} {\bibinfo {author} {\bibfnamefont {H.}~\bibnamefont
  {Gies}}\ and\ \bibinfo {author} {\bibfnamefont {L.}~\bibnamefont
  {Zambelli}},\ }\href {\doibase 10.1103/PhysRevD.92.025016} {\bibfield
  {journal} {\bibinfo  {journal} {Phys. Rev.}\ }\textbf {\bibinfo {volume}
  {D92}},\ \bibinfo {pages} {025016} (\bibinfo {year} {2015})},\ \Eprint
  {http://arxiv.org/abs/1502.05907} {arXiv:1502.05907 [hep-ph]} \BibitemShut
  {NoStop}%
%%CITATION = ARXIV:1502.05907;%%
\bibitem [{\citenamefont {Gies}\ and\ \citenamefont
  {Zambelli}(2017)}]{Gies:2016kkk}%
  \BibitemOpen
  \bibfield  {author} {\bibinfo {author} {\bibfnamefont {H.}~\bibnamefont
  {Gies}}\ and\ \bibinfo {author} {\bibfnamefont {L.}~\bibnamefont
  {Zambelli}},\ }\href {\doibase 10.1103/PhysRevD.96.025003} {\bibfield
  {journal} {\bibinfo  {journal} {Phys. Rev.}\ }\textbf {\bibinfo {volume}
  {D96}},\ \bibinfo {pages} {025003} (\bibinfo {year} {2017})},\ \Eprint
  {http://arxiv.org/abs/1611.09147} {arXiv:1611.09147 [hep-ph]} \BibitemShut
  {NoStop}%
%%CITATION = ARXIV:1611.09147;%%
\bibitem [{\citenamefont {Reichert}\ \emph {et~al.}(2018)\citenamefont
  {Reichert}, \citenamefont {Eichhorn}, \citenamefont {Gies}, \citenamefont
  {Pawlowski}, \citenamefont {Plehn},\ and\ \citenamefont
  {Scherer}}]{Reichert:2017puo}%
  \BibitemOpen
  \bibfield  {author} {\bibinfo {author} {\bibfnamefont {M.}~\bibnamefont
  {Reichert}}, \bibinfo {author} {\bibfnamefont {A.}~\bibnamefont {Eichhorn}},
  \bibinfo {author} {\bibfnamefont {H.}~\bibnamefont {Gies}}, \bibinfo {author}
  {\bibfnamefont {J.~M.}\ \bibnamefont {Pawlowski}}, \bibinfo {author}
  {\bibfnamefont {T.}~\bibnamefont {Plehn}}, \ and\ \bibinfo {author}
  {\bibfnamefont {M.~M.}\ \bibnamefont {Scherer}},\ }\href {\doibase
  10.1103/PhysRevD.97.075008} {\bibfield  {journal} {\bibinfo  {journal} {Phys.
  Rev.}\ }\textbf {\bibinfo {volume} {D97}},\ \bibinfo {pages} {075008}
  (\bibinfo {year} {2018})},\ \Eprint {http://arxiv.org/abs/1711.00019}
  {arXiv:1711.00019 [hep-ph]} \BibitemShut {NoStop}%
%%CITATION = ARXIV:1711.00019;%%
\bibitem [{\citenamefont {Ellwanger}\ \emph {et~al.}(1996)\citenamefont
  {Ellwanger}, \citenamefont {Hirsch},\ and\ \citenamefont
  {Weber}}]{Ellwanger:1995qf}%
  \BibitemOpen
  \bibfield  {author} {\bibinfo {author} {\bibfnamefont {U.}~\bibnamefont
  {Ellwanger}}, \bibinfo {author} {\bibfnamefont {M.}~\bibnamefont {Hirsch}}, \
  and\ \bibinfo {author} {\bibfnamefont {A.}~\bibnamefont {Weber}},\ }\href
  {\doibase 10.1007/s002880050073} {\bibfield  {journal} {\bibinfo  {journal}
  {Z. Phys.}\ }\textbf {\bibinfo {volume} {C69}},\ \bibinfo {pages} {687}
  (\bibinfo {year} {1996})},\ \Eprint {http://arxiv.org/abs/hep-th/9506019}
  {arXiv:hep-th/9506019 [hep-th]} \BibitemShut {NoStop}%
%%CITATION = HEP-TH/9506019;%%
\bibitem [{\citenamefont {Litim}\ and\ \citenamefont
  {Pawlowski}(1998)}]{Litim:1998qi}%
  \BibitemOpen
  \bibfield  {author} {\bibinfo {author} {\bibfnamefont {D.~F.}\ \bibnamefont
  {Litim}}\ and\ \bibinfo {author} {\bibfnamefont {J.~M.}\ \bibnamefont
  {Pawlowski}},\ }\href {\doibase 10.1016/S0370-2693(98)00761-8} {\bibfield
  {journal} {\bibinfo  {journal} {Phys.Lett.}\ }\textbf {\bibinfo {volume}
  {B435}},\ \bibinfo {pages} {181} (\bibinfo {year} {1998})},\ \Eprint
  {http://arxiv.org/abs/hep-th/9802064} {arXiv:hep-th/9802064 [hep-th]}
  \BibitemShut {NoStop}%
%%CITATION = HEP-TH/9802064;%%
\bibitem [{\citenamefont {Litim}\ and\ \citenamefont
  {Sannino}(2014)}]{Litim:2014uca}%
  \BibitemOpen
  \bibfield  {author} {\bibinfo {author} {\bibfnamefont {D.~F.}\ \bibnamefont
  {Litim}}\ and\ \bibinfo {author} {\bibfnamefont {F.}~\bibnamefont
  {Sannino}},\ }\href {\doibase 10.1007/JHEP12(2014)178} {\bibfield  {journal}
  {\bibinfo  {journal} {JHEP}\ }\textbf {\bibinfo {volume} {12}},\ \bibinfo
  {pages} {178} (\bibinfo {year} {2014})},\ \Eprint
  {http://arxiv.org/abs/1406.2337} {arXiv:1406.2337 [hep-th]} \BibitemShut
  {NoStop}%
%%CITATION = ARXIV:1406.2337;%%
\bibitem [{\citenamefont {Bond}\ and\ \citenamefont
  {Litim}(2016)}]{Bond:2016dvk}%
  \BibitemOpen
  \bibfield  {author} {\bibinfo {author} {\bibfnamefont {A.~D.}\ \bibnamefont
  {Bond}}\ and\ \bibinfo {author} {\bibfnamefont {D.~F.}\ \bibnamefont
  {Litim}},\ }\href@noop {} {\  (\bibinfo {year} {2016})},\ \Eprint
  {http://arxiv.org/abs/1608.00519} {arXiv:1608.00519 [hep-th]} \BibitemShut
  {NoStop}%
%%CITATION = ARXIV:1608.00519;%%
\bibitem [{\citenamefont {Codello}\ \emph {et~al.}(2016)\citenamefont
  {Codello}, \citenamefont {Langæble}, \citenamefont {Litim},\ and\
  \citenamefont {Sannino}}]{Codello:2016muj}%
  \BibitemOpen
  \bibfield  {author} {\bibinfo {author} {\bibfnamefont {A.}~\bibnamefont
  {Codello}}, \bibinfo {author} {\bibfnamefont {K.}~\bibnamefont {Langæble}},
  \bibinfo {author} {\bibfnamefont {D.~F.}\ \bibnamefont {Litim}}, \ and\
  \bibinfo {author} {\bibfnamefont {F.}~\bibnamefont {Sannino}},\ }\href
  {\doibase 10.1007/JHEP07(2016)118} {\bibfield  {journal} {\bibinfo  {journal}
  {JHEP}\ }\textbf {\bibinfo {volume} {07}},\ \bibinfo {pages} {118} (\bibinfo
  {year} {2016})},\ \Eprint {http://arxiv.org/abs/1603.03462} {arXiv:1603.03462
  [hep-th]} \BibitemShut {NoStop}%
%%CITATION = ARXIV:1603.03462;%%
\bibitem [{\citenamefont {Bajc}\ and\ \citenamefont
  {Sannino}(2016)}]{Bajc:2016efj}%
  \BibitemOpen
  \bibfield  {author} {\bibinfo {author} {\bibfnamefont {B.}~\bibnamefont
  {Bajc}}\ and\ \bibinfo {author} {\bibfnamefont {F.}~\bibnamefont {Sannino}},\
  }\href@noop {} {\  (\bibinfo {year} {2016})},\ \Eprint
  {http://arxiv.org/abs/1610.09681} {arXiv:1610.09681 [hep-th]} \BibitemShut
  {NoStop}%
%%CITATION = ARXIV:1610.09681;%%
\bibitem [{\citenamefont {Pendleton}\ and\ \citenamefont
  {Ross}(1981)}]{Pendleton:1980as}%
  \BibitemOpen
  \bibfield  {author} {\bibinfo {author} {\bibfnamefont {B.}~\bibnamefont
  {Pendleton}}\ and\ \bibinfo {author} {\bibfnamefont {G.~G.}\ \bibnamefont
  {Ross}},\ }\href {\doibase 10.1016/0370-2693(81)90017-4} {\bibfield
  {journal} {\bibinfo  {journal} {Phys. Lett.}\ }\textbf {\bibinfo {volume}
  {98B}},\ \bibinfo {pages} {291} (\bibinfo {year} {1981})}\BibitemShut
  {NoStop}%
%%CITATION = PHLTA,98B,291;%%
\bibitem [{\citenamefont {Maas}(2013)}]{Maas:2012tj}%
  \BibitemOpen
  \bibfield  {author} {\bibinfo {author} {\bibfnamefont {A.}~\bibnamefont
  {Maas}},\ }\href {\doibase 10.1142/S0217732313501034} {\bibfield  {journal}
  {\bibinfo  {journal} {Mod. Phys. Lett.}\ }\textbf {\bibinfo {volume} {A28}},\
  \bibinfo {pages} {1350103} (\bibinfo {year} {2013})},\ \Eprint
  {http://arxiv.org/abs/1205.6625} {arXiv:1205.6625 [hep-lat]} \BibitemShut
  {NoStop}%
%%CITATION = ARXIV:1205.6625;%%
\bibitem [{\citenamefont {Maas}\ and\ \citenamefont
  {Mufti}(2015)}]{Maas:2014pba}%
  \BibitemOpen
  \bibfield  {author} {\bibinfo {author} {\bibfnamefont {A.}~\bibnamefont
  {Maas}}\ and\ \bibinfo {author} {\bibfnamefont {T.}~\bibnamefont {Mufti}},\
  }\href {\doibase 10.1103/PhysRevD.91.113011} {\bibfield  {journal} {\bibinfo
  {journal} {Phys. Rev.}\ }\textbf {\bibinfo {volume} {D91}},\ \bibinfo {pages}
  {113011} (\bibinfo {year} {2015})},\ \Eprint {http://arxiv.org/abs/1412.6440}
  {arXiv:1412.6440 [hep-lat]} \BibitemShut {NoStop}%
%%CITATION = ARXIV:1412.6440;%%
\bibitem [{\citenamefont {Maas}\ \emph {et~al.}(2017)\citenamefont {Maas},
  \citenamefont {Sondenheimer},\ and\ \citenamefont {Törek}}]{Maas:2017xzh}%
  \BibitemOpen
  \bibfield  {author} {\bibinfo {author} {\bibfnamefont {A.}~\bibnamefont
  {Maas}}, \bibinfo {author} {\bibfnamefont {R.}~\bibnamefont {Sondenheimer}},
  \ and\ \bibinfo {author} {\bibfnamefont {P.}~\bibnamefont {Törek}},\
  }\href@noop {} {\  (\bibinfo {year} {2017})},\ \Eprint
  {http://arxiv.org/abs/1709.07477} {arXiv:1709.07477 [hep-ph]} \BibitemShut
  {NoStop}%
%%CITATION = ARXIV:1709.07477;%%
\bibitem [{\citenamefont {Maas}(2017)}]{Maas:2017wzi}%
  \BibitemOpen
  \bibfield  {author} {\bibinfo {author} {\bibfnamefont {A.}~\bibnamefont
  {Maas}},\ }\href@noop {} {\  (\bibinfo {year} {2017})},\ \Eprint
  {http://arxiv.org/abs/1712.04721} {arXiv:1712.04721 [hep-ph]} \BibitemShut
  {NoStop}%
%%CITATION = ARXIV:1712.04721;%%
\bibitem [{\citenamefont {Wilson}\ and\ \citenamefont
  {Kogut}(1974)}]{Wilson:1973jj}%
  \BibitemOpen
  \bibfield  {author} {\bibinfo {author} {\bibfnamefont {K.~G.}\ \bibnamefont
  {Wilson}}\ and\ \bibinfo {author} {\bibfnamefont {J.~B.}\ \bibnamefont
  {Kogut}},\ }\href {\doibase 10.1016/0370-1573(74)90023-4} {\bibfield
  {journal} {\bibinfo  {journal} {Phys. Rept.}\ }\textbf {\bibinfo {volume}
  {12}},\ \bibinfo {pages} {75} (\bibinfo {year} {1974})}\BibitemShut {NoStop}%
%%CITATION = PRPLC,12,75;%%
\bibitem [{\citenamefont {Wegner}\ and\ \citenamefont
  {Houghton}(1973)}]{Wegner:1972ih}%
  \BibitemOpen
  \bibfield  {author} {\bibinfo {author} {\bibfnamefont {F.~J.}\ \bibnamefont
  {Wegner}}\ and\ \bibinfo {author} {\bibfnamefont {A.}~\bibnamefont
  {Houghton}},\ }\href {\doibase 10.1103/PhysRevA.8.401} {\bibfield  {journal}
  {\bibinfo  {journal} {Phys. Rev.}\ }\textbf {\bibinfo {volume} {A8}},\
  \bibinfo {pages} {401} (\bibinfo {year} {1973})}\BibitemShut {NoStop}%
%%CITATION = PHRVA,A8,401;%%
\bibitem [{\citenamefont {Wetterich}(1993)}]{Wetterich:1992yh}%
  \BibitemOpen
  \bibfield  {author} {\bibinfo {author} {\bibfnamefont {C.}~\bibnamefont
  {Wetterich}},\ }\href {\doibase 10.1016/0370-2693(93)90726-X} {\bibfield
  {journal} {\bibinfo  {journal} {Phys. Lett.}\ }\textbf {\bibinfo {volume}
  {B301}},\ \bibinfo {pages} {90} (\bibinfo {year} {1993})}\BibitemShut
  {NoStop}%
%%CITATION = PHLTA,B301,90;%%
\bibitem [{\citenamefont {Ellwanger}(1994)}]{Ellwanger:1993mw}%
  \BibitemOpen
  \bibfield  {author} {\bibinfo {author} {\bibfnamefont {U.}~\bibnamefont
  {Ellwanger}},\ }\bibfield  {booktitle} {\emph {\bibinfo {booktitle}
  {{Proceedings, Workshop on Quantum field theoretical aspects of high energy
  physics: Bad Frankenhausen, Germany, September 20-24, 1993}}},\ }\href
  {\doibase 10.1007/BF01555911} {\bibfield  {journal} {\bibinfo  {journal} {Z.
  Phys.}\ }\textbf {\bibinfo {volume} {C62}},\ \bibinfo {pages} {503} (\bibinfo
  {year} {1994})},\ \bibinfo {note} {[,206(1993)]},\ \Eprint
  {http://arxiv.org/abs/hep-ph/9308260} {arXiv:hep-ph/9308260 [hep-ph]}
  \BibitemShut {NoStop}%
%%CITATION = HEP-PH/9308260;%%
\bibitem [{\citenamefont {Morris}(1994{\natexlab{a}})}]{Morris:1993qb}%
  \BibitemOpen
  \bibfield  {author} {\bibinfo {author} {\bibfnamefont {T.~R.}\ \bibnamefont
  {Morris}},\ }\href {\doibase 10.1142/S0217751X94000972} {\bibfield  {journal}
  {\bibinfo  {journal} {Int. J. Mod. Phys.}\ }\textbf {\bibinfo {volume}
  {A9}},\ \bibinfo {pages} {2411} (\bibinfo {year} {1994}{\natexlab{a}})},\
  \Eprint {http://arxiv.org/abs/hep-ph/9308265} {arXiv:hep-ph/9308265 [hep-ph]}
  \BibitemShut {NoStop}%
%%CITATION = HEP-PH/9308265;%%
\bibitem [{\citenamefont {Bonini}\ \emph {et~al.}(1993)\citenamefont {Bonini},
  \citenamefont {D'Attanasio},\ and\ \citenamefont
  {Marchesini}}]{Bonini:1992vh}%
  \BibitemOpen
  \bibfield  {author} {\bibinfo {author} {\bibfnamefont {M.}~\bibnamefont
  {Bonini}}, \bibinfo {author} {\bibfnamefont {M.}~\bibnamefont {D'Attanasio}},
  \ and\ \bibinfo {author} {\bibfnamefont {G.}~\bibnamefont {Marchesini}},\
  }\href {\doibase 10.1016/0550-3213(93)90588-G} {\bibfield  {journal}
  {\bibinfo  {journal} {Nucl. Phys.}\ }\textbf {\bibinfo {volume} {B409}},\
  \bibinfo {pages} {441} (\bibinfo {year} {1993})},\ \Eprint
  {http://arxiv.org/abs/hep-th/9301114} {arXiv:hep-th/9301114 [hep-th]}
  \BibitemShut {NoStop}%
%%CITATION = HEP-TH/9301114;%%
\bibitem [{\citenamefont {Berges}\ \emph {et~al.}(2002)\citenamefont {Berges},
  \citenamefont {Tetradis},\ and\ \citenamefont {Wetterich}}]{Berges:2000ew}%
  \BibitemOpen
  \bibfield  {author} {\bibinfo {author} {\bibfnamefont {J.}~\bibnamefont
  {Berges}}, \bibinfo {author} {\bibfnamefont {N.}~\bibnamefont {Tetradis}}, \
  and\ \bibinfo {author} {\bibfnamefont {C.}~\bibnamefont {Wetterich}},\ }\href
  {\doibase 10.1016/S0370-1573(01)00098-9} {\bibfield  {journal} {\bibinfo
  {journal} {Phys. Rept.}\ }\textbf {\bibinfo {volume} {363}},\ \bibinfo
  {pages} {223} (\bibinfo {year} {2002})},\ \Eprint
  {http://arxiv.org/abs/hep-ph/0005122} {arXiv:hep-ph/0005122 [hep-ph]}
  \BibitemShut {NoStop}%
%%CITATION = HEP-PH/0005122;%%
\bibitem [{\citenamefont {Pawlowski}(2007)}]{Pawlowski:2005xe}%
  \BibitemOpen
  \bibfield  {author} {\bibinfo {author} {\bibfnamefont {J.~M.}\ \bibnamefont
  {Pawlowski}},\ }\href {\doibase 10.1016/j.aop.2007.01.007} {\bibfield
  {journal} {\bibinfo  {journal} {Annals Phys.}\ }\textbf {\bibinfo {volume}
  {322}},\ \bibinfo {pages} {2831} (\bibinfo {year} {2007})},\ \Eprint
  {http://arxiv.org/abs/hep-th/0512261} {arXiv:hep-th/0512261 [hep-th]}
  \BibitemShut {NoStop}%
%%CITATION = HEP-TH/0512261;%%
\bibitem [{\citenamefont {Gies}(2012)}]{Gies:2006wv}%
  \BibitemOpen
  \bibfield  {author} {\bibinfo {author} {\bibfnamefont {H.}~\bibnamefont
  {Gies}},\ }\bibfield  {booktitle} {\emph {\bibinfo {booktitle} {{ECT* School
  on Renormalization Group and Effective Field Theory Approaches to Many-Body
  Systems Trento, Italy, February 27-March 10, 2006}}},\ }\href {\doibase
  10.1007/978-3-642-27320-9_6} {\bibfield  {journal} {\bibinfo  {journal}
  {Lect. Notes Phys.}\ }\textbf {\bibinfo {volume} {852}},\ \bibinfo {pages}
  {287} (\bibinfo {year} {2012})},\ \Eprint
  {http://arxiv.org/abs/hep-ph/0611146} {arXiv:hep-ph/0611146 [hep-ph]}
  \BibitemShut {NoStop}%
%%CITATION = HEP-PH/0611146;%%
\bibitem [{\citenamefont {Delamotte}(2012)}]{Delamotte:2007pf}%
  \BibitemOpen
  \bibfield  {author} {\bibinfo {author} {\bibfnamefont {B.}~\bibnamefont
  {Delamotte}},\ }\href {\doibase 10.1007/978-3-642-27320-9_2} {\bibfield
  {journal} {\bibinfo  {journal} {Lect. Notes Phys.}\ }\textbf {\bibinfo
  {volume} {852}},\ \bibinfo {pages} {49} (\bibinfo {year} {2012})},\ \Eprint
  {http://arxiv.org/abs/cond-mat/0702365} {arXiv:cond-mat/0702365
  [cond-mat.stat-mech]} \BibitemShut {NoStop}%
%%CITATION = COND-MAT/0702365;%%
\bibitem [{\citenamefont {Braun}(2012)}]{Braun:2011pp}%
  \BibitemOpen
  \bibfield  {author} {\bibinfo {author} {\bibfnamefont {J.}~\bibnamefont
  {Braun}},\ }\href {\doibase 10.1088/0954-3899/39/3/033001} {\bibfield
  {journal} {\bibinfo  {journal} {J. Phys.}\ }\textbf {\bibinfo {volume}
  {G39}},\ \bibinfo {pages} {033001} (\bibinfo {year} {2012})},\ \Eprint
  {http://arxiv.org/abs/1108.4449} {arXiv:1108.4449 [hep-ph]} \BibitemShut
  {NoStop}%
%%CITATION = ARXIV:1108.4449;%%
\bibitem [{\citenamefont {Gies}\ \emph {et~al.}(2013)\citenamefont {Gies},
  \citenamefont {Rechenberger}, \citenamefont {Scherer},\ and\ \citenamefont
  {Zambelli}}]{Gies:2013pma}%
  \BibitemOpen
  \bibfield  {author} {\bibinfo {author} {\bibfnamefont {H.}~\bibnamefont
  {Gies}}, \bibinfo {author} {\bibfnamefont {S.}~\bibnamefont {Rechenberger}},
  \bibinfo {author} {\bibfnamefont {M.~M.}\ \bibnamefont {Scherer}}, \ and\
  \bibinfo {author} {\bibfnamefont {L.}~\bibnamefont {Zambelli}},\ }\href
  {\doibase 10.1140/epjc/s10052-013-2652-y} {\bibfield  {journal} {\bibinfo
  {journal} {Eur. Phys. J.}\ }\textbf {\bibinfo {volume} {C73}},\ \bibinfo
  {pages} {2652} (\bibinfo {year} {2013})},\ \Eprint
  {http://arxiv.org/abs/1306.6508} {arXiv:1306.6508 [hep-th]} \BibitemShut
  {NoStop}%
%%CITATION = ARXIV:1306.6508;%%
\bibitem [{\citenamefont {Eichhorn}\ and\ \citenamefont
  {Scherer}(2014)}]{Eichhorn:2014qka}%
  \BibitemOpen
  \bibfield  {author} {\bibinfo {author} {\bibfnamefont {A.}~\bibnamefont
  {Eichhorn}}\ and\ \bibinfo {author} {\bibfnamefont {M.~M.}\ \bibnamefont
  {Scherer}},\ }\href {\doibase 10.1103/PhysRevD.90.025023} {\bibfield
  {journal} {\bibinfo  {journal} {Phys. Rev.}\ }\textbf {\bibinfo {volume}
  {D90}},\ \bibinfo {pages} {025023} (\bibinfo {year} {2014})},\ \Eprint
  {http://arxiv.org/abs/1404.5962} {arXiv:1404.5962 [hep-ph]} \BibitemShut
  {NoStop}%
%%CITATION = ARXIV:1404.5962;%%
\bibitem [{\citenamefont {Jakovac}\ \emph
  {et~al.}(2016{\natexlab{a}})\citenamefont {Jakovac}, \citenamefont
  {Kaposvari},\ and\ \citenamefont {Patkos}}]{Jakovac:2015iqa}%
  \BibitemOpen
  \bibfield  {author} {\bibinfo {author} {\bibfnamefont {A.}~\bibnamefont
  {Jakovac}}, \bibinfo {author} {\bibfnamefont {I.}~\bibnamefont {Kaposvari}},
  \ and\ \bibinfo {author} {\bibfnamefont {A.}~\bibnamefont {Patkos}},\
  }\bibfield  {booktitle} {\emph {\bibinfo {booktitle} {{Proceedings, Gribov-85
  Memorial Workshop on Theoretical Physics of XXI Century: Chernogolovka,
  Russia, June 7-20, 2015}}},\ }\href {\doibase 10.1142/S0217751X16450421}
  {\bibfield  {journal} {\bibinfo  {journal} {Int. J. Mod. Phys.}\ }\textbf
  {\bibinfo {volume} {A31}},\ \bibinfo {pages} {1645042} (\bibinfo {year}
  {2016}{\natexlab{a}})},\ \Eprint {http://arxiv.org/abs/1510.05782}
  {arXiv:1510.05782 [hep-th]} \BibitemShut {NoStop}%
%%CITATION = ARXIV:1510.05782;%%
\bibitem [{\citenamefont {Jakovac}\ \emph
  {et~al.}(2016{\natexlab{b}})\citenamefont {Jakovac}, \citenamefont
  {Kaposvari},\ and\ \citenamefont {Patkos}}]{Jakovac:2015kka}%
  \BibitemOpen
  \bibfield  {author} {\bibinfo {author} {\bibfnamefont {A.}~\bibnamefont
  {Jakovac}}, \bibinfo {author} {\bibfnamefont {I.}~\bibnamefont {Kaposvari}},
  \ and\ \bibinfo {author} {\bibfnamefont {A.}~\bibnamefont {Patkos}},\ }\href
  {\doibase 10.1142/S0217732317500110} {\bibfield  {journal} {\bibinfo
  {journal} {Mod. Phys. Lett.}\ }\textbf {\bibinfo {volume} {A32}},\ \bibinfo
  {pages} {1750011} (\bibinfo {year} {2016}{\natexlab{b}})},\ \Eprint
  {http://arxiv.org/abs/1508.06774} {arXiv:1508.06774 [hep-th]} \BibitemShut
  {NoStop}%
%%CITATION = ARXIV:1508.06774;%%
\bibitem [{\citenamefont {Vacca}\ and\ \citenamefont
  {Zambelli}(2015)}]{Vacca:2015nta}%
  \BibitemOpen
  \bibfield  {author} {\bibinfo {author} {\bibfnamefont {G.~P.}\ \bibnamefont
  {Vacca}}\ and\ \bibinfo {author} {\bibfnamefont {L.}~\bibnamefont
  {Zambelli}},\ }\href {\doibase 10.1103/PhysRevD.91.125003} {\bibfield
  {journal} {\bibinfo  {journal} {Phys. Rev.}\ }\textbf {\bibinfo {volume}
  {D91}},\ \bibinfo {pages} {125003} (\bibinfo {year} {2015})},\ \Eprint
  {http://arxiv.org/abs/1503.09136} {arXiv:1503.09136 [hep-th]} \BibitemShut
  {NoStop}%
%%CITATION = ARXIV:1503.09136;%%
\bibitem [{\citenamefont {Borchardt}\ \emph {et~al.}(2016)\citenamefont
  {Borchardt}, \citenamefont {Gies},\ and\ \citenamefont
  {Sondenheimer}}]{Borchardt:2016xju}%
  \BibitemOpen
  \bibfield  {author} {\bibinfo {author} {\bibfnamefont {J.}~\bibnamefont
  {Borchardt}}, \bibinfo {author} {\bibfnamefont {H.}~\bibnamefont {Gies}}, \
  and\ \bibinfo {author} {\bibfnamefont {R.}~\bibnamefont {Sondenheimer}},\
  }\href {\doibase 10.1140/epjc/s10052-016-4300-9} {\bibfield  {journal}
  {\bibinfo  {journal} {Eur. Phys. J.}\ }\textbf {\bibinfo {volume} {C76}},\
  \bibinfo {pages} {472} (\bibinfo {year} {2016})},\ \Eprint
  {http://arxiv.org/abs/1603.05861} {arXiv:1603.05861 [hep-ph]} \BibitemShut
  {NoStop}%
%%CITATION = ARXIV:1603.05861;%%
\bibitem [{\citenamefont {Jakovác}\ \emph {et~al.}(2017)\citenamefont
  {Jakovác}, \citenamefont {Kaposvári},\ and\ \citenamefont
  {Patkós}}]{Jakovac:2017nsi}%
  \BibitemOpen
  \bibfield  {author} {\bibinfo {author} {\bibfnamefont {A.}~\bibnamefont
  {Jakovác}}, \bibinfo {author} {\bibfnamefont {I.}~\bibnamefont
  {Kaposvári}}, \ and\ \bibinfo {author} {\bibfnamefont {A.}~\bibnamefont
  {Patkós}},\ }\href {\doibase 10.1103/PhysRevD.96.076018} {\bibfield
  {journal} {\bibinfo  {journal} {Phys. Rev.}\ }\textbf {\bibinfo {volume}
  {D96}},\ \bibinfo {pages} {076018} (\bibinfo {year} {2017})},\ \Eprint
  {http://arxiv.org/abs/1703.00831} {arXiv:1703.00831 [hep-ph]} \BibitemShut
  {NoStop}%
%%CITATION = ARXIV:1703.00831;%%
\bibitem [{\citenamefont {Gies}\ \emph {et~al.}(2017)\citenamefont {Gies},
  \citenamefont {Sondenheimer},\ and\ \citenamefont
  {Warschinke}}]{Gies:2017zwf}%
  \BibitemOpen
  \bibfield  {author} {\bibinfo {author} {\bibfnamefont {H.}~\bibnamefont
  {Gies}}, \bibinfo {author} {\bibfnamefont {R.}~\bibnamefont {Sondenheimer}},
  \ and\ \bibinfo {author} {\bibfnamefont {M.}~\bibnamefont {Warschinke}},\
  }\href {\doibase 10.1140/epjc/s10052-017-5312-9} {\bibfield  {journal}
  {\bibinfo  {journal} {Eur. Phys. J.}\ }\textbf {\bibinfo {volume} {C77}},\
  \bibinfo {pages} {743} (\bibinfo {year} {2017})},\ \Eprint
  {http://arxiv.org/abs/1707.04394} {arXiv:1707.04394 [hep-ph]} \BibitemShut
  {NoStop}%
%%CITATION = ARXIV:1707.04394;%%
\bibitem [{\citenamefont {Gies}\ and\ \citenamefont
  {Sondenheimer}(2017)}]{Gies:2017ajd}%
  \BibitemOpen
  \bibfield  {author} {\bibinfo {author} {\bibfnamefont {H.}~\bibnamefont
  {Gies}}\ and\ \bibinfo {author} {\bibfnamefont {R.}~\bibnamefont
  {Sondenheimer}},\ }in\ \href
  {http://inspirehep.net/record/1616100/files/arXiv:1708.04305.pdf} {\emph
  {\bibinfo {booktitle} {{Higgs cosmology Newport Pagnell, Buckinghamshire, UK,
  March 27-28, 2017}}}}\ (\bibinfo {year} {2017})\ \Eprint
  {http://arxiv.org/abs/1708.04305} {arXiv:1708.04305 [hep-ph]} \BibitemShut
  {NoStop}%
%%CITATION = ARXIV:1708.04305;%%
\bibitem [{\citenamefont {Litim}(2000)}]{Litim:2000ci}%
  \BibitemOpen
  \bibfield  {author} {\bibinfo {author} {\bibfnamefont {D.~F.}\ \bibnamefont
  {Litim}},\ }\href {\doibase 10.1016/S0370-2693(00)00748-6} {\bibfield
  {journal} {\bibinfo  {journal} {Phys. Lett.}\ }\textbf {\bibinfo {volume}
  {B486}},\ \bibinfo {pages} {92} (\bibinfo {year} {2000})},\ \Eprint
  {http://arxiv.org/abs/hep-th/0005245} {arXiv:hep-th/0005245 [hep-th]}
  \BibitemShut {NoStop}%
%%CITATION = HEP-TH/0005245;%%
\bibitem [{\citenamefont {Litim}(2001)}]{Litim:2001up}%
  \BibitemOpen
  \bibfield  {author} {\bibinfo {author} {\bibfnamefont {D.~F.}\ \bibnamefont
  {Litim}},\ }\href {\doibase 10.1103/PhysRevD.64.105007} {\bibfield  {journal}
  {\bibinfo  {journal} {Phys. Rev.}\ }\textbf {\bibinfo {volume} {D64}},\
  \bibinfo {pages} {105007} (\bibinfo {year} {2001})},\ \Eprint
  {http://arxiv.org/abs/hep-th/0103195} {arXiv:hep-th/0103195 [hep-th]}
  \BibitemShut {NoStop}%
%%CITATION = HEP-TH/0103195;%%
\bibitem [{\citenamefont {Pawlowski}\ and\ \citenamefont
  {Rennecke}(2014)}]{Pawlowski:2014zaa}%
  \BibitemOpen
  \bibfield  {author} {\bibinfo {author} {\bibfnamefont {J.~M.}\ \bibnamefont
  {Pawlowski}}\ and\ \bibinfo {author} {\bibfnamefont {F.}~\bibnamefont
  {Rennecke}},\ }\href {\doibase 10.1103/PhysRevD.90.076002} {\bibfield
  {journal} {\bibinfo  {journal} {Phys. Rev.}\ }\textbf {\bibinfo {volume}
  {D90}},\ \bibinfo {pages} {076002} (\bibinfo {year} {2014})},\ \Eprint
  {http://arxiv.org/abs/1403.1179} {arXiv:1403.1179 [hep-ph]} \BibitemShut
  {NoStop}%
%%CITATION = ARXIV:1403.1179;%%
\bibitem [{\citenamefont {Coleman}\ and\ \citenamefont
  {Weinberg}(1973)}]{Coleman:1973jx}%
  \BibitemOpen
  \bibfield  {author} {\bibinfo {author} {\bibfnamefont {S.~R.}\ \bibnamefont
  {Coleman}}\ and\ \bibinfo {author} {\bibfnamefont {E.~J.}\ \bibnamefont
  {Weinberg}},\ }\href {\doibase 10.1103/PhysRevD.7.1888} {\bibfield  {journal}
  {\bibinfo  {journal} {Phys. Rev.}\ }\textbf {\bibinfo {volume} {D7}},\
  \bibinfo {pages} {1888} (\bibinfo {year} {1973})}\BibitemShut {NoStop}%
%%CITATION = PHRVA,D7,1888;%%
\bibitem [{\citenamefont {Jackiw}(1974)}]{Jackiw:1974cv}%
  \BibitemOpen
  \bibfield  {author} {\bibinfo {author} {\bibfnamefont {R.}~\bibnamefont
  {Jackiw}},\ }\href {\doibase 10.1103/PhysRevD.9.1686} {\bibfield  {journal}
  {\bibinfo  {journal} {Phys. Rev.}\ }\textbf {\bibinfo {volume} {D9}},\
  \bibinfo {pages} {1686} (\bibinfo {year} {1974})}\BibitemShut {NoStop}%
%%CITATION = PHRVA,D9,1686;%%
\bibitem [{\citenamefont {Morris}(1996)}]{Morris:1996nx}%
  \BibitemOpen
  \bibfield  {author} {\bibinfo {author} {\bibfnamefont {T.~R.}\ \bibnamefont
  {Morris}},\ }\href {\doibase 10.1103/PhysRevLett.77.1658} {\bibfield
  {journal} {\bibinfo  {journal} {Phys. Rev. Lett.}\ }\textbf {\bibinfo
  {volume} {77}},\ \bibinfo {pages} {1658} (\bibinfo {year} {1996})},\ \Eprint
  {http://arxiv.org/abs/hep-th/9601128} {arXiv:hep-th/9601128 [hep-th]}
  \BibitemShut {NoStop}%
%%CITATION = HEP-TH/9601128;%%
\bibitem [{\citenamefont {O'Dwyer}\ and\ \citenamefont
  {Osborn}(2008)}]{ODwyer:2007brp}%
  \BibitemOpen
  \bibfield  {author} {\bibinfo {author} {\bibfnamefont {J.}~\bibnamefont
  {O'Dwyer}}\ and\ \bibinfo {author} {\bibfnamefont {H.}~\bibnamefont
  {Osborn}},\ }\href {\doibase 10.1016/j.aop.2007.10.005} {\bibfield  {journal}
  {\bibinfo  {journal} {Annals Phys.}\ }\textbf {\bibinfo {volume} {323}},\
  \bibinfo {pages} {1859} (\bibinfo {year} {2008})},\ \Eprint
  {http://arxiv.org/abs/0708.2697} {arXiv:0708.2697 [hep-th]} \BibitemShut
  {NoStop}%
%%CITATION = ARXIV:0708.2697;%%
\bibitem [{\citenamefont {Bridle}\ and\ \citenamefont
  {Morris}(2016)}]{Bridle:2016nsu}%
  \BibitemOpen
  \bibfield  {author} {\bibinfo {author} {\bibfnamefont {I.~H.}\ \bibnamefont
  {Bridle}}\ and\ \bibinfo {author} {\bibfnamefont {T.~R.}\ \bibnamefont
  {Morris}},\ }\href@noop {} {\  (\bibinfo {year} {2016})},\ \Eprint
  {http://arxiv.org/abs/1605.06075} {arXiv:1605.06075 [hep-th]} \BibitemShut
  {NoStop}%
%%CITATION = ARXIV:1605.06075;%%
\bibitem [{\citenamefont {Borchardt}\ and\ \citenamefont
  {Knorr}(2015)}]{Borchardt:2015rxa}%
  \BibitemOpen
  \bibfield  {author} {\bibinfo {author} {\bibfnamefont {J.}~\bibnamefont
  {Borchardt}}\ and\ \bibinfo {author} {\bibfnamefont {B.}~\bibnamefont
  {Knorr}},\ }\href {\doibase 10.1103/PhysRevD.91.105011} {\bibfield  {journal}
  {\bibinfo  {journal} {Phys. Rev.}\ }\textbf {\bibinfo {volume} {D91}},\
  \bibinfo {pages} {105011} (\bibinfo {year} {2015})},\ \Eprint
  {http://arxiv.org/abs/1502.07511} {arXiv:1502.07511 [hep-th]} \BibitemShut
  {NoStop}%
%%CITATION = ARXIV:1502.07511;%%
\bibitem [{\citenamefont {Borchardt}\ and\ \citenamefont
  {Knorr}(2016)}]{Borchardt:2016pif}%
  \BibitemOpen
  \bibfield  {author} {\bibinfo {author} {\bibfnamefont {J.}~\bibnamefont
  {Borchardt}}\ and\ \bibinfo {author} {\bibfnamefont {B.}~\bibnamefont
  {Knorr}},\ }\href {\doibase 10.1103/PhysRevD.94.025027} {\bibfield  {journal}
  {\bibinfo  {journal} {Phys. Rev.}\ }\textbf {\bibinfo {volume} {D94}},\
  \bibinfo {pages} {025027} (\bibinfo {year} {2016})},\ \Eprint
  {http://arxiv.org/abs/1603.06726} {arXiv:1603.06726 [hep-th]} \BibitemShut
  {NoStop}%
%%CITATION = ARXIV:1603.06726;%%
\bibitem [{\citenamefont {Borchardt}\ and\ \citenamefont
  {Eichhorn}(2016)}]{Borchardt:2016kco}%
  \BibitemOpen
  \bibfield  {author} {\bibinfo {author} {\bibfnamefont {J.}~\bibnamefont
  {Borchardt}}\ and\ \bibinfo {author} {\bibfnamefont {A.}~\bibnamefont
  {Eichhorn}},\ }\href {\doibase 10.1103/PhysRevE.94.042105} {\bibfield
  {journal} {\bibinfo  {journal} {Phys. Rev.}\ }\textbf {\bibinfo {volume}
  {E94}},\ \bibinfo {pages} {042105} (\bibinfo {year} {2016})},\ \Eprint
  {http://arxiv.org/abs/1606.07449} {arXiv:1606.07449 [cond-mat.stat-mech]}
  \BibitemShut {NoStop}%
%%CITATION = ARXIV:1606.07449;%%
\bibitem [{\citenamefont {Heilmann}\ \emph {et~al.}(2015)\citenamefont
  {Heilmann}, \citenamefont {Hellwig}, \citenamefont {Knorr}, \citenamefont
  {Ansorg},\ and\ \citenamefont {Wipf}}]{Heilmann:2014iga}%
  \BibitemOpen
  \bibfield  {author} {\bibinfo {author} {\bibfnamefont {M.}~\bibnamefont
  {Heilmann}}, \bibinfo {author} {\bibfnamefont {T.}~\bibnamefont {Hellwig}},
  \bibinfo {author} {\bibfnamefont {B.}~\bibnamefont {Knorr}}, \bibinfo
  {author} {\bibfnamefont {M.}~\bibnamefont {Ansorg}}, \ and\ \bibinfo {author}
  {\bibfnamefont {A.}~\bibnamefont {Wipf}},\ }\href {\doibase
  10.1007/JHEP02(2015)109} {\bibfield  {journal} {\bibinfo  {journal} {JHEP}\
  }\textbf {\bibinfo {volume} {02}},\ \bibinfo {pages} {109} (\bibinfo {year}
  {2015})},\ \Eprint {http://arxiv.org/abs/1409.5650} {arXiv:1409.5650
  [hep-th]} \BibitemShut {NoStop}%
%%CITATION = ARXIV:1409.5650;%%
\bibitem [{\citenamefont {Fischer}\ and\ \citenamefont
  {Gies}(2004)}]{Fischer:2004uk}%
  \BibitemOpen
  \bibfield  {author} {\bibinfo {author} {\bibfnamefont {C.~S.}\ \bibnamefont
  {Fischer}}\ and\ \bibinfo {author} {\bibfnamefont {H.}~\bibnamefont {Gies}},\
  }\href {\doibase 10.1088/1126-6708/2004/10/048} {\bibfield  {journal}
  {\bibinfo  {journal} {JHEP}\ }\textbf {\bibinfo {volume} {10}},\ \bibinfo
  {pages} {048} (\bibinfo {year} {2004})},\ \Eprint
  {http://arxiv.org/abs/hep-ph/0408089} {arXiv:hep-ph/0408089 [hep-ph]}
  \BibitemShut {NoStop}%
%%CITATION = HEP-PH/0408089;%%
\bibitem [{\citenamefont {Boyd}(2000)}]{Boyd:ChebyFourier}%
  \BibitemOpen
  \bibfield  {author} {\bibinfo {author} {\bibfnamefont {J.~P.}\ \bibnamefont
  {Boyd}},\ }\href@noop {} {\emph {\bibinfo {title} {{Chebyshev and Fourier
  Spectral Methods}}}},\ \bibinfo {edition} {2nd}\ ed.\ (\bibinfo  {publisher}
  {Dover Publications},\ \bibinfo {year} {2000})\BibitemShut {NoStop}%
\bibitem [{\citenamefont {Robson}\ and\ \citenamefont
  {Prytz}(1993)}]{Robson:1993}%
  \BibitemOpen
  \bibfield  {author} {\bibinfo {author} {\bibfnamefont {R.}~\bibnamefont
  {Robson}}\ and\ \bibinfo {author} {\bibfnamefont {A.}~\bibnamefont {Prytz}},\
  }\href@noop {} {\bibfield  {journal} {\bibinfo  {journal} {Australian Journal
  of Physics}\ }\textbf {\bibinfo {volume} {46}} (\bibinfo {year}
  {1993})}\BibitemShut {NoStop}%
\bibitem [{\citenamefont {Ansorg}\ \emph {et~al.}(2003)\citenamefont {Ansorg},
  \citenamefont {Kleinwachter},\ and\ \citenamefont {Meinel}}]{Ansorg:2003br}%
  \BibitemOpen
  \bibfield  {author} {\bibinfo {author} {\bibfnamefont {M.}~\bibnamefont
  {Ansorg}}, \bibinfo {author} {\bibfnamefont {A.}~\bibnamefont
  {Kleinwachter}}, \ and\ \bibinfo {author} {\bibfnamefont {R.}~\bibnamefont
  {Meinel}},\ }\href {\doibase 10.1051/0004-6361:20030618} {\bibfield
  {journal} {\bibinfo  {journal} {Astron. Astrophys.}\ }\textbf {\bibinfo
  {volume} {405}},\ \bibinfo {pages} {711} (\bibinfo {year} {2003})},\ \Eprint
  {http://arxiv.org/abs/astro-ph/0301173} {arXiv:astro-ph/0301173 [astro-ph]}
  \BibitemShut {NoStop}%
%%CITATION = ASTRO-PH/0301173;%%
\bibitem [{\citenamefont {Macedo}\ and\ \citenamefont
  {Ansorg}(2014)}]{Macedo:2014bfa}%
  \BibitemOpen
  \bibfield  {author} {\bibinfo {author} {\bibfnamefont {R.~P.}\ \bibnamefont
  {Macedo}}\ and\ \bibinfo {author} {\bibfnamefont {M.}~\bibnamefont
  {Ansorg}},\ }\href {\doibase 10.1016/j.jcp.2014.07.040} {\bibfield  {journal}
  {\bibinfo  {journal} {J. Comput. Phys.}\ }\textbf {\bibinfo {volume} {276}},\
  \bibinfo {pages} {357} (\bibinfo {year} {2014})},\ \Eprint
  {http://arxiv.org/abs/1402.7343} {arXiv:1402.7343 [physics.comp-ph]}
  \BibitemShut {NoStop}%
%%CITATION = ARXIV:1402.7343;%%
\bibitem [{\citenamefont {Morris}(1998)}]{Morris:1998da}%
  \BibitemOpen
  \bibfield  {author} {\bibinfo {author} {\bibfnamefont {T.~R.}\ \bibnamefont
  {Morris}},\ }\bibfield  {booktitle} {\emph {\bibinfo {booktitle}
  {{Nonperturbative QCD: Structure of the QCD vacuum}}},\ }\href {\doibase
  10.1143/PTPS.131.395} {\bibfield  {journal} {\bibinfo  {journal} {Prog.
  Theor. Phys. Suppl.}\ }\textbf {\bibinfo {volume} {131}},\ \bibinfo {pages}
  {395} (\bibinfo {year} {1998})},\ \Eprint
  {http://arxiv.org/abs/hep-th/9802039} {arXiv:hep-th/9802039 [hep-th]}
  \BibitemShut {NoStop}%
%%CITATION = HEP-TH/9802039;%%
\bibitem [{\citenamefont {Morris}(1994{\natexlab{b}})}]{Morris:1994ki}%
  \BibitemOpen
  \bibfield  {author} {\bibinfo {author} {\bibfnamefont {T.~R.}\ \bibnamefont
  {Morris}},\ }\href {\doibase 10.1016/0370-2693(94)90700-5} {\bibfield
  {journal} {\bibinfo  {journal} {Phys. Lett.}\ }\textbf {\bibinfo {volume}
  {B334}},\ \bibinfo {pages} {355} (\bibinfo {year} {1994}{\natexlab{b}})},\
  \Eprint {http://arxiv.org/abs/hep-th/9405190} {arXiv:hep-th/9405190 [hep-th]}
  \BibitemShut {NoStop}%
%%CITATION = HEP-TH/9405190;%%
\bibitem [{\citenamefont {Codello}(2012)}]{Codello:2012sc}%
  \BibitemOpen
  \bibfield  {author} {\bibinfo {author} {\bibfnamefont {A.}~\bibnamefont
  {Codello}},\ }\href {\doibase 10.1088/1751-8113/45/46/465006} {\bibfield
  {journal} {\bibinfo  {journal} {J. Phys.}\ }\textbf {\bibinfo {volume}
  {A45}},\ \bibinfo {pages} {465006} (\bibinfo {year} {2012})},\ \Eprint
  {http://arxiv.org/abs/1204.3877} {arXiv:1204.3877 [hep-th]} \BibitemShut
  {NoStop}%
%%CITATION = ARXIV:1204.3877;%%
\bibitem [{\citenamefont {Codello}\ and\ \citenamefont
  {D'Odorico}(2013)}]{Codello:2012ec}%
  \BibitemOpen
  \bibfield  {author} {\bibinfo {author} {\bibfnamefont {A.}~\bibnamefont
  {Codello}}\ and\ \bibinfo {author} {\bibfnamefont {G.}~\bibnamefont
  {D'Odorico}},\ }\href {\doibase 10.1103/PhysRevLett.110.141601} {\bibfield
  {journal} {\bibinfo  {journal} {Phys. Rev. Lett.}\ }\textbf {\bibinfo
  {volume} {110}},\ \bibinfo {pages} {141601} (\bibinfo {year} {2013})},\
  \Eprint {http://arxiv.org/abs/1210.4037} {arXiv:1210.4037 [hep-th]}
  \BibitemShut {NoStop}%
%%CITATION = ARXIV:1210.4037;%%
\bibitem [{\citenamefont {Lee}\ and\ \citenamefont
  {Weisberger}(1974)}]{Lee:1974gua}%
  \BibitemOpen
  \bibfield  {author} {\bibinfo {author} {\bibfnamefont {B.~W.}\ \bibnamefont
  {Lee}}\ and\ \bibinfo {author} {\bibfnamefont {W.~I.}\ \bibnamefont
  {Weisberger}},\ }\href {\doibase 10.1103/PhysRevD.10.2530} {\bibfield
  {journal} {\bibinfo  {journal} {Phys. Rev.}\ }\textbf {\bibinfo {volume}
  {D10}},\ \bibinfo {pages} {2530} (\bibinfo {year} {1974})}\BibitemShut
  {NoStop}%
%%CITATION = PHRVA,D10,2530;%%
\bibitem [{\citenamefont {Zanusso}\ \emph {et~al.}(2010)\citenamefont
  {Zanusso}, \citenamefont {Zambelli}, \citenamefont {Vacca},\ and\
  \citenamefont {Percacci}}]{Zanusso:2009bs}%
  \BibitemOpen
  \bibfield  {author} {\bibinfo {author} {\bibfnamefont {O.}~\bibnamefont
  {Zanusso}}, \bibinfo {author} {\bibfnamefont {L.}~\bibnamefont {Zambelli}},
  \bibinfo {author} {\bibfnamefont {G.~P.}\ \bibnamefont {Vacca}}, \ and\
  \bibinfo {author} {\bibfnamefont {R.}~\bibnamefont {Percacci}},\ }\href
  {\doibase 10.1016/j.physletb.2010.04.043} {\bibfield  {journal} {\bibinfo
  {journal} {Phys. Lett.}\ }\textbf {\bibinfo {volume} {B689}},\ \bibinfo
  {pages} {90} (\bibinfo {year} {2010})},\ \Eprint
  {http://arxiv.org/abs/0904.0938} {arXiv:0904.0938 [hep-th]} \BibitemShut
  {NoStop}%
%%CITATION = ARXIV:0904.0938;%%
\bibitem [{\citenamefont {Vacca}\ and\ \citenamefont
  {Zanusso}(2010)}]{Vacca:2010mj}%
  \BibitemOpen
  \bibfield  {author} {\bibinfo {author} {\bibfnamefont {G.~P.}\ \bibnamefont
  {Vacca}}\ and\ \bibinfo {author} {\bibfnamefont {O.}~\bibnamefont
  {Zanusso}},\ }\href {\doibase 10.1103/PhysRevLett.105.231601} {\bibfield
  {journal} {\bibinfo  {journal} {Phys. Rev. Lett.}\ }\textbf {\bibinfo
  {volume} {105}},\ \bibinfo {pages} {231601} (\bibinfo {year} {2010})},\
  \Eprint {http://arxiv.org/abs/1009.1735} {arXiv:1009.1735 [hep-th]}
  \BibitemShut {NoStop}%
%%CITATION = ARXIV:1009.1735;%%
\bibitem [{\citenamefont {Braun}\ \emph {et~al.}(2014)\citenamefont {Braun},
  \citenamefont {Fister}, \citenamefont {Pawlowski},\ and\ \citenamefont
  {Rennecke}}]{Braun:2014ata}%
  \BibitemOpen
  \bibfield  {author} {\bibinfo {author} {\bibfnamefont {J.}~\bibnamefont
  {Braun}}, \bibinfo {author} {\bibfnamefont {L.}~\bibnamefont {Fister}},
  \bibinfo {author} {\bibfnamefont {J.~M.}\ \bibnamefont {Pawlowski}}, \ and\
  \bibinfo {author} {\bibfnamefont {F.}~\bibnamefont {Rennecke}},\ }\href@noop
  {} {\  (\bibinfo {year} {2014})},\ \Eprint {http://arxiv.org/abs/1412.1045}
  {arXiv:1412.1045 [hep-ph]} \BibitemShut {NoStop}%
%%CITATION = ARXIV:1412.1045;%%
\bibitem [{\citenamefont {Knorr}(2016)}]{Knorr:2016sfs}%
  \BibitemOpen
  \bibfield  {author} {\bibinfo {author} {\bibfnamefont {B.}~\bibnamefont
  {Knorr}},\ }\href {\doibase 10.1103/PhysRevB.94.245102} {\bibfield  {journal}
  {\bibinfo  {journal} {Phys. Rev.}\ }\textbf {\bibinfo {volume} {B94}},\
  \bibinfo {pages} {245102} (\bibinfo {year} {2016})},\ \Eprint
  {http://arxiv.org/abs/1609.03824} {arXiv:1609.03824 [cond-mat.str-el]}
  \BibitemShut {NoStop}%
%%CITATION = ARXIV:1609.03824;%%
\end{thebibliography}%

\end{document}